\documentclass[fleqn,usenatbib]{mnras}

\usepackage{newtxtext,newtxmath}
\usepackage[T1]{fontenc}

\DeclareRobustCommand{\VAN}[3]{#2}
\let\VANthebibliography\thebibliography
\def\thebibliography{\DeclareRobustCommand{\VAN}[3]{##3}\VANthebibliography}

\usepackage{graphicx}	% Including figure files
\usepackage{amsmath}	% Advanced maths commands
\usepackage{booktabs}
\usepackage{textgreek}
\usepackage{threeparttable}
\def    \apjl  		{\rm {ApJL}}
\def    \apj  		{\rm {ApJ}}
\def    \mnras  	{\rm {MNRAS}}
\def    \araa  		{\rm {ARA\& A}}
\def    \apjl  		{\rm {ApJL}}

\def	\cm		{\,{\rm {cm}}}
\def	\K		{\,{\rm K}}

\def \bea {\begin{eqnarray}}
\def \ena {\end{eqnarray}}                  

\def    \ba     {\bf  a}
\def    \bB     {\bf  B}

\def	\ba	{{\bf a}}

\def	\bB	{{\bf B}}

\def	\bJ	{{\bf J}}

\def	\cm	{\,{\rm cm}}

\def	\erg	{\,{\rm erg}}

\def	\K	{{\rm K}}

\def	\pc	{\,{\rm pc}}

\def	\s	{\,{\rm s}}

\def	\ba			{\boldsymbol{a}}

\DeclareUnicodeCharacter{0301}{\'{e}}

\title[Grain alignment and polarization in evolved stars]{Numerical modeling of thermal dust polarization from aligned grains in the envelope of evolved stars with updated POLARIS}

\author[Truong et al.]{Bao Truong,$^{1,2}$\thanks{Email: baotruong@kasi.re.kr}
Thiem Hoang,$^{1,2}$\thanks{Email: thiemhoang@kasi.re.kr}
Nguyen Chau Giang,$^{1,2}$
Pham Ngoc Diep,$^{3}$
\newauthor
Dieu D. Nguyen,$^{4}$
Nguyen Bich Ngoc$^{3,5}$
\\
$^{1}$Korea Astronomy and Space Science Institute, 776 Daedeokdae-ro, Yuseong-gu, Daejeon 34055, Republic of Korea\\
$^{2}$Department of Astronomy and Space Science, University of Science and Technology, 217 Gajeong-ro, Yuseong-gu, Daejeon 34113, Republic of Korea\\
$^{3}$Department of Astrophysics, Vietnam National Space Center, Vietnam Academy of Science and Technology, 18 Hoang Quoc Viet, Hanoi, Vietnam\\
$^{4}$Université de Lyon 1, Ens de Lyon, CNRS, Centre de Recherche Astrophysique de Lyon (CRAL) UMR5574, F-69230 Saint-Genis-Laval, France\\
$^{5}$Graduate University of Science and Technology, Vietnam Academy of Science and Technology, 18 Hoang Quoc Viet, Hanoi, Vietnam
}
\date{Accepted 2024 December 03. Received 2024 November 27; in original form 2023 August 22}

\pubyear{2024}

\begin{document}
\label{firstpage}
\pagerange{\pageref{firstpage}--\pageref{lastpage}}
\maketitle

% Abstract of the paper
\begin{abstract}
Magnetic fields are thought to influence the formation and evolution of circumstellar envelopes around evolved stars. Thermal dust polarization from aligned grains is a promising tool for probing magnetic fields and dust properties in these environments; however, a quantitative study on the dependence of thermal dust polarization on the physical properties of dust and magnetic fields for these circumstellar environments is still lacking. In this paper, we first perform the numerical modeling of thermal dust polarization in the IK Tau envelope using the magnetically enhanced radiative torque (MRAT) alignment mechanism implemented in our updated POLARIS code, accounting for the effect of grain drift relative to the gas. Despite experiencing grain drift and high gas density $n_{\rm gas} > 10^6\,\rm  cm^{-3}$, the minimum grain size required for efficient MRAT alignment of silicate grains is $\sim 0.007 - 0.05\,\rm\mu m$ due to strong stellar radiation fields. Ordinary paramagnetic grains can achieve perfect alignment by MRAT in the inner envelope of $r < 500\,\rm au$ due to stronger magnetic fields of $B\sim10$ mG - 1G, producing the polarization degree of $\sim10\%$. The polarization degree can be enhanced to $\sim20-40\%$ for superparamagnetic grains with embedded iron inclusions. The magnetic field geometry affects the resulting polarization degree due to the projection effect. We investigate the effect of rotational disruption by RATs (RAT-D) and find that the RAT-D effect decreases the dust polarization degree due to the decrease in the maximum grain size. Our modeling results motivate further observational studies at far-infrared/sub-millimeter to constrain the properties of magnetic fields and dust in evolved star's envelopes.

\end{abstract}

% Select between one and six entries from the list of approved keywords.
% Don't make up new ones.

%%%%%%%%%%%%%%%%%%%%%%%%%%%%%%%%%%%%%%%%%%%%%%%%%%
\section{Introduction}
Magnetic fields (B-fields) are essential for understanding various astrophysical processes in the Universe. In the late-stage evolution of low- and intermediate-mass stars (i.e., $M_{\ast} \sim 0.5 - 8\,M_{\odot}$, see \citealt{Hofner2018}) as Asymptotic Giant Branch (AGB), magnetic fields likely play a crucial role in the mass-loss mechanism and the formation of circumstellar envelopes (CSEs) (\citealt{Soker2002}; \citealt{Thirumalai2013}; \citealt{Hofner2018}). Theoretical models showed that the oscillating Alfvén waves by stellar magnetic fields could efficiently levitate stellar materials above the atmosphere and, later on, stellar winds are produced by radiation pressure on the newly formed grains (\citealt{Airapetian2000}; \citealt{Airapetian2010}; \citealt{Cranmer2011}). Additionally, magnetic fields in a close binary system during post-AGB phases could shape the dusty winds from a spherical structure to a bipolar structure before being exploded as planetary nebulae (PNe) (e.g., \citealt{Garcia1999}; \citealt{Vlemmings2006}; \citealt{Nordhaus2007}), significantly contributing to explaining the ``breaking asymmetry" as shown in pre-PNe/PNe observations (\citealt{Sabin2007}; \citealt{Sabin2015}). 

Magnetic fields in the envelopes of AGB/post-AGB stars have been mainly studied through the Zeeman splitting of maser lines. The polarized maser emissions from molecules such as $\rm SiO, H_2O$, and OH were observed in the compact regions of AGB envelopes (\citealt{Boboltz2005}; \citealt{Ferreira2013}) and PNe (\citealt{Vlemmings2006}). The profiles of polarized maser lines could be used for measuring the magnetic field strength and constraining the configuration of magnetic fields in CSEs (see in \citealt{Vlemmings2019}). However, the maser lines strongly depend on the radiative or collisional pumping in the local very dense clumps in the envelopes, which resolve small portions of AGB outflows and only reveal small-scale components of the magnetic fields (\citealt{Soker2002}; \citealt{Vlemmings2019}). Also, the stellar magnetic fields could be measured via non-maser molecular line emissions by the Goldreich-Kylafis effect (\citealt{Vlemmings2012}; \citealt{Huang2020}); however, there are uncertainties in the polarization direction, which could either be parallel or perpendicular to the magnetic field lines (\citealt{Kylafis1983}). Therefore, it is essential to measure magnetic fields on a global scale to understand their effects on stellar winds in the evolved star's envelopes.  

Polarized thermal dust emission from magnetically aligned grains is expected to be a powerful tool for mapping magnetic fields in large-scale CSEs of evolved stars (see a review of \citealt{Scicluna2020}). Following the principle of magnetic alignment of spinning grains with their longest axes perpendicular to the local magnetic fields, the magnetic field can be mapped from thermal dust polarization (see \citealt{Hildebrand1988}). Thermal dust polarization in post-AGB and PNe objects has been observed at sub-millimeter (sub-mm) by the James Clerk Maxwell Telescope (JCMT), Combined Array for Research in Millimeter-wave Astronomy (CARMA), and Atacama Large Millimeter Array (ALMA) such as NGC 7027, NGC 6537, NGC 6302 (e.g., \citealt{Greaves2002}; \citealt{Sabin2007}) and OH 231.8+4.2 (e.g., \citealt{Sabin2015}; \citealt{Sabin2020}), revealing different magnetic field geometries and probing their roles in shaping these nebulae. \cite{Vlemmings2017} first observed thermal dust polarization by aligned grains within the evolved star's envelopes as VY CMa by ALMA at 178 GHz. \cite{Andersson2022} observed the polarized thermal emission from carbonaceous grains at far-infrared (far-IR) wavelengths in the envelope of IRC + 10216 by the Stratospheric Observatory for Infrared Astronomy (SOFIA)/High-resolution Airborne Wideband Camera-plus (HAWC+).

Furthermore, thermal dust polarization is vital for exploring dust properties in AGB/post-AGB envelopes. Thermal dust polarization at far-IR/sub-mm wavelengths opens the possibility of trace large grains of $a > 0.1\,\rm\mu m$ (e.g., \citealt{Vlemmings2017}; \citealt{Andersson2022}), while optical/near-IR polarization by scattered light could characterize the existence of smaller grains of $a < 0.1\,\rm\mu m$ (e.g., \citealt{Khouri2016}; \citealt{Adam2019}). In addition, the properties of dust polarization reveal the physical properties of circumstellar dust in AGB envelopes, e.g., grain shape, size distribution, chemical compositions, and their porosity (see \citealt{Hensley2021}), which extends our knowledge of dust enrichment in the interstellar medium (ISM) through stellar ejecta (see the review of \citealt{Tielens2005} and \citealt{DeBeck2019}). However, a quantitative model of thermal dust polarization induced by aligned grains is still lacking in this circumstellar environment, which prevents the use of dust polarimetry as a powerful diagnostic of circumstellar dust and magnetic fields.

The modern theory of grain alignment implies that the alignment process of dust grains, in general, includes two main stages: (1) the alignment of the grain axis of maximum moment of inertia, $\bf a_1$, with the grain angular momentum, $\bf J$, (so-called internal alignment), and (2) the alignment of the angular momentum $\bf J$ with a preferred direction in space such as radiation fields $\bf k$ or magnetic fields $\bf B$ (so-called external alignment) (see, e.g., \citealt{Hoang2022}). The former process has been studied extensively for interstellar grains \citep{Purcell1979} and grains in protostellar environments \citep{Hoang2022}. For the latter, it is recently established that the leading theory of grain alignment is based on RAdiative Torques (hereafter RATs, see \citealt{Dolginov1976}; \citealt{Draine1996}; \citealt{Lazarian2007}). The RAT theory has been developed with quantitative predictions for various environmental conditions (\citealt{Hoang2014}) and dust compositions (\citealt{Hoang2016a}) and successfully tested with observations from the diffuse ISM and molecular clouds (MCs) (see reviews of \citealt{Andersson2015, Tram2022}). The RAT theory has also been advanced by unifying the effects of RATs and magnetic relaxation, which is known as the Magnetically Enhanced RAdiative Torque (MRAT) mechanism (\citealt{Lazarian2008}; \citealt{Hoang2016a}). Numerical calculations for grain alignment by the MRAT theory in \cite{Hoang2016a} demonstrated that grains with embedded iron clusters can achieve perfect magnetic alignment. 

In addition, for an environment close to an intense radiation source such as AGB envelopes and PNe, dust properties can be modified by a mechanism called RAdiative Torque Disruption (hereafter RAT-D), first introduced by \cite{Hoang2019}. The basic idea is that the induced RATs from strong radiation fields can spin up large grains to suprathermal rotation so that grains would be disrupted into smaller fragments when the centrifugal stress exceeds the binding energy of the grain material. As a result, the RAT-D mechanism reduces the abundance of large grains in local environments, constraining the maximum size at which grains can survive by RAT-D and modifying the original grain size distribution. The effects of RAT-D mechanism on the dust properties and observed extinction/polarization were numerically studied for different astrophysical environments such as cosmic transients (\citealt{Hoang2020}; \citealt{Giang2020}; \citealt{Giang2022Blanet}), massive protostars (\citealt{Hoang2021}), and in the evolved star's envelopes (\citealt{Tram2020}; \citealt{Truong2022}). 

Numerical modeling by \cite{Lee2020} showed that the RAT-D mechanism reduces the degree of thermal dust polarization due to the removal of large grains near intense radiation sources. The evidence of RAT-D was found in polarimetric observations in star-forming MCs (e.g., \citealt{Ngoc2021}; \citealt{Tram2021_Ophiuchi}; \citealt{Tram2021_30Doradus}; \citealt{Thuong2022}). Yet, the modeling in \cite{Lee2020} did not consider the effect of magnetic fields and radiative transfer, which is not applicable for CSEs with complex magnetic field geometry. An accurate model of thermal dust polarization from aligned grains in the envelopes of evolved stars can be achieved only when taking into account both grain alignment and disruption induced by RATs, combined with the realistic magnetic field configuration and radiative transfer. 

\cite{Reissl2016} first developed the POLArized RadIation Simulator (POLARIS) - a three-dimensional simulation code describing multi-wavelength polarization in local environments through the combination of the basic RAT alignment physics and the radiative transfer of dust emission and polarization. The code was successfully used for modeling the RAT alignment and reproducing the observed polarization in MCs (\citealt{Seifried2019}; \citealt{Seifried2020}; \citealt{Reissl2021}) and diffuse ISM (\citealt{Reissl2020}), in which the gas density is lower with $n_{\rm gas} \sim 10^2 - 10^6 \cm^{-3}$. However, in very dense regions such as protostellar environments and evolved star's envelopes with $n_{\rm gas} > 10^6 \cm^{-3}$, the physics of grain alignment is quite complicated due to stronger gas randomization. \cite{Hoang2022} conducted a detailed theoretical study for grain alignment in protostellar environments using the RAT paradigm. The authors concluded that the high dust polarization level of up to $20 - 40\%$ observed in protostellar environments by ALMA (\citealt{Hull2014}; \citealt{Cox2018}) could only be reproduced by the MRAT mechanism for grains having iron inclusions \citep{Hoang2016a}. 

\cite{Giang2023} first implemented the MRAT mechanism into POLARIS and successfully applied it in modeling synthetic dust polarization in protostellar cores and disks. The updated POLARIS is potentially applied in evolved star's CSEs, in which strong stellar magnetic fields of $\sim$ mG-G (\citealt{Vlemmings2017}; \citealt{Vlemmings2019}) could produce efficient internal and external alignment by MRAT (\citealt{Hoang2016a}) and then be expected to produce a high degree of thermal dust polarization. Thus, by measuring the level of thermal dust polarization, we may be able to constrain the abundance of iron, which is expected to be present inside circumstellar grains in AGB envelopes (e.g., \citealt{Jones1990}; \citealt{Hofner2022}). The code is currently being upgraded (Giang et al., in preparation) by incorporating the RAT-D to model dust polarization from regions of intense radiation fields, such as massive protostellar environments, photodissociation regions, and cosmic transients.

The aim of this paper is to present a detailed modeling of grain alignment and disruption induced by RATs and thermal dust polarization for different dust and magnetic field properties in CSEs. The key difference in the physical conditions of CSEs from the ISM and star-forming regions is that dust grains rapidly drift through the gas due to radiation pressure from the AGB stars (see, e.g., \citealt{Hofner2018}). Thus, we will first include the effect of grain drift on the rotational damping of circumstellar grains (\citealt{Roberge1995}; \citealt{HoangTram2019}; \citealt{Tram2020}) into the updated POLARIS code of \cite{Giang2023} that incorporated the latest effects of grain alignment and disruption by RATs. For our study, we choose a specific target of O-rich AGB stars, namely IK Tauri (a.k.a IK Tau or NML Tauri), which is a Mira variable star pulsating in a period of $\sim 470$ days (\citealt{Wing1973}; \citealt{Hale1997}), in the late evolutionary phase of a low-mass star of $M_{\ast} \sim 1.5\,M_{\odot}$ (\citealt{Gobrecht2016}). We will quantify the properties of both internal and external alignment of grains induced by RATs across the IK Tau's CSE. We will model the polarized dust emission from aligned grains in the envelopes of IK Tau using our updated POLARIS code. This provides numerical predictions for future polarimetric observations in evolved stars at far-IR/sub-mm wavelengths, which would help better understand the properties of dust and magnetic fields. In addition to the RAT effects from stellar radiation, the relative grain drift to the gas could induce MEchanical Torques (METs) onto irregular grains, possibly producing the spin-up and alignment with respect to the direction of the stellar magnetic fields $\bf B$ or the stellar wind $\bf v$ (see \citealt{LazHoang.2007b,Das2016}; \citealt{HoangChoLazarian2018}; \citealt{Hoang2022}; \citealt{Reissl2023}). The possibility of spin-up and alignment processes by METs in AGB envelopes is discussed in this study and will be implemented in the updated POLARIS code of \cite{Giang2023} in subsequent studies to investigate further the MET alignment for various astrophysical environments.

The rest of this paper is structured as follows. Section \ref{sec:Intial_condition} describes the model set-up based on the physical conditions of the IK Tau envelope and the magnetic field morphology. In Section \ref{sec:modeling_polaris}, we will describe the processes of modeling synthetic dust polarization using POLARIS. The numerical results of grain alignment and disruption and the modeling results of dust polarization are presented in Section \ref{sec:modeling_result}. We will discuss the implications of our results for interpreting polarimetric observations, studying magnetic fields, and constraining dust properties in evolved star's envelopes in Section \ref{sec:discussion}. Our main findings are summarized in Section \ref{sec:summary}.

\section{Initial conditions of IK Tau's circumstellar envelope}
\label{sec:Intial_condition}

\subsection{Gas structure}
\label{sec:CSE_input}

During the AGB phases, stellar materials are blown out and accelerated above the escape velocity by strong shocks produced by stellar activities in the photosphere (i.e., global pulsation, convection, and magnetic Alfvén waves, see \cite{Hofner2018} for a review). When reaching a certain distance $r_{\rm c}$ at which the gas temperature drops to the condensation temperature $T_{\rm c}$, these materials condense and form first solid particles (i.e., dust grains). The newly formed grains can be accelerated by the intense radiation pressure from the central AGB star. These grains collide with the gas, transferring their momentum to the gas and producing the winds that expand radially at a constant speed $v_{\rm exp}$ (so-called dust-driven or radiation-driven winds, see \citealt{Hofner2008};\citealt{Hofner2016};\citealt{Tram2020}).

\begin{figure*}
    \centering
    \includegraphics[width = 1\textwidth]{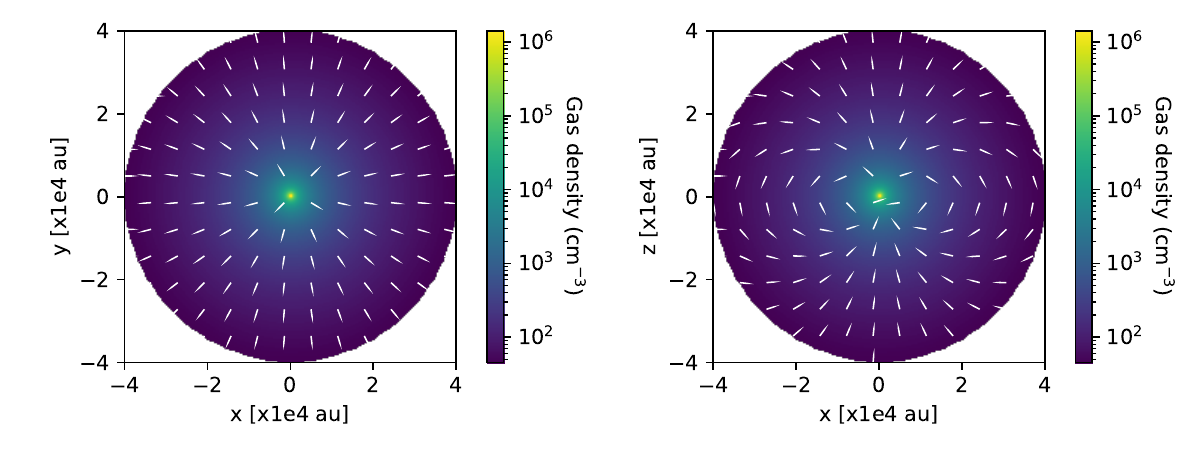}
    \includegraphics[width = 0.48\textwidth]{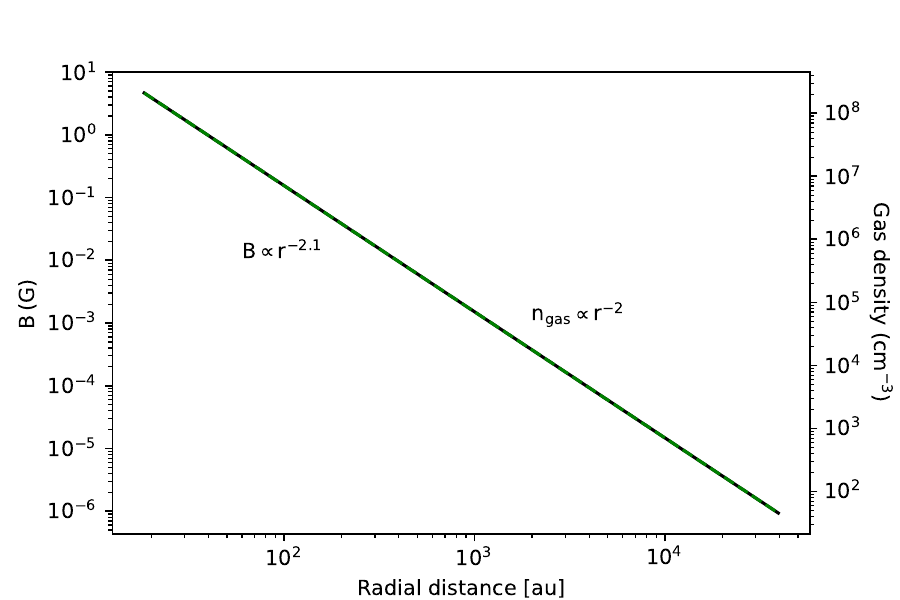}
    \caption{The structure of IK Tau's magnetic fields observed in the $xy$-plane (upper left panel) and the $xz$-plane (upper right panel). The lower panel illustrates the strength of stellar magnetic fields with $B \varpropto r^{-2.1}$ (solid black line) and the gas density profile in the IK Tau CSE (dashed green line).}
    \label{fig:Mag_geometry}
\end{figure*}

In this work, we assume the IK Tau CSE has a spherical structure based on optical and sub-millimeter observations (\citealt{Marvel2005}; \citealt{Kim2010}; \citealt{Adam2019}). The density of hydrogen gas (i.e., $m_{\rm gas} = m_{\rm H} = 1\,\rm amu$) in the envelope at a distance $r$ from the central star is calculated as (see \citealt{Tram2020}) 
\begin{equation}
\label{eq:gas_density}
\begin{split}
 n_{\rm gas}(r) &= \frac{\dot{M}}{4 \pi r^2 m_{\rm gas} v_{\rm exp} } \\
&\simeq 4.5 \times 10^{5}  \, \left(\frac{\dot{M}}{10^{-5} M_{\odot}\, \rm yr^{-1}}\right) \, \left(\frac{10 \rm\,km \,s^{-1}}{v_{\rm exp}}\right) \\
&\times \left(\frac{100 \, \rm au}{r}\right)^2 \, \rm cm^{-3}, \\
\end{split}
\end{equation}
where $\dot{M}$ is the mass-loss rate; adopting $\dot{M} = 4.5 \times 10^{-6} \, \rm M_{\odot} \, \rm yr^{-1}$ and $v_{\rm exp} = 24\,\rm km\,s^{-1}$ for the IK Tau CSE (see Table \ref{tab:IK_Tau}). The gas density decreases significantly with $\sim r^{-2}$ from $n_{\rm gas} \sim 10^8\,\cm^{-3}$ at $r = 18\,\rm au$ to $n_{\rm gas} \sim 10^2\,\cm^{-3}$ at $r \sim 10000\,\rm au$. The gas density profile is plotted by a dashed green line in the lower panel of Figure \ref{fig:Mag_geometry}.

For numerical calculations, we construct the model of the IK Tau CSE with a symmetric structure on a three-dimensional spherical grid of $N_{\rm r} \times N_{\theta} \times N_{\phi} = 151 \times 61 \times 61$, where $N_{\rm r}$, $N_{\theta}$ and $N_{\phi}$ are the number of grid cells in radial, polar and azimuthal directions. The inner boundary is set at $18\,\rm au$, where circumstellar dust is first condensed and formed within the IK Tau CSE. The outer boundary is set at $40000\,\rm au$, where circumstellar winds interact with the ISM (see in \citealt{Decin2010}). The characteristics of the IK Tau CSE are described in Table \ref{tab:IK_Tau}.

\begin{table}
    \centering
    \caption{Modeling parameters used in POLARIS}
    \label{tab:IK_Tau}
    \begin{tabular}{l c c}
         \toprule
         Parameters & Values & References \\
         \toprule
         \multicolumn{3}{c}{\bf Circumstellar envelope model} \\
         \toprule
          Inner boundary ($R_{\rm in}$) & $18\,\rm au$ & (1) \\
          Outer boundary ($R_{\rm out}$) & $40000\,\rm au$ & (1) \\
          Mass-loss rate ($\dot{M}$) & $4.5 \times 10^{-6} \, M_{\odot} \, \rm yr^{-1}$ & (1) \\
          Expansion velocity ($v_{\rm exp}$) & $24\,\rm km\,s^{-1}$ & (1); (2)   \\
          \toprule
          \multicolumn{3}{c}{\bf Stellar radiation source} \\
          \toprule
          Stellar luminosity ($L_{\ast}$) & $8724\,L_{\odot}$ & (3) \\
          Stellar radius ($R_{\ast}$) & $608\,R_{\odot}$ & (3) \\
          Stellar mass ($M_{\ast}$) & $1.5\,M_{\odot}$ & (1); (4) \\
          Effective temperature ($T_{\ast}$) & $2234\,\K$ & (3) \\
          Distance to observer ($d_{\ast}$) & $280\,\pc$ & (3)\\
          \toprule
          \multicolumn{3}{c}{\bf Dust model} \\
          \toprule
          Chemical composition & silicate & (5); (6) \\
          Minimum grain size ($a_{\rm min}$) & $5\,\rm nm$ & (7) \\
          Maximum grain size ($a_{\rm max}$) & $0.5\,\rm\mu m$ & (8); (9) \\
          Size distribution & $Ca^{-3.5}$ & (9); (10)  \\
          Dust-to-gas mass ratio ($M_{\rm d/g}$) & 0.005 & (5) \\
          Grain axial ratio ($s$) & 0.5 & (7) \\
          Iron fraction ($f_{p}$) & 0.01 - 0.1 &  (10) \\
          Volume filling factor ($\phi_{\rm sp}$) & 0.005 &  (10) \\
          Level of iron clusters ($N_{\rm cl}$) & $10 - 10^3$ &  (10) \\
         \bottomrule
    \end{tabular}
    \begin{tablenotes}
      \small
      \item \textbf{References:} (1) \cite{Decin2010}; (2) \cite{Maercker2016}; (3) \cite{Adam2019}; (4) \cite{Fox1985}; (5) \cite{Gobrecht2016}; (6) \cite{Decin2018}; (7) \cite{Reissl2016}; (8) \cite{Norris2012}; (9) \cite{Scicluna2015};
      (10) \cite{Hoang2016a}
    \end{tablenotes}
\end{table}

\subsection{Grain drift relative to the gas}

The absorption and scattering of stellar photons induces radiation pressure on circumstellar grains and pushes them outward. These grains can move faster than the ambient gas within the expanding outflows and, subsequently, experience strong drag force by the gas. The balance between radiative force and drag force establishes the terminal velocity of grains relative to the gas, i.e., causing the grain drift (see, e.g., \citealt{Gilman1972}; \citealt{Hofner2018}; \citealt{Tram2020}). Let $v_{\rm drift}$ be the drift velocity of grains through the gas. For the star with luminosity $L_{\ast}$, the drift velocity of the grain of size $a$ is given by \citep{Tram2020}
\begin{equation}
    v_{\rm drift}^2(a) = \frac{1}{2}\left[\left[\left(\frac{2v_{\rm exp}}{\dot{M} c} \bar{Q}_{\rm rp}(a)L_{\ast}\right)^2 + v_{\rm ther}^4\right]^{0.5} - v_{\rm ther}^2\right],\label{eq:vd}
\end{equation}
where 
\begin{equation}
    \bar{Q}_{\rm rp}(a) = \frac{\int_0^{\infty} Q_{\rm rp}(a, \lambda)u_{\lambda} d\lambda}{\int_0^{\infty} u_{\lambda} d\lambda} = \frac{\int_0^{\infty} Q_{\rm rp}(a, \lambda)u_{\lambda} d\lambda}{u_{\rm rad}}
\end{equation}
is the average radiation pressure cross-section efficiency over the stellar radiation energy density $u_{\lambda}$, with $Q_{\rm rp}(a, \lambda) = Q_{\rm abs}(a, \lambda) + Q_{\rm sca}(a, \lambda)(1 - \langle \cos\theta \rangle)$ for a grain size $a$, and $v_{\rm ther} = (2k_{\rm B}T_{\rm gas}/m_{\rm gas})^{1/2}$ is the thermal gas speed. $Q_{\rm abs}$ and $Q_{\rm sca}$ are the absorption and scattering cross-section efficiency, and $ \langle \cos\theta \rangle$ is the mean cosine of the scattering angle, which were pre-calculated in the POLARIS code for silicate dust (\citealt{Reissl2016}).

\begin{figure*}
    \centering
    \includegraphics[width = 0.48\textwidth]{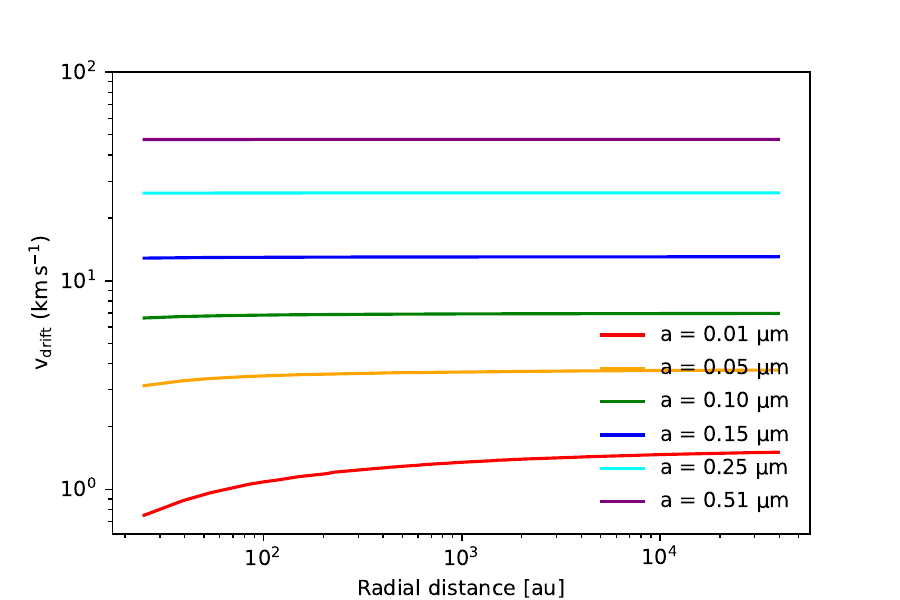}
    \includegraphics[width = 0.48\textwidth]{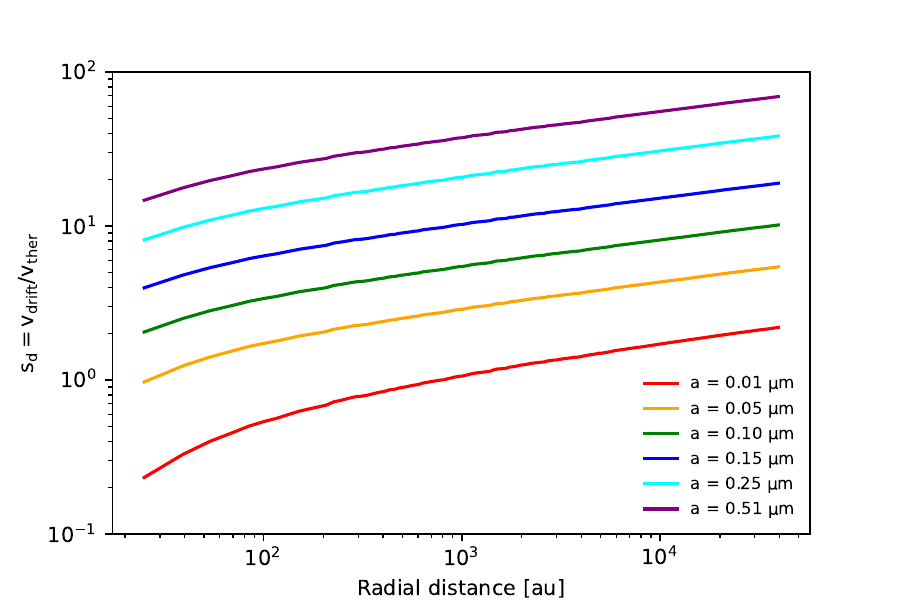} 
    \caption{The grain drift velocity $v_{\rm drift}$ (left panel) and the drift parameter $s_{\rm d} = v_{\rm drift}/v_{\rm ther}$ (right panel) vs. the radial distance of the IK Tau envelope for various grain sizes $a = 0.01 - 0.5\,\rm\mu m$.}
    \label{fig:vd_grain}
\end{figure*}

Figure \ref{fig:vd_grain} shows the grain drift velocity and the dimensionless drift parameter $s_{\rm d} = v_{\rm drift}/v_{\rm ther}$ over the radial distance of the IK Tau CSE for grain sizes $a = 0.01 - 0.5\,\rm\mu m$. The drift velocity increases with increasing envelope distance. Small grains $a < 0.01\,\rm\mu m$ undergo subsonic to transonic drift with $v_{\rm drift} \sim 1 - 3\,\rm km \, s^{-1}$ ($s_{\rm d} \sim 0.2 - 1$). Large grains $a > 0.05\,\rm \mu m$, on the other hand, experience stronger drift and move to supersonic speed up to $20 - 50\,\rm km\, s^{-1}$ ($s_{\rm d} \gg 1$). The calculation of the grain drift velocity as a function of grain size is first included in the POLARIS code. The grain drift relative to the grain will influence grain alignment and the thermal dust polarization, which will be discussed in Section \ref{sec:method_RATA}.

\subsection{Magnetic field morphology}
The precise determination of the global magnetic field morphology of the IK Tau envelope is uncertain. For numerical modeling of grain alignment and dust polarization, here we adopt the model of a dipole stellar magnetic field for the IK Tau envelope, which is presented in the upper panels of Figure \ref{fig:Mag_geometry}. The white segments present the pattern of stellar magnetic field lines in the $xy$-plane (upper left panel) and the $xz$-plane (upper right panel). The stellar magnetic fields generally reveal the radial pattern when observed in the $xy$-plane. Meanwhile, the magnetic fields show the morphology of a dipole structure along the $z$-direction when observed in the $xz$-plane.

The dependence of stellar magnetic field strength on the envelope distance is assumed to follow $B \varpropto r^{-2.1}$, based on the results derived from SiO, and $\rm H_2O$ maser polarization by \cite{Boboltz2005} and \cite{Ferreira2013}, as plotted by a dashed black line in the lower panel of Figure \ref{fig:Mag_geometry}. The B-field strength is considerably high, about 1 - 3 G in the inner region of $r \sim 18\,\rm au$ near the central star. Moving to the outer region of $r \sim 10000\,\rm au$, the B-field strength decreases to $\sim 10\,\mu\rm G$ - close to the typical magnetic field strength of the diffuse ISM (\citealt{Crutcher2012}).

\section{Modeling of grain alignment and dust polarization with updated POLARIS}
\label{sec:modeling_polaris}
Considering the profiles of gas density and stellar magnetic fields as modeling inputs, we perform synthetic dust continuum polarization calculation by the updated POLARIS code (\citealt{Reissl2016}; \citealt{Giang2023}). In this section, we briefly describe the simulation processes in POLARIS, including (1) the radiative transfer calculation of dust heating by the Monte Carlo technique, (2) the measurement of aligned grains based on RAT alignment theories, (3) and the polarized radiative transfer calculation for modeling thermal polarized emission from aligned grains in AGB envelopes.

\subsection{Dust heating radiative transfer calculation}
\label{sec:Dust_RT}
Firstly, we investigate the dust heating and cooling processes by numerically solving radiative transfer using POLARIS. For the main radiation source, we consider circumstellar grains receive photons mostly from the central AGB star. The stellar radiation of IK Tau can be described as a blackbody with the effective temperature of $2234\,\K$ and the bolometric luminosity of $L_{\ast} = 8724\,L_{\odot}$ (\citealt{Adam2019}, see Table \ref{tab:IK_Tau}). Note that besides the impact of stellar radiation, circumstellar dust can be affected by interstellar radiation fields (ISRF) from the outer boundary of the IK Tau envelope, which is negligible and is not considered in our calculation.

For the dust model, we consider the composition of O-rich materials with $100\,\%$ of silicate grains presented in the IK Tau envelope. Our assumption of pure silicate grains is based on the nature of O-rich AGB envelopes as primary sources of silicate dust in the ISM environment (see the review of \citealt{Tielens2005}). There is also the existence of alumina and other metal oxide grains in the O-rich envelopes (e.g., \citealt{Velhoelst2006}; \citealt{Velhoelst2009}; \citealt{Gobrecht2016}). However, their contribution to the total dust abundance in the envelope is quite low, especially for high mass-loss rate AGB stars of $\dot{M} \sim 10^{-6} - 10^{-5} \, \rm M_{\odot} \, \rm yr^{-1}$ as IK Tau with roughly $5-10\%$ (see, e.g., \citealt{Jones2014}; \citealt{Suh2016}; \citealt{Gobrecht2016}; \citealt{Suh2020}). The effect of metal oxide grains is then negligible in our numerical calculations.

In terms of grain shape and grain size, we assume the oblate shape with the ratio of the semi-minor and semi-major axes of grains $s = 1/2$. The choice of oblate shape with $s = 1/2$ is pre-existed in the original POLARIS code (\citealt{Reissl2016}). The size varies in the range from $a_{\rm min} = 5\,\rm nm$  to $a_{\rm max} = 0.5\,\rm\mu m$. The choice of the minimum grain size is pre-existed in the POLARIS code. The consideration of the maximum grain size is motivated by the presence of large grains $a > 0.3\,\rm\mu m$ in the AGB envelopes, taken from the optical/near-IR polarization by scattering light (see. e.g., \citealt{Norris2012}; \citealt{Scicluna2015}; \citealt{Ohnaka2016}; \citealt{Khouri2020}). The grain size distribution (GSD) is assumed to be followed by the standard Mathis-Rumpl-Nordsieck (MRN) size distribution as $dn \varpropto Ca^{-3.5}da$, where $C$ is the normalization constant derived from the total dust-to-gas mass ratio $M_{\rm d/g}$ (\citealt{Mathis1977}). Our works assume the dust-to-mass ratio of $0.005$ in the entire envelope of IK Tau (\citealt{Gobrecht2016}), which is lower than the standard $M_{\rm d/g} = 0.01$ in the diffuse ISM (\citealt{Cardelli1989}). The assumption of MRN-like distribution is widely used for fitting the optical/near-IR polarimetric observations (\citealt{Scicluna2015}; \citealt{Khouri2016}).\footnote{In \cite{Dominik1989} and \cite{VandeSande2020}, the original slope of grain size distribution could be much steeper up to $\eta \sim -5$, implying the possibility of small grain abundance formed in AGB circumstellar environments. The variation of grain size distribution directly impacts the alignment efficiency and the resulting thermal dust polarization degree, which is justified in Appendix \ref{sec:appendix_GSD}.}. This GSD can be modified by the RAT-D from stellar radiation and will be discussed in Section \ref{sec:theory_ratd}. Details about the radiation source and the dust model in the IK Tau CSE are presented in Table \ref{tab:IK_Tau}.

Given the radial distribution of gas density (see Section \ref{sec:CSE_input}) and the dust profile, POLARIS could model the light propagation from the main source in the dusty environment as AGB envelopes. In each grid cell of the model space, the code simulates the interaction between incident photons and circumstellar dust using the Monte Carlo technique (\citealt{Lucy1999}). Consequently, the code spontaneously updates the grain temperature $T_{\rm d}$, followed by the balance of radiation absorption and thermal dust emission (\citealt{Lucy1999}; \citealt{Reissl2016}; \citealt{Giang2023}). POLARIS stores the radiation energy density $u_{\lambda}$ and the mean propagation directions of photons from the processes of radiation absorption, dust emission, and dust scattering in each grid cell, giving the anisotropic degree of radiation $\gamma_{\lambda}$ (\citealt{Bethell2007}; \citealt{Reissl2016}) required to model RAT alignment in the following step.

\subsection{Modeling grain alignment and disruption driven by RATs}
\label{sec:grain_modeling}
\subsubsection{Magnetic properties of grains}
Iron is one of the most abundant elements in the universe and is expected to be present during the formation and evolution of circumstellar grains in AGB envelopes (\citealt{Jones1990}; \citealt{Hofner2022}). Iron atoms could be first synthesized in the explosive ejecta of core-collapse supernovae (CCSNe) and also type Ia supernovae; later on, they are injected into the ISM and molecular clouds, which subsequently collapse to form young stars (e.g., \citealt{Tielens2005}; \citealt{Dwek2016}). In the late phases of stellar evolution, these pre-existed iron atoms would be incorporated into newly-formed dust grains within the outflows of both C-rich and O-rich AGB stars (\citealt{Gail1999}; \citealt{Kemper2002}; \citealt{Velhoelst2009}). Nevertheless, the exact forms of embedded irons and their level inside circumstellar grains in AGB envelopes remain unclear. Iron atoms can be diffusely distributed within dust grains, making them paramagnetic (PM) material. Iron atoms can also be embedded in the form of iron clusters (\citealt{Jones1967}), in which each iron cluster can be considered as a large magnetic moment, and grains become superparamagnetic (SPM). \cite{Hoang2016a} showed that the presence of iron inclusions plays an essential role in the interaction of dust and the ambient magnetic fields, described by the magnetic susceptibility for different magnetic materials (see the review in \citealt{Hoang2022}). Generally, SPM grains obtain higher magnetic susceptibility than PM grains, and the magnetic susceptibility increases with increasing the number of iron atoms per cluster, $N_{\rm cl}$, which varies from $20$ to $10^5$ (\citealt{Jones1967}). However, grains could experience such a very fast rotation $\omega > 10^8 \rm rad\,\s^{-1}$ induced by RATs with a faster timescale than the magnetic response timescale, causing the suppression of magnetic susceptibility even for SPM grains with high $N_{\rm cl} > 10^3$ (\citealt{Hoang2016b}; \citealt{Lazarian2019}).

In the updated version of POLARIS by \cite{Giang2023}, the calculation of PM and SPM grains' magnetic susceptibility was included in the grain alignment modeling. Here, we apply in circumstellar grains in the IK Tau envelope, considering the model of PM materials with $f_{\rm p} = 0.01$ and $f_{\rm p} = 0.1$, corresponding to the configurations of $1\,\%$ and $10\,\%$ of Fe atoms randomly distributed inside grains, respectively. For SPM grains, we consider the increasing levels of iron inclusions $N_{\rm cl} = 10 - 10^3$ with a fixed volume filling factor of Fe atoms in each cluster $\phi_{\rm sp} = 0.005$ (\citealt{Hoang2016a}), corresponding to $1.67\,\%$ of Fe abundance present in the form of iron clusters in the grain. The magnetic properties of PM and SPM grains are shown in Table \ref{tab:IK_Tau}.

\subsubsection{Internal alignment by Barnett relaxation}
When silicate grains of PM (or SPM) materials in O-rich AGB envelopes rotate at a certain angular speed $\bf \Omega$, spins of unpaired electrons inside grains will be aligned along $\bf \Omega$, which produces a net magnetic moment for the grain by the Barnett effect (see \citealt{Barnett1915}). Theoretical studies of \cite{Purcell1979} showed that the induced magnetic moment for a magnetic rotating grain could dissipate the grain rotational energy to the minimum energy level at which grains can stably rotate around the axis of maximum inertia moment, $\bf a_{1}$ (i.e., internal alignment with $\bJ \parallel \bf \Omega \parallel {\bf a}_1$, see the illustration in \cite {Hoang2022} and references therein). As a result, grains that have the Barnett magnetic relaxation faster than the randomization by gas collisions can have the alignment of $\ba_{1}$ with $\bJ$ (so-called {\it right} internal alignment). However, grains that have slow internal relaxation may align with their major or minor axes along $\bJ$ (see \citealt{Hoang2016a} and \citealt{Hoang2022}). In general, large grains tend to have slow internal relaxation because of their large inertia moment, but this issue can be solved if grains have a higher rotational speed or high levels of iron inclusions (e.g., \citealt{Hoang2016a}; \citealt{Hoang2022}; \citealt{Giang2023}). There are other mechanisms, such as nuclear relaxation (i.e., the Barnett relaxation induced by nuclear magnetism, see \citealt{LazarianDraine1999}) and inelastic relaxation (i.e., the internal relaxation induced by the deformation of grain materials, see \citealt{LazarianEfroimsky1999}), which have a longer timescale in comparison to the timescale of Barnett relaxation for sub-micron sized grains $a < 0.5\,\rm\mu m$ of PM and SPM materials (\citealt{Hoang2022}) and do not include in this numerical modeling. These mechanisms are dominant for very large grains (VLGs) of $a > 10\,\rm\mu m$ presented in protostellar environments (see \citealt{Hoang2022}). These mechanisms are also favored for diamagnetic grains such as carbonaceous grains. For instance, the internal alignment by nuclear relaxation is significant for hydrogenated amorphous carbon (HAC) grains having nuclear paramagnetism due to hydrogen protons, while the inelastic relaxation is responsible for the internal alignment of both HAC and graphite grains (see \citealt{Hoang2023}).

The internal alignment properties of PM and SPM grains were modeled in the updated version of POLARIS by \cite{Giang2023}. The code first computes the range of sizes for grains having fast internal relaxation [$a_{\rm min, aJ}$ - $a_{\rm max, aJ}$], following the condition of $\tau_{\rm BR} = \tau_{\rm gas}$, then determines the internal alignment degree of grains for the polarized RT modeling in the third simulation. The states of internal alignment depend on internal relaxation processes and the rotational properties of grains, which will be described in the following sections.

\subsubsection{External alignment with magnetic fields}
As grains are magnetized by the Barnett effect, the induced magnetic moment will interact with the ambient magnetic field $\bf B$, producing a magnetic torque that causes the precession of the grain angular momentum $\bf J$ around $\bf B$ (i.e., Larmor precession, see the illustration in \citealt{Hoang2022}). Grains are considered to be stably aligned with B-fields if the Larmor precession happens faster than the gas randomization, at which the magnetic field is considered as the preferred axis of alignment (so-called magnetic alignment, see \cite{Hoang2022} and references therein).

The external magnetic alignment can happen due to the dissipation of rotational energy caused by magnetic relaxation when $\bf J$ makes some angle with $\bf B$, which was first studied by \citealt{DavidGreenstein1951} for paramagnetic grains (so-called David-Greenstein mechanism). The efficiency of magnetic relaxation is characterized by the magnetic relaxation parameter $\delta_{\rm m} = \tau_{\rm gas}/\tau_{\rm mag}$, which was included in the updated POLARIS (\citealt{Giang2023}). The code can calculate the maximum size $a^{\rm DG}_{\rm max,JB}$ at which magnetic relaxation is still faster than gas randomization (i.e., $\delta_{\rm m} > 1$). Generally, large grains can have slower magnetic relaxation than small grains, and SPM grains can achieve more efficient magnetic relaxation than PM grains because of higher magnetic susceptibility. However, \cite{Hoang2016a} showed that even for SPM grains with $\delta_{\rm m} \gg 1$, magnetic relaxation itself can induce the external alignment but with a rather low alignment degree due to the absence of grain suprathermal rotation.

\subsubsection{Radiative Torque Alignment (RAT-A) Paradigm}
\label{sec:method_RATA}
The interaction of stellar radiation with a helical grain can induce RATs due to the difference in scattering/absorption between left- and right-handed circularly polarized photons (\citealt{Dolginov1976}; \citealt{Lazarian2007}). Here, we briefly review the paradigm of magnetic alignment induced by RATs, which is essential for predicting grain alignment and polarization in AGB circumstellar environments. 

The RATs induced by strong radiation from the central source can spin up circumstellar grains and achieve the maximum angular velocity $\Omega \sim \Omega_{\rm RAT}$, which is given by
\begin{equation}
\label{eq:angular_RAT}
    \Omega_{\rm RAT} = \frac{\Gamma_{\rm RAT}\cos{\psi}\tau_{\rm damp}}{I_{\parallel}},
\end{equation}
where $I_{\parallel} = 8\pi \rho a^{5} / 15$ is the moment of inertia in the minor axis of grains $\bf {a_1}$, $\psi$ is the angle between the directions of the magnetic field $\bf B$ and radiation field $\bf k$ and $\tau_{\rm damp}$ is the grain rotational damping time (e.g., \citealt{Hoang2014,Giang2023}). Above, $\Gamma_{\rm RAT}$ is the radiative torque acting on the grain size $a$ by stellar radiation, which is defined as
\begin{equation}
\label{eq:radiative_torque}
    \Gamma_{\rm RAT} = \int_{\lambda_{\rm min}}^{\lambda_{\rm max}} \pi a^2\gamma_{\lambda}u_{\lambda}\left(\frac{\lambda}{2\pi}\right)\,Q_{\Gamma}\,d\lambda,
\end{equation}
where $Q_{\Gamma}$ is the RAT efficiency depending on the grain size $a$ and the wavelength of incident photons $\lambda$ (see the numerical calculation in \citealt{Lazarian2007}). $u_{\lambda}$ and $\gamma_{\lambda}$ are the spectral energy density and the anisotropic degree derived from the first simulation in Section \ref{sec:Dust_RT}, respectively. 

In the CSEs of AGB stars, the grain rotational damping can arise from gas-grain collisions (both gas random collisions and relative gas-grain drift), and infrared emission. The total rotational damping timescale is given by
\bea
\tau_{\rm damp} = \frac{\tau_{\rm gas}}{1  + F_{\rm sd}+ F_{\rm IR}},\label{eq:tau_damp}
\ena
where $\tau_{\rm gas}$ is the damping by gas random collisions described as
\begin{equation}
\label{eq:tau_gas}
    \tau_{\rm gas} \simeq 8.74 \times 10^{4}\,a_{-5}\hat{\rho}\left(\frac{30\,\rm cm^{-3}}{n_{\rm H}}\right)\left(\frac{100\,\rm K}{T_{\rm gas}}\right)^{1/2}\,\rm yr,
\end{equation}
where $a_{-5} = a/(10^{-5} \rm cm)$ and $\hat{\rho} = \rho/(3.3\,\rm g\,\rm cm^{-3})$; $F_{\rm sd}$ is the dimensionless damping coefficient caused by the grain drift relative to the gas as a function of the drift parameter $s_{\rm d}$ as (see \citealt{Roberge1995}; \citealt{HoangTram2019})
\begin{equation}
\label{eq:F_sd}
    F_{\rm sd} = M_0 = \left(\frac{\sqrt{\pi}}{4 s_d}\right)[2(1 + s_d^2)\textrm{erf}(s_d) + \mathcal{P}(3/2 , s_d^2)],
\end{equation}
where $\mathcal{P}$ is the incomplete gamma function; and $F_{\rm IR}$ is the dimensionless damping coefficient due to IR re-emission
\begin{equation}
\label{eq:F_IR}
    F_{\rm IR} \simeq \left(\frac{0.4U^{2/3}}{a_{-5}}\right)\left(\frac{30\,\rm cm^{-3}}{n_{\rm H}}\right)\left(\frac{100\,\rm K}{T_{\rm gas}}\right)^{1/2},
\end{equation}
where $U$ is the normalized average radiation energy density $u_{\rm rad} = \int u_{\lambda}d\lambda$ over the energy density of ISRF in the solar neighborhood with $u_{\rm ISRF}=8.6\times 10^{-13}\erg\cm^{-3}$ \citep{Mathis1983}.

\begin{figure}
    % \centering
    % \includegraphics[width = 0.48\linewidth]{AGB_POLARIS/Figure/Drift_vel/taudamp_drift.pdf}
    \includegraphics[width = 1\linewidth]{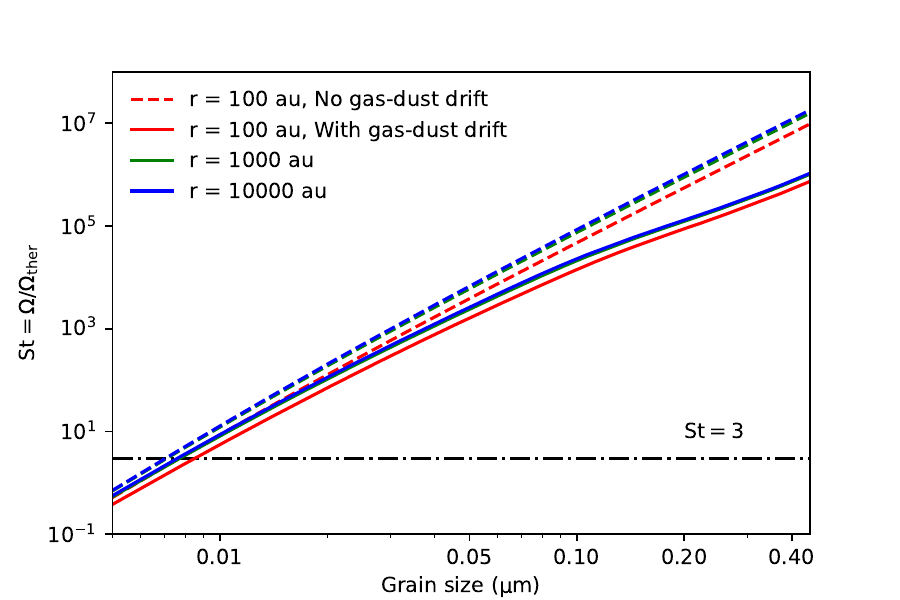}
    \caption{The suprathermal rotation parameter $St = \Omega/\Omega_{\rm ther}$ over the grain size at r = 100, 1000, and 10000 au in the IK Tau envelope for two cases: with (solid lines) and without (dashed lines) the grain drift. The lower limit $St = 3$ at which grains are aligned with $\bB$ by RATs is presented by a dotted-dashed black line.}
    \label{fig:RAT_vd}
\end{figure}

Figure \ref{fig:RAT_vd} shows the suprathermal rotation parameter $St = \Omega/\Omega_{\rm ther}$ induced by RATs with respect to the grain sizes at r = 100, 1000, and 10000 au in the polar region of the IK Tau's CSE ($\bf k \parallel \bf B$, $\cos{\psi} = 1$) between two cases: with (solid lines) and without (dashed lines) the effect of grain drift. One can see that grain sizes $a > 0.01\,\rm \mu m$ achieve suprathermal speed $St \gg 1$ due to intense radiation from the central star. The grain drift relative to the gas provides the additional rotational damping (see Equation \ref{eq:tau_damp}), then lowers the rotation rate by RATs. This effect dominates for large grains $a > 0.05\,\rm\mu m$ as they undergo supersonic drift (see Figure \ref{fig:vd_grain}).

Besides the spin-up effect, RATs can cause the alignment of the grain angular momentum $\bf J$ with $\bf B$ by the alignment torque components (\citealt{Lazarian2007}; \citealt{Hoang2008}; \citealt{Hoang2016a}). Circumstellar grains can be aligned with $\bf B$ when grains rotate suprathermally defined by $\Omega_{\rm RAT} \geq 3\Omega_{\rm ther}$. The minimum alignment size, denoted by $a_{\rm align}$, is set at which $\Omega_{\rm RAT} = 3\Omega_{\rm ther}$. As shown in Figure $\ref{fig:RAT_vd}$, grains of size $a > a_{\rm align} \sim 0.007 - 0.01\,\rm\mu m$ can have efficient magnetic alignment by RATs, even for large grains $a > 0.05\,\rm\mu m$ experiencing strong grain drift within the AGB outflows. Note that RATs can also produce the radiative precession of ${\bf J}$ along the radiation field direction $\bf k$, and irregular grains could achieve the alignment with respect to the radiation fields $\bf k$ (i.e., k-RAT) if the radiative precession is faster than the Larmon precession and gas randomization (i.e., $\tau_{\rm K} < \tau_{\rm Lar} < \tau_{\rm gas}$, see \citealt{Lazarian2007}; \citealt{Hoang2022}; \citealt{Tram2022}). For AGB envelopes with the strong magnetic field strength $B \sim $ 10 mG - 1G (Figure \ref{fig:Mag_geometry}), the magnetic alignment is expected to be more dominant than the radiative alignment. Therefore, the contribution of radiative alignment to the net polarization is minor in this environment and is not considered in this numerical modeling. Its effect will be discussed in detail in Section \ref{sec:discusson_mag_alignment}.

The studies of \cite{Hoang2008} showed that a fraction $f_{\rm high-J}$ of grains can be spun up by RATs and achieve suprathermal rotation (i.e., high-J attractors with $\Omega > \Omega_{\rm ther}$). Then, they can achieve efficient internal and external alignment and be stably aligned with B-fields (see in \citealt{Hoang2022}). Meanwhile, the remaining portion $1 - f_{\rm high-J}$ of grains can be spun down by RATs to subthermal speed (i.e., low-J attractors with $\Omega \sim \Omega_{\rm ther}$). Consequently, they have inefficient internal alignment due to thermal fluctuation of electron spins inside grains (\citealt{Purcell1979}) and inefficient external alignment with B-fields. The value of $f_{\rm high-J}$ can range from 0.2 to 0.7 for irregular grains (\citealt{Herranen2021}).

Theoretical and numerical studies of \cite{Hoang2016a} illustrated the combined effect of magnetic relaxation and grain alignment induced by RATs could increase the RAT alignment efficiency and lead to the perfect external alignment with $\bf B$ (i.e., MRAT mechanism). This mechanism is more effective for PM grains located in strong magnetic field environments or SPM grains with iron inclusions due to enhanced magnetic relaxation. This results in the high abundance of grains at high-J attractors with $f_{\rm high-J} = 1$.

\cite{Giang2023} first considered the enhanced RAT alignment by magnetic relaxation in the updated POLARIS. The code considered the dependence of $f_{\rm high-J}$ on the magnetic relaxation parameter (see Equation 28 in \citealt{Giang2023}), and thus, determined the maximum grain sizes having $f_{\rm high-J} = 0.5$ (i.e., $a_{\rm max, JB}^{\rm DG-0.5}$) and $f_{\rm high-J} = 1$ (i.e., $a_{\rm max, JB}^{\rm DG-1}$). The code took into account the nature of grains at high-J and low-J attractors in the calculation of aligned grain sizes and the alignment degree (see the flowchart in \citealt{Giang2023}). In terms of internal alignment, within the critical sizes [$a^{\rm high-J}_{\rm min, aJ}$ - $a^{\rm high-J}_{\rm max, aJ}$] for grains at high-J attractors and [$a^{\rm low-J}_{\rm min, aJ}$ - $a^{\rm low-J}_{\rm max, aJ}$] for grains at low-J attractors, we set the alignment degree with $Q_{\rm X, high-J} = 0.15$ and $Q_{\rm X, low-J} = 0.05$, respectively. In terms of external alignment, the code numerically calculated the minimum size of aligned grains $a_{\rm align}$. The maximum alignment size is determined by the Larmor precession, $a^{\rm Lar}_{\rm max, JB}$, given by $\tau_{\rm Lar}  \leq \tau_{\rm gas}/10$.  Within the critical size of [$a_{\rm align}$ - $a^{\rm Lar}_{\rm max, JB}$], PM and SPM grains aligned at high-J and low-J attractors have perfect external alignment, with the alignment degree $Q_{\rm J, high-J} = Q_{\rm J, low-J} = 1$. The effect of grain drift relative to the gas on rotational damping and the grain alignment in AGB envelopes is taken into account and first implemented in the updated POLARIS code.

Noticeably, there is a possible magnetic alignment by METs caused by the strong grain drift relative to the gas \citep{LazHoang.2007b,HoangChoLazarian2018}. \cite{HoangChoLazarian2018} numerically studied the METs induced by the grain drift for grains with different irregular shapes and showed that it can produce the same spin-up and alignment components onto these grains and help them align with the ambient B-fields (see also \citealt{Hoang2022}; \citealt{Reissl2023}). For an environment experiencing intense radiation fields as AGB envelopes, the MET alignment is expected to have a lower efficiency compared to the RAT alignment, which is not included in this modeling and will be further discussed about its contribution to grain alignment in AGB circumstellar environments in Section \ref{sec:discussion_drift}.

\subsubsection{Rotational disruption by RATs (RAT-D)}
\label{sec:theory_ratd}
While being spun up to suprathermal rotation by RATs, circumstellar grains experience centrifugal stress pointing outward from the center of the grain mass. If grains reach a critical velocity at which the centrifugal stress exceeds the binding energy of grain material, they are instantaneously disrupted by RATs (i.e., Radiative Torque Disruption (RAT-D) mechanism, see \citealt{Hoang2019}; \citealt{Hoang2021}). The critical velocity $\Omega_{\rm disr}$ at which RAT-D happens is therefore calculated as
\begin{equation}
\label{eq:omega_crit}
    \Omega_{\rm disr} = \frac{2}{a}\,\left(\frac{S_{\rm max}}{\rho}\right)^{1/2} \simeq 3.6\times10^{9}a_{-5}^{-1}\hat{\rho}^{-1/2}S_{\rm max,9}^{1/2}\,\rm rad\,s^{-1},
\end{equation}
where $S_{\rm max,9} = S_{\rm max}/(10^{9}\,\rm erg\,\rm cm^{-3})$ is the maximum tensile strength of grains. The value of $S_{\rm max}$ depends on the grain internal structure, with $S_{\rm max} < 10^9\,\erg\,\cm^{-3}$ for porous/composite grains and $S_{\rm max} \geq 10^9\,\erg\,\cm^{-3}$ for compact/crystalline grains (see \citealt{Hoang2019}).

As discussed in \cite{Hoang2019}, the RAT-D effect disrupts large grains into smaller species, modifying the original grain size distribution $dn \varpropto Ca^{\eta}da$ with $\eta < -3.5$, enhancing the abundance of small grains (see \citealt{GiangSuper2020}; \citealt{Giang2020}; \citealt{Giang2022Blanet}). The mechanism is effective as grains are exposed to strong radiation from evolved stars and considered in interpreting stellar observations (\citealt{Tram2020}; \citealt{Truong2022}). The mechanism is more efficient for grains at high-J attractors, while grains at low-J attractors are retained (see in \citealt{Lazarian2021}). We include the RAT-D effect in the main modeling by considering the calculation of the maximum grain size after disruption $a_{\rm disr}$ and the grain size distribution with the modified slope $\eta$. We also consider the dependence of the RAT-D effect on the rotational properties of grains at high-J or low-J attractors in calculating thermal dust polarization, which is currently developed in the newest version of POLARIS (Giang et al., in preparation).

\subsection{Polarized radiative transfer calculation}
Lastly, synthetic dust polarization from aligned grains is calculated by solving a set of radiative transfer equations of polarized radiation described by the Stokes parameters $[I, Q, U, V]^{T}$, where $I$ presents the total emission intensity, $Q$ and $U$ present the linear polarized intensities, and $V$ presents the circularly polarized intensity. The polarized radiative transfer equations take into account the extinction, emission, and polarization properties observed in the plane-of-sky (see \citealt{Martin1974}), as well as the degree of both internal and external alignment (\citealt{Greenberg1968}). Given the information of radiation from the first simulation (Section \ref{sec:Dust_RT}), grain properties from grain alignment modeling (Section \ref{sec:grain_modeling}) and the conditions of the IK Tau envelope as gas density and magnetic field morphology (Section \ref{sec:Intial_condition}), POLARIS (\citealt{Reissl2016}) applies the ray-tracing technique to trace the light propagation through grid cells to the observer and uses the Runge-Kutta Felhberg (RFK45) method to solve the full radiative transfer of Stokes parameters. The final results are the modeled degree of linear polarization $p(\%)$ described as
\begin{equation}
\label{eq:pol_degree}
    p(\%) = 100\frac{\sqrt{Q^2 + U^2}}{I},
\end{equation}
and the polarization angle (PA)
\begin{equation}
\label{eq:pol_angle}
    PA = \frac{1}{2}\arctan\left(\frac{Q}{U}\right).
\end{equation}

The updated version of POLARIS by \cite{Giang2023} improved the polarized radiative transfer calculations by incorporating the modern MRAT theory (\citealt{Hoang2016a}; \citealt{Hoang2022}). The effects of internal alignment for grains at high-J and low-J attractors were considered in calculating the Rayleigh reduction factor $R$ (see Section 4.2.5 in \citealt{Giang2023}). The roles of embedded iron in enhancing RAT alignment were also considered in updating dust polarization calculations, allowing us to examine the MRAT effects for different magnetic properties of grains on dust polarization in AGB envelopes.

\section{Numerical Results}
\label{sec:modeling_result}
\subsection{Radiation and dust temperature distribution}
\begin{figure}
    %\centering
    \includegraphics[width = 1\linewidth]{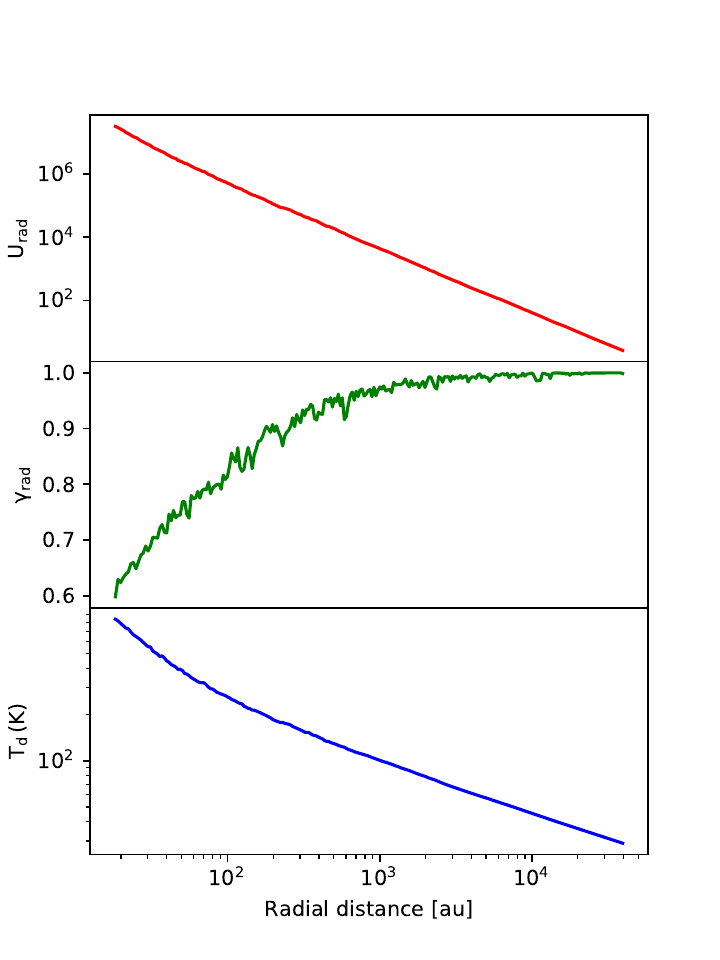} 
    \caption{The distribution of radiation strength $U_{\rm rad}$ (top panel), the average of anisotropy degree $\gamma_{\rm rad}$ (middle panel) and dust temperature $T_{\rm d}$ (bottom panel) with respect to the envelope distance.}
    \label{fig:Rad_fig}
\end{figure}

Figure \ref{fig:Rad_fig} illustrates the numerical results of stellar radiation distribution as a function of the envelope distance as a result of the interaction between circumstellar dust and stellar photon, including radiation strength $U_{\rm rad}$ (top panel), the average anisotropy degree $\gamma_{\rm rad}$ (middle panel) and the average dust temperature $T_{\rm d}$ (bottom panel). Generally, circumstellar dust is heated by the luminous stellar radiation in the innermost envelope of $\rm 18\, au$ to $\sim 700\,\K$. The radiation is anisotropic (i.e., $\gamma_{\rm rad} \sim 0.6$) due to the strong scattering between circumstellar dust and incident photons from the central star as well as the thermal emission of hot dust in the very dense envelope. At a larger distance of $ r > \rm 18\, au$, both stellar radiation $U_{\rm rad}$ and dust temperature $T_{\rm d}$ decrease with increasing the radial distance and achieve $T_{\rm d} < 100\,\K$ in the outer region of the IK Tau CSE. Moreover, the radiation fields become unidirectional (i.e., $\gamma_{\rm rad} \sim 1$), resulting from the weak interaction between thermal dust emission and circumstellar dust in the outer envelope.

\subsection{Grain alignment and disruption size distribution}
From the modeled stellar radiation and dust temperature, we investigate their effects on grain alignment and disruption induced by RATs. In this section, we present our modeling results of RAT alignment and rotational disruption paradigm in the IK Tau envelope, with the incorporated impacts of magnetic properties of dust grains for PM and SPM grains with iron inclusions.

\subsubsection{Critical sizes for external grain alignment by RATs $a_{\rm align}$, $a_{\rm max, JB}^{\rm Lar}$}
\label{sec:Result_amaxJB}
\begin{figure*}
    \centering
    \includegraphics[width = 1\linewidth]{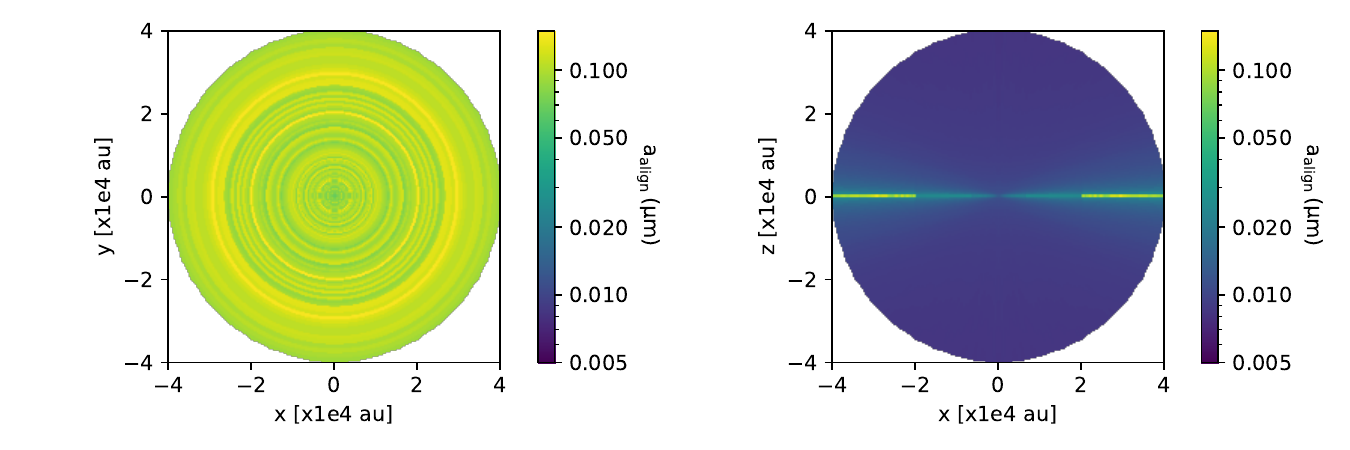}
    \includegraphics[width = 0.48\linewidth]{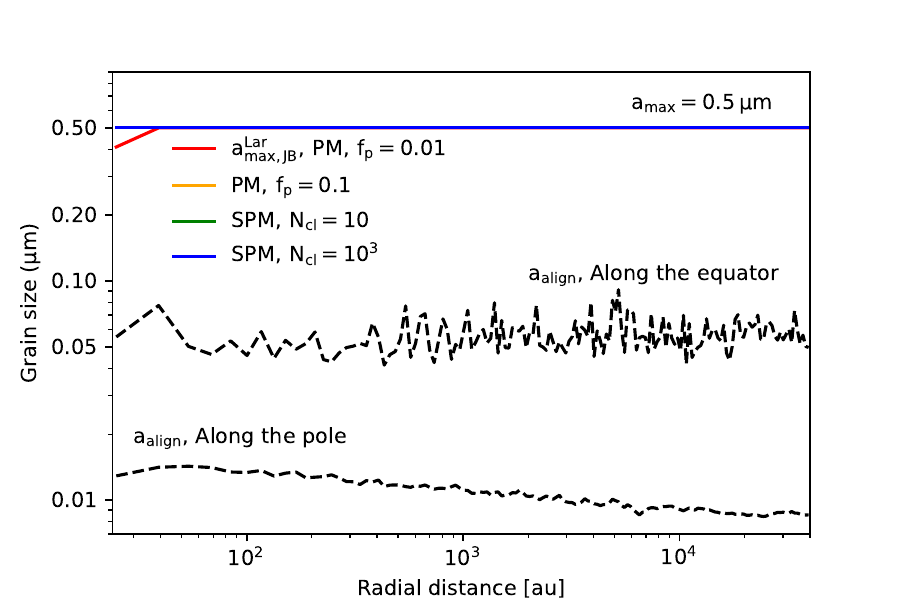}
    \caption{Upper panels: Maps of the minimum aligned size $a_{\rm align}$ calculated in the same $xy$-plane (left panel) and $xz$-plane (right panel). As being affected by stellar magnetic field geometry, the external alignment efficiency is reduced at the equator due to $\bf k \perp B$, resulting in larger $a_{\rm align}$. Lower panel: The variation of the minimum $a_{\rm align}$ (dashed black lines) and the maximum aligned sizes $a_{\rm max,JB}^{\rm Lar}$ (solid color lines) as a function of the envelope distance for both PM and SPM grains with different levels of iron inclusions $N_{\rm cl}$, assuming $\phi = 0.005$ and dust-to-gas ratio $M_{\rm d/g} = 0.005$. Being exposed to luminous stellar radiation fields, most circumstellar dust could effectively achieve magnetic alignment by RATs.}
    \label{fig:abar_maxJB_fig}
\end{figure*}

Figure \ref{fig:abar_maxJB_fig} shows the calculation of the minimum and maximum sizes having external alignment driven by RATs, denoted by $a_{\rm align}$ and $a_{\rm max,JB}^{\rm Lar}$. We consider the major effects of magnetic properties for both PM and SPM grains with increasing levels of iron inclusions from $N_{\rm cl} = 10$ to $N_{\rm cl} = 10^3$. 

The upper panels of Figure \ref{fig:abar_maxJB_fig} present the resulting maps of minimum aligned size $a_{\rm align}$ calculated in the $xy$-plane (left panel) and the $xz$-plane (right panel). Despite the effect of grain drift on rotational damping (Figure \ref{fig:RAT_vd}), circumstellar grains could easily achieve suprathermal rotation by strong stellar radiation fields from IK Tau and be effectively aligned with B-fields, with a considerably small value of $a_{\rm align} \sim 0.007 - 0.05\,\rm \mu m$. However, $a_{\rm align}$ is higher along the equator with $a_{\rm align} \sim 0.05\,\rm\mu m$ (the lower panel of Figure \ref{fig:abar_maxJB_fig}, dashed black line). This is caused by the dipole field components perpendicular to radiation fields (i.e., $\bf k \perp B$) at the equator, leading to the reduced RAT alignment efficiency and larger $a_{\rm align}$. While in the polar region, the external grain alignment becomes more efficient since $\bf k \parallel B$, resulting in small $a_{\rm align} \sim 7 - 10\,\rm nm$. The grain alignment size decreases from $a_{\rm align} \sim 10\,\rm nm$ at $r \sim 100\,\rm au$ to $a_{\rm align} \sim 7\,\rm nm$ at $r \sim 10000\,\rm au$ due to the reduced gas density within the IK Tau CSE.

For the maximum aligned size $a_{\rm max,JB}^{\rm Lar}$ (the lower panel of Figure \ref{fig:abar_maxJB_fig}, solid color lines), most circumstellar dust could have fast Larmor precession with shorter timescale than the gas damping timescale. Consequently, they could achieve magnetic alignment by RATs efficiently with $a_{\rm max,JB}^{\rm Lar} = a_{\rm max} = 0.5\,\rm\mu m$, even for the cases of PM and SPM grains.

\subsubsection{Critical sizes for internal grain alignment by Barnett relaxation $a_{\rm max, aJ}^{\rm high-J}$, $a_{\rm max, aJ}^{\rm low-J}$}
\label{sec:Result_amaxaJ}
\begin{figure}
   % \centering
    \includegraphics[width = 1\linewidth]{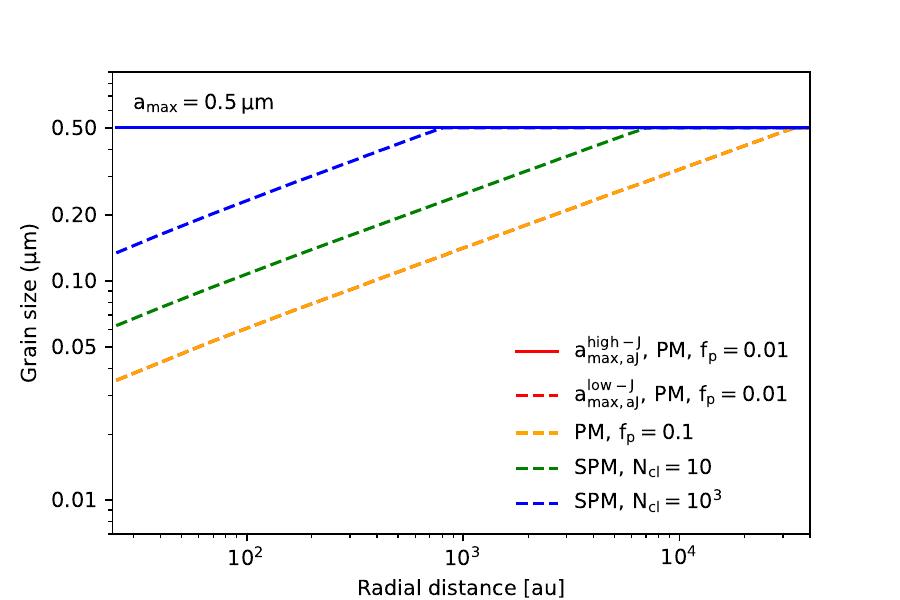} 
    \caption{The variation of the maximum sizes of grains at high-J (solid color lines) and low-J attractors (dashed color lines) having fast internal alignment by Barnett relaxation, denoted by $a_{\rm max, aJ}^{\rm high-J}$ and $a_{\rm max, aJ}^{\rm low-J}$, for PM and SPM grains with various values of $N_{\rm cl}$. Most grains at high-J attractors could have fast internal alignment with $a_{\rm max,aJ}^{\rm high-J} = a_{\rm max} = 0.5\,\rm\mu m$. Meanwhile, grains at low-J obtain thermal rotation by gas. Thus, $a_{\rm max, aJ}^{\rm low-J}$ increases due to the decrease in gas randomization within the envelope distance.}
    \label{fig:abar_maxJa_fig}
\end{figure}

Figure \ref{fig:abar_maxJa_fig} shows the critical sizes having internal alignment for grains at high-J and low-J attractors, with the maximum size of $a_{\rm max, aJ}^{\rm high-J}$ (solid color lines) and $a_{\rm max, aJ}^{\rm low-J}$ (dashed color lines), considering different magnetic properties of PM and SPM grains. Under the impact of strong stellar radiation fields, grains at high-J can achieve fast internal alignment, leading to $a_{\rm max,aJ}^{\rm high-J} = a_{\rm max} = 0.5\,\rm\mu m$.

For grains at low-J attractors, one obtains $a_{\rm max, aJ}^{\rm low-J}$ decreasing with increasing envelope distance, resulting from the reduction of gas randomization followed by the decrease in gas density. In addition, the increase in magnetic susceptibility for SPM grains with high $N_{\rm cl}$ causes the increase in $a_{\rm max, aJ}^{\rm low-J}$ and the extension in the fast internal alignment region from the outer region of $r \sim 10000\,\rm au$ to the inner region of $r \sim 100\,\rm au$ of the IK Tau envelope.

\subsubsection{Critical sizes for grain alignment by MRAT mechanism $a_{\rm max, JB}^{\rm DG-0.5}$, $a_{\rm max, JB}^{\rm DG-1}$}
\label{sec:Result_aDG}
\begin{figure}
    % \centering
    \includegraphics[width = 1\linewidth]{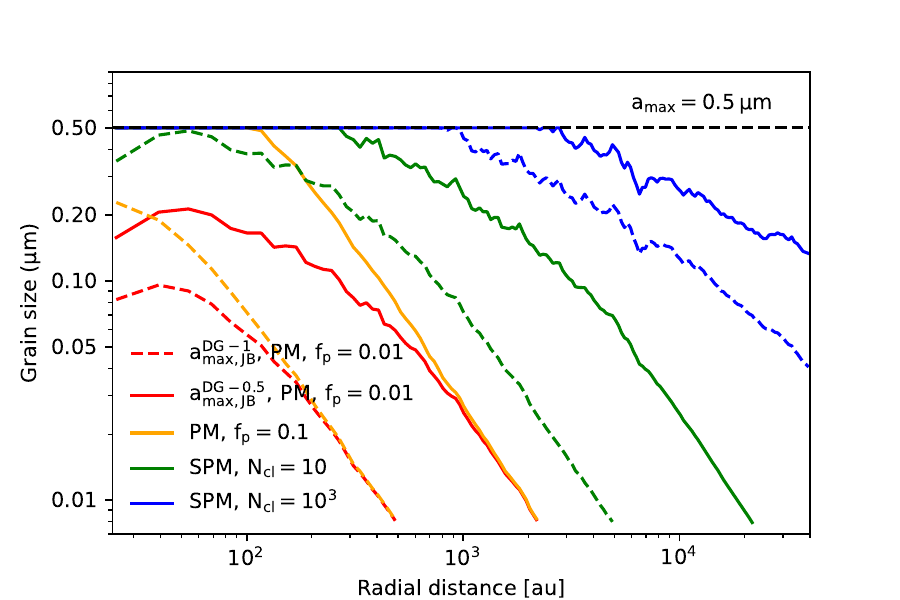} 
    \caption{The variation of the maximum grain sizes having external alignment by MRAT mechanism, denoted by $a_{\rm max, JB}^{\rm DG-1}$ (dashed color lines) and $a_{\rm max, JB}^{\rm DG-0.5}$ (solid color lines), for PM and SPM grains. Under the effect of strong stellar magnetic fields, the values of $a_{\rm max, JB}^{\rm DG-1}$ and $a_{\rm max, JB}^{\rm DG-0.5}$ are higher in the inner envelope $r < 500\,\rm au$. The impact of the MRAT mechanism extends from the inner to the outer region of the IK Tau envelope for SPM grains with increasing $N_{\rm cl}$.}
    \label{fig:a_MRAT_fig}
\end{figure}

Figure \ref{fig:a_MRAT_fig} illustrates the critical aligned sizes $a_{\rm max, JB}^{\rm DG-1}$ (dashed color lines) and $a_{\rm max, JB}^{\rm DG-0.5}$ (solid color lines) considering the contribution of magnetic relaxation in enhancing RAT alignment (i.e., MRAT mechanism). Even for PM grains with $f_p = 0.01$ and $f_p = 0.1$ (red and yellow lines), the alignment can be induced by strong stellar magnetic fields of 10 mG - 1 G within the inner region of $r < 500\,\rm au$, and subsequently, the effect of magnetic relaxation on grain alignment becomes efficient. At a larger envelope distance $r > 500\,\rm au$, both values of $a_{\rm max, JB}^{\rm DG-1}$ and $a_{\rm max, JB}^{\rm DG-0.5}$ drop significantly, corresponding to the decrease in the magnetic field strength to < 10 mG (see the lower panel of Figure \ref{fig:Mag_geometry}).

For SPM grains with $N_{\rm cl} = 10$ (green lines), the contribution of iron inclusions helps the MRAT mechanism to happen at $r > 500\,\rm au$. By increasing the level of iron inclusions to $N_{\rm cl} = 10^3$ (blue lines), the active MRAT region can extend to the outer envelope of $r > 10000\,\rm au$.

\subsubsection{Critical disruption sizes by RAT-D}
\begin{figure*}
    \centering
    \includegraphics[width = 0.48\linewidth]{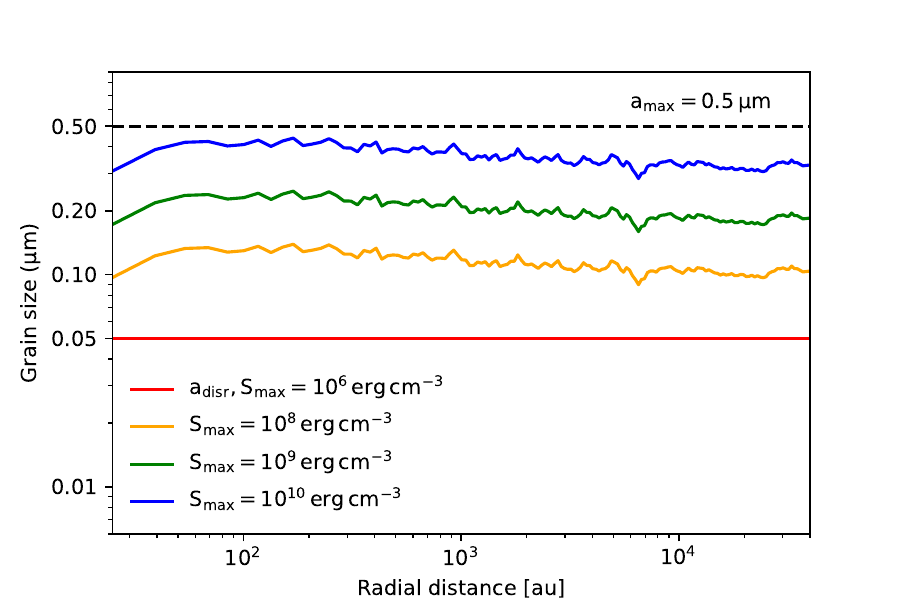}
    \includegraphics[width = 0.48\linewidth]{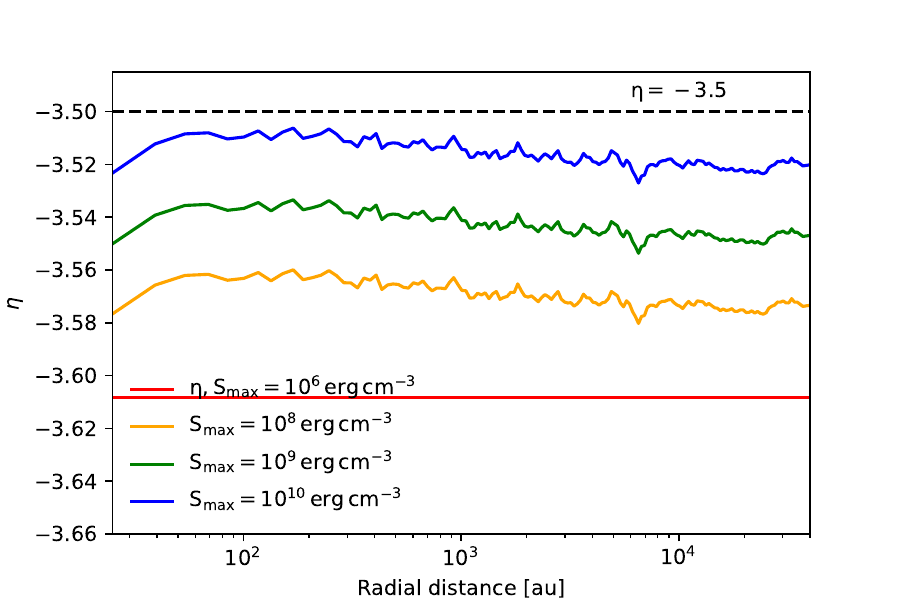} 
    \caption{The variation of grain disruption size $a_{\rm disr}$ (left panel) and slope of grain size distribution $\eta$ (right panel) with respect to the envelope distance for grains with $S_{\rm max} = 10^6 - 10^{10}\,\erg\,\cm^{-3}$. Radiative torques from stellar radiation could disrupt large grains into smaller species with $a_{\rm disr} < a_{\rm max} = 0.5\,\rm\mu m$ and steeper slope $\eta < -3.5$.}
    \label{fig:Disr_fig}
\end{figure*}

Figure \ref{fig:Disr_fig} shows the maximum grain size after being disrupted by RATs $a_{\rm disr}$ (i.e., RAT-D mechanism, left panel) and the modified slope of GSD $\eta$ (right panel), assuming the tensile strength of $S_{\rm max} = 10^6 - 10^{10}\,\rm erg\,cm^{-3}$. The impact of RATs from intense stellar radiation causes the fragmentation of large grains, resulting in the enhanced distribution of smaller grains with steeper $\eta < -3.5$ and smaller $a_{\rm disr} < a_{\rm max} = 0.5\,\rm\mu m$. 

The variation of both $a_{\rm disr}$ and $\eta$ within the envelope distance is associated with the joint effect of the spin-up by RATs and the \textbf{slow-down} by gas randomization and grain drift. As shown in Figure \ref{fig:Disr_fig}, in the innermost region of $r \sim 18\,\rm au$, most grains are disrupted by RAT-D with small $a_{\rm disr}$ and $\eta$. Then, these values slightly increase as a result of high gas density (i.e., $n_{\rm gas} > 10^7\,\cm^{-3}$), causing efficient rotational damping by gas randomization. At a further distance of $r > 100\,\rm au$, the considerable decrease in gas density causes the enhanced rotational disruption by stellar radiation, resulting in a steady decline in both values of grain disruption size $a_{\rm disr}$ and the slope of GSD $\eta$. 

Figure \ref{fig:Disr_fig} also shows the dependency of RAT-D efficiency on grain internal structure. Porous grains with $S_{\rm max} < 10^9 \,\rm erg\,cm^{-3}$ are strongly being disrupted by RAT-D, and thus, an amount of smaller grains is enhanced (i.e., low $a_{\rm disr}$ and $\eta$). By contrast, grains with compact structure (i.e., $S_{\rm max} > 10^9 \,\rm erg\,cm^{-3}$) are difficult to fragment by RAT-D, leading to the survival of large grains with higher values of $a_{\rm disr}$ and $\eta$.

\subsection{Dust polarization maps}
\label{sec:Dust_pol_map}
From critical aligned sizes constrained by RAT alignment, we model synthetic dust polarization with the implementation of the updated Rayleigh factor based on grain alignment properties in the updated version of POLARIS (see \citealt{Giang2023}). This section presents synthetic results of thermal dust polarization observed in far-IR/sub-mm wavelengths, considering the major effects on the polarization map: (1) inclination viewing angles, (2) magnetic properties of grains, and (3) rotational disruption by RATs.  

\subsubsection{Multi-wavelength polarization}
\begin{figure*}
    \centering
    \includegraphics[width = 1\linewidth]{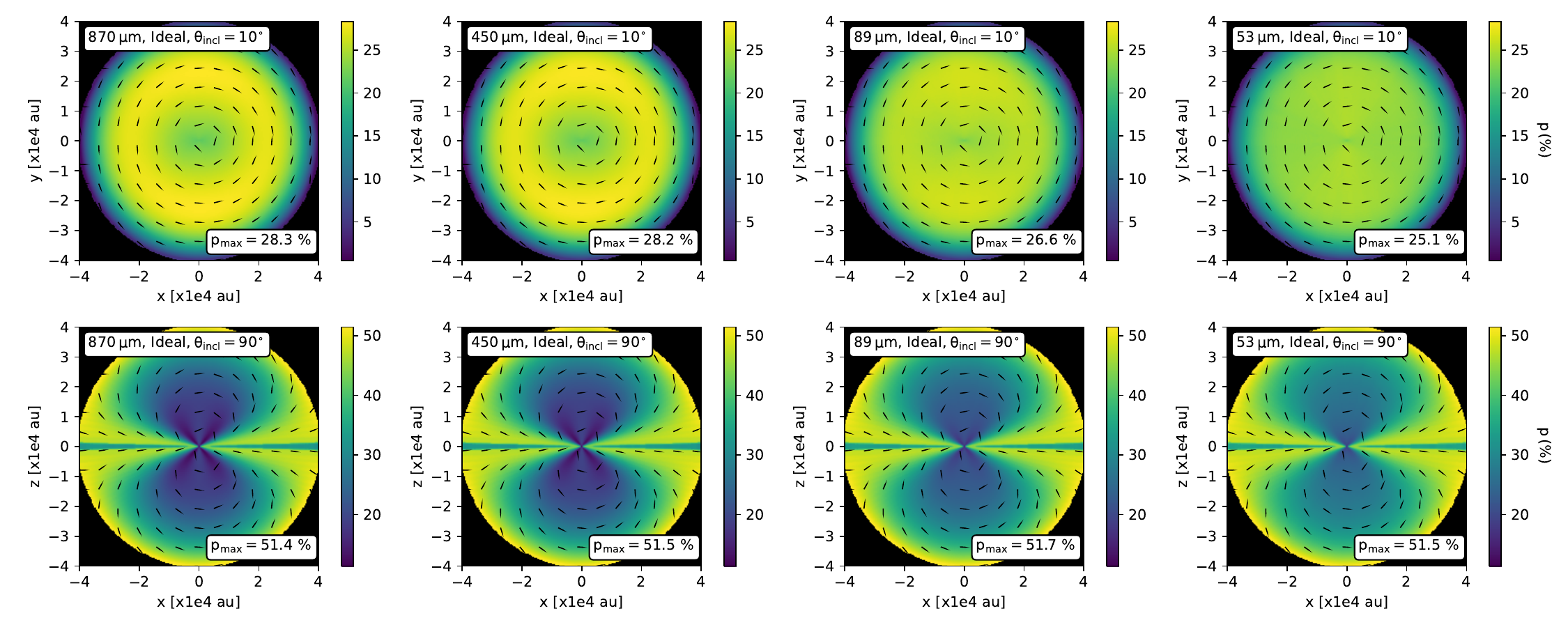}  
    \caption{Multi-wavelength dust continuum polarization maps ($\rm \lambda = 53 - 870\,\mu m$) produced by the Ideal model (i.e., $\rm a > a_{align}$ could align with B-fields) observed in face-on (i.e., $\rm \theta_{incl} = 10^{\circ}$, top panels) and edge-on views (i.e., $\rm \theta_{incl} = 90^{\circ}$, bottom panels). The color code illustrates the polarization fraction $p(\%)$, while black quiver lines show the pattern of polarization vector $\bf P$. The polarization degree shifts from thermal dust emission by cold dust in the outer envelope at longer wavelengths to thermal dust emission by warm dust in the inner envelope at shorter wavelengths.}
    \label{fig:Dustpol_wave_fig}
\end{figure*}

Figure \ref{fig:Dustpol_wave_fig} illustrates synthetic polarization maps of the entire IK Tau CSE of $80000\,\rm au $ at optically thin wavelengths ranging from $53\,\rm\mu m$ to $870\,\rm\mu m$. The large-scale maps are captured by a plane detector having $250 \times 250$ pixels placed at 280 pc from the central star to an observer on Earth, providing a spatial resolution of 320 au. The color code represents the polarization degree $p(\%)$, while black segments represent the pattern of polarization vector $\bf P$. We first consider the Ideal model in which all grains larger than the minimum aligned sizes $a_{\rm align}$ could achieve fast internal relaxation (i.e., $f_{\rm high-J} = 1$) and perfectly align with stellar magnetic fields.

One can see that at sub-mm wavelengths (e.g., $870\,\rm\mu m$ and $450\,\rm\mu m$), the polarization degree is higher in the outer region of $r > 10000\,\rm au$, which is mainly produced by the thermal emission of cold dust of $T_{\rm d} < 100\,\K$ (see the dust temperature profile in Figure \ref{fig:Rad_fig}). At shorter wavelengths (e.g., $\lambda=53\,\rm\mu m$ and $89\,\rm\mu m$), the maximum polarization degree shifts to the inner envelope of $r < 10000\,\rm au$, indicating the thermal emission by warm dust of $T_{\rm d} > 100\,\K$ in this region. 

\subsubsection{Effects of inclination angle}
An important note is the major impact of inclination angles (i.e., $\theta_{\rm incl}$, the angle between the observed line-of-sight toward the source and the $z$-axis) on the observed stellar magnetic fields in the plane-of-sky, which directly affects the calculation of dust polarization in AGB envelopes. Figure \ref{fig:Dustpol_wave_fig} shows this effect on the dust polarization maps, both on polarization vector and polarization degree measured at optically thin wavelengths, assuming viewed at two inclination angles: $\theta_{\rm incl} = 10^{\circ}$ and $\theta_{\rm incl} = 90^{\circ}$.

The upper panels of Figure \ref{fig:Dustpol_wave_fig} show the polarization maps observed in a face-on view (i.e., $\theta_{\rm incl} = 10^{\circ}$) along $x$- and $y$-directions. The polarization vectors indicate a toroidal structure associated with the radial structure of stellar magnetic fields in the $xy$-plane when rotating to 90 degrees (see the upper left panel of Figure \ref{fig:Mag_geometry}). The polarization degree distributes uniformly with $p \sim 20\,\%$ due to the projected components of B-fields mostly lying in the plane-of-sky, enhancing the polarization fraction in observation.

The lower panels of Figure \ref{fig:Dustpol_wave_fig} illustrate the resulting map in an edge-on-view (i.e., $\theta_{\rm incl} = 90^{\circ}$) along $x$- and $z$-directions. The polarization vectors present a dipole structure along the $x$-axis, which is perpendicular to the dipole pattern of stellar magnetic fields in the $xz$-plane (see the upper right panel of Figure \ref{fig:Mag_geometry}). The change in the curvature of dipole fields results in the variation in the projected B-fields in the plane-of-sky. Hence, the polarization degree is higher at the equator and the outermost polar region of the IK Tau envelope of $r \sim 30000\,\rm au$ with $p \sim 50\,\%$. In the inner region of $r < 30000\,\rm au$, $p$ is lower with $\sim 20\,\%$, which is the impact of perpendicular components of stellar magnetic fields that reduces the resulting polarization degree.

\subsubsection{Effects of iron inclusions}
As discussed in the previous sections, dust polarization at far-IR/sub-mm regimes could trace the global magnetic fields of IK Tau with $\bf P \perp B$ as most grains have perfect internal and external alignment in the Ideal model. We then consider the Realistic models for PM and SPM grains with increasing levels of iron inclusion.

\begin{figure*}
    \centering
    \includegraphics[width = 1\textwidth]{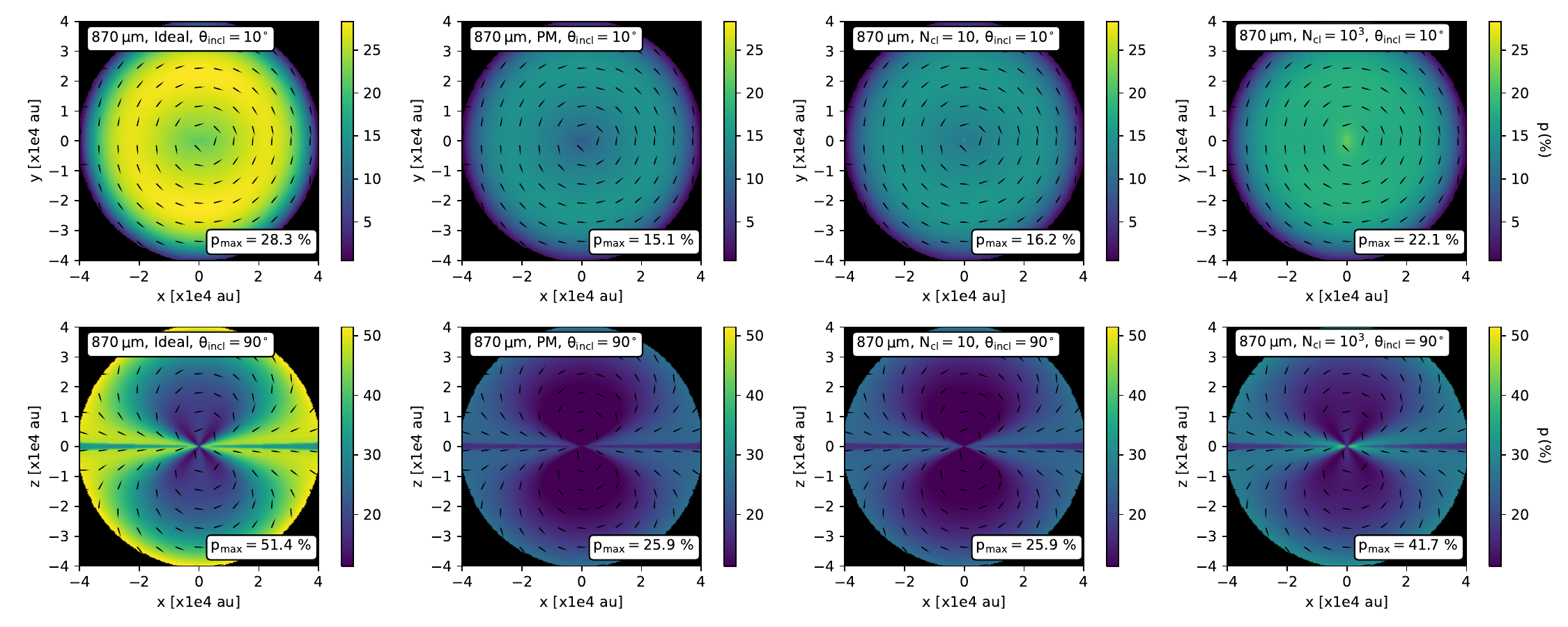}  
    \caption{Synthetic dust polarization maps observed at $\rm 870\,\mu m$ with two inclination angles: $\rm \theta_{incl} = 10^{\circ}$ and $\rm \theta_{incl} = 90^{\circ}$, in the envelope scale, produced by PM and SPM grains with different levels of $N_{\rm cl}$. The polarization patterns change from the toroidal to the dipole structure as increasing $\rm \theta_{incl}$. The polarization degree is lower for the case of PM grains, and it increases with increasing $N_{\rm cl}$ for SPM grains due to higher magnetic susceptibility.}
    \label{fig:Dustpol_mag_scale10000_fig}
\end{figure*}

Figure \ref{fig:Dustpol_mag_scale10000_fig} shows the large-scale map of the IK Tau envelope at $870\,\rm \mu m$ in the face-on (upper panels) and edge-on observations (lower panels), considering the Realistic models of PM grains with $f_p = 0.1$ (second column), and SPM grains with $N_{\rm cl} = 10$ (third column) and $N_{\rm cl} = 10^3$ (last column). For PM grains, the polarization fraction is lower with $p \sim 15 - 25\,\%$ due to low magnetic susceptibility. For SPM grains, the contribution of iron inclusions enhances the magnetic susceptibility. As a result, the polarization degree increases to $p \sim 20 - 40\,\%$ for SPM grains with $N_{\rm cl} = 10^3$.

\begin{figure*}
    \centering
    \includegraphics[width = 1\textwidth]{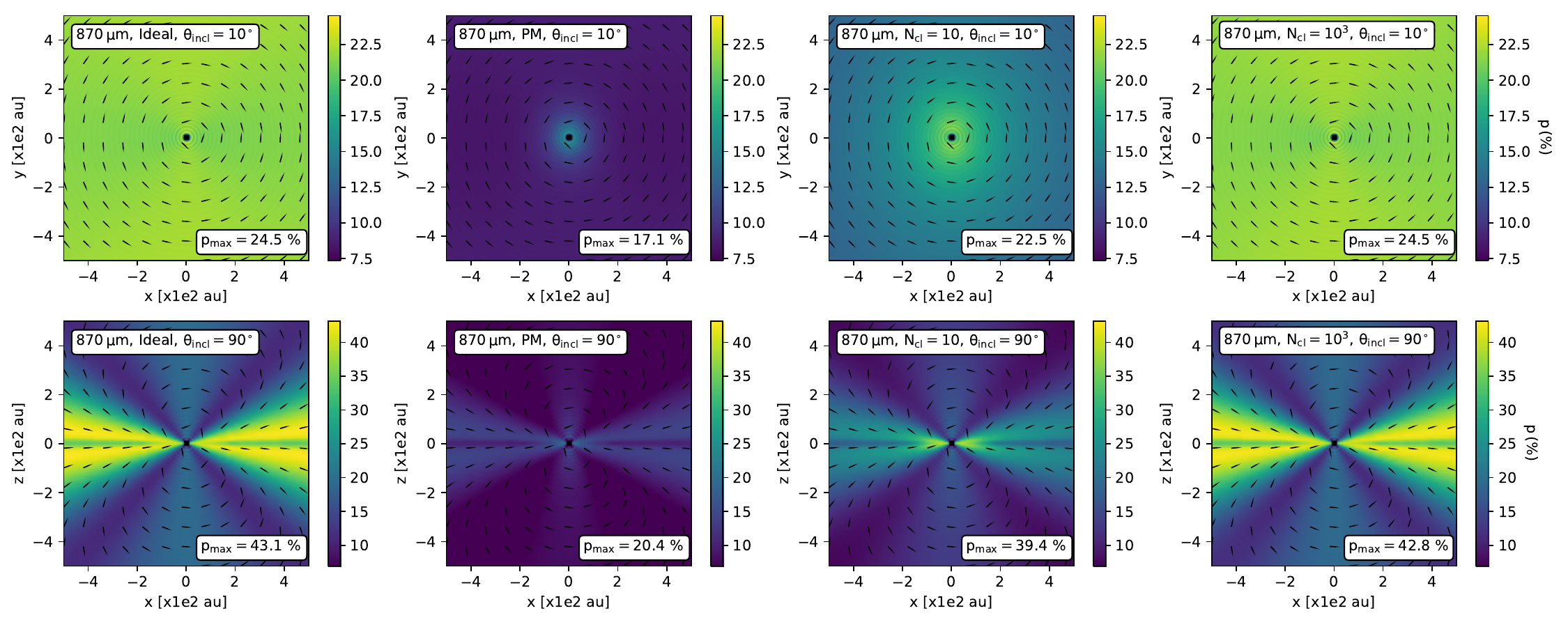}
    \caption{Similar to Figure \ref{fig:Dustpol_mag_scale10000_fig} but in the small central region scale of 1000 au. The polarization degree extends from the inner region of  $r < 100\,\rm au$ to the outer region of $r > 500\,\rm au$ for PM and SPM grains with higher levels of iron inclusion as a result of the increased contribution of the MRAT mechanism (see Figure \ref{fig:a_MRAT_fig}).}
    \label{fig:Dustpol_mag_scale1000_fig}
\end{figure*}

Figure \ref{fig:Dustpol_mag_scale1000_fig} shows similar, but zooming in the small-scale region of $1000\,\rm au$. For PM grains, the higher $p$ is mostly concentrated within the inner envelope of $r < 100\,\rm au$, which is a result of the enhanced MRAT alignment due to strong stellar magnetic field strength (see Figure \ref{fig:a_MRAT_fig}). For SPM grains, the incorporation of iron into dust grains causes the extension of the active MRAT region as increasing $N_{\rm cl}$. Consequently, the polarization degree extends to the outer envelope of $r > 500\,\rm au$.

\subsubsection{Effects of RAT-D}
Later, we consider the implication of the RAT-D mechanism for observing dust polarization in AGB envelopes. The modification of grain size distribution by RAT-D could considerably impact the resulting polarization maps, which will be described in this section.

\begin{figure*}
    \centering
    \includegraphics[width = 1\textwidth]{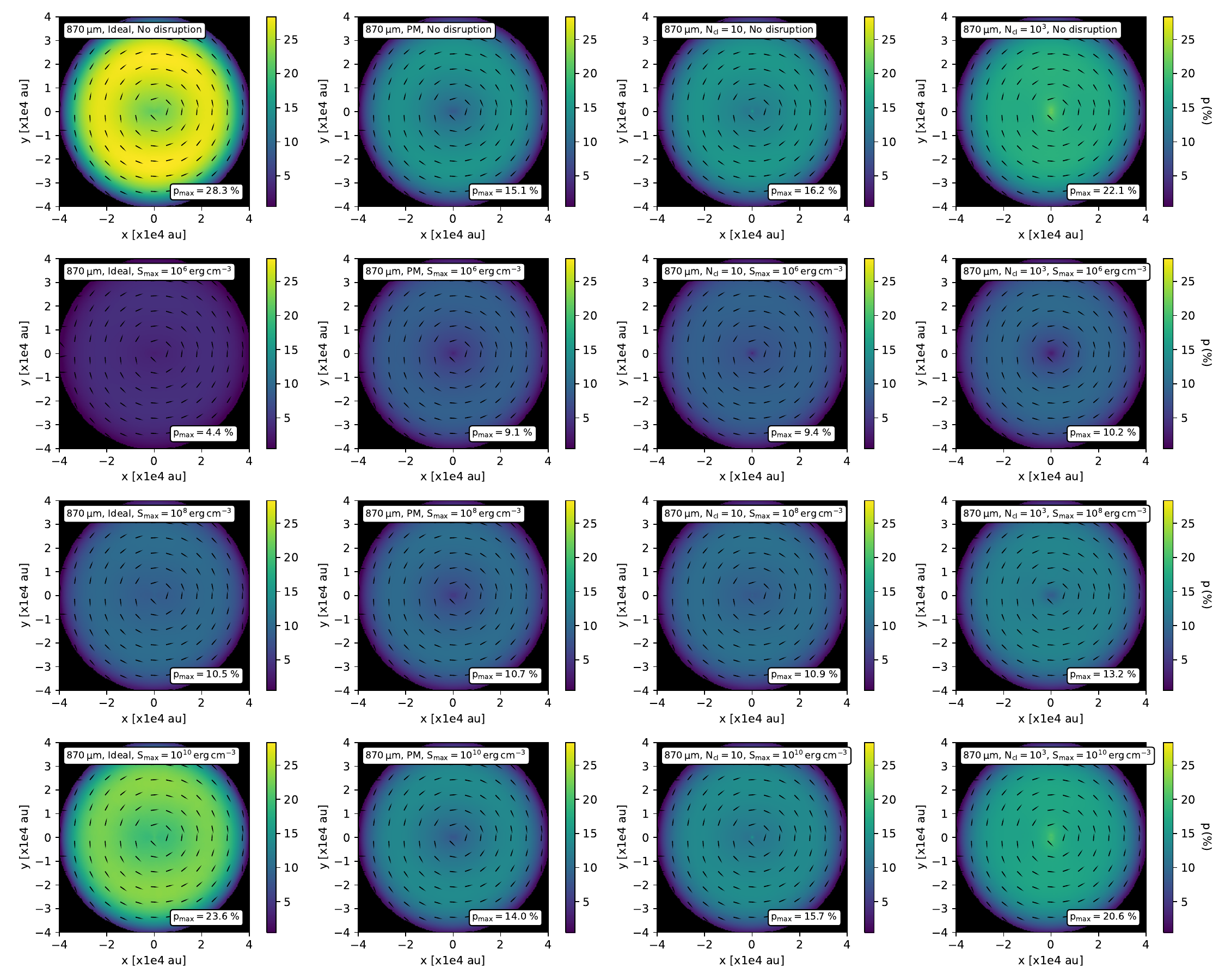}  
    \caption{Synthetic dust polarization maps observed in the face-on view (i.e., $\theta_{\rm incl} = 10^{\circ}$) at $\rm 870\,\mu m$ produced by the Ideal and Realistic models for PM and SPM grains with $N_{\rm cl} = 10 - 10^3$, considering two configurations: no disruption (first row) and disruption for different grain internal structures with $S_{\rm max} = 10^6\,\erg\,\cm^{-3}$ (second row), $S_{\rm max} = 10^8\,\erg\,\cm^{-3}$ (third row) and $S_{\rm max} = 10^{10}\,\erg\,\cm^{-3}$ (last row). The impact of the RAT-D mechanism effectively reduces the overall polarization degree in all cases of grain alignment models, especially for grains with lower $S_{\rm max}$.}
    \label{fig:Dustpol_RATD_10}
\end{figure*}

\begin{figure*}
    \centering
    \includegraphics[width = 1\textwidth]{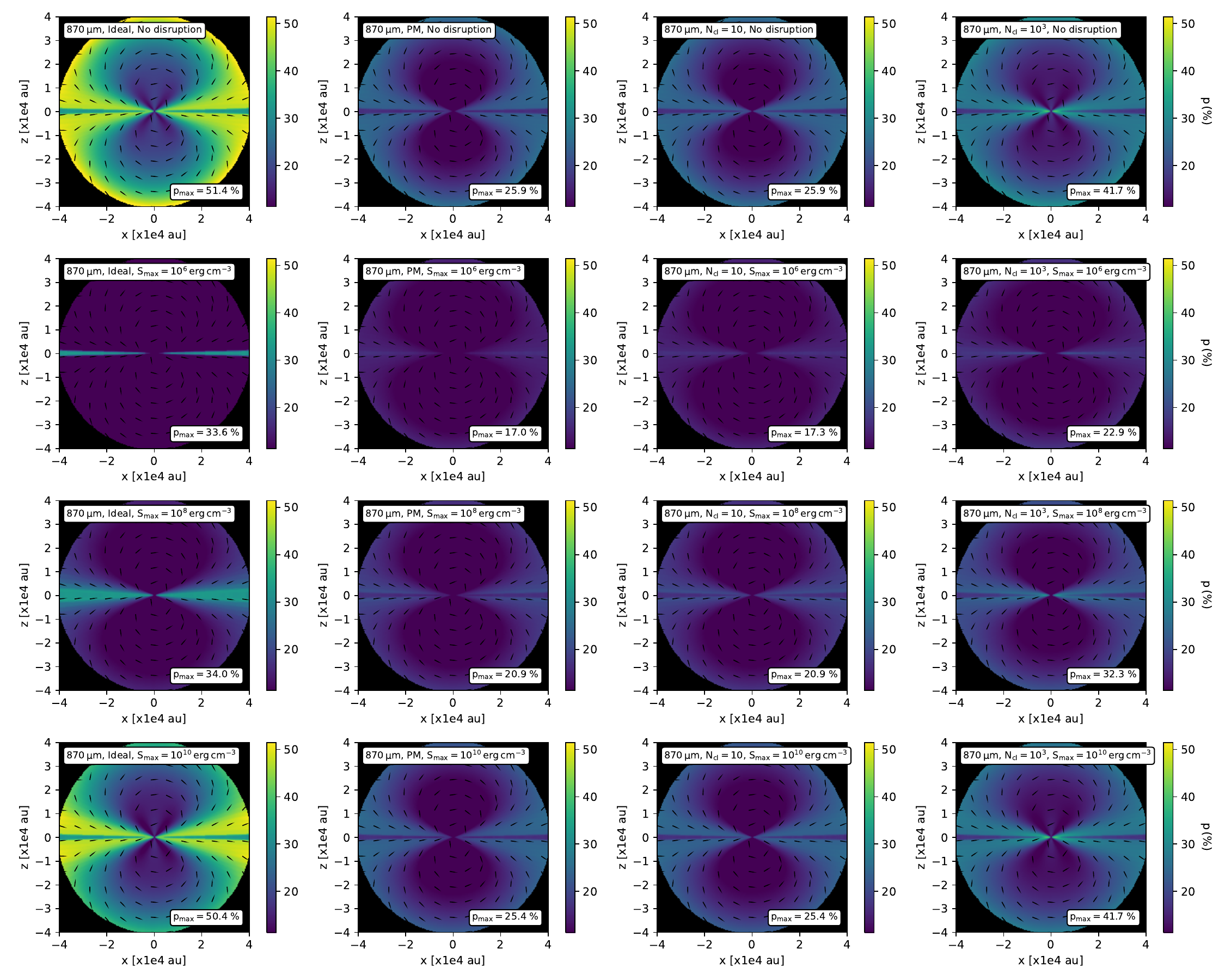}  
    \caption{Same as Figure \ref{fig:Dustpol_RATD_10} in the edge-on view with $\theta_{\rm incl} = 90^{\circ}$. The effect of RAT-D is more significant in the polar region with lower $p$ than in the equatorial region due to $\bf k \perp B$.}
    \label{fig:Dustpol_RATD_90}
\end{figure*}

Figure \ref{fig:Dustpol_RATD_10} and \ref{fig:Dustpol_RATD_90} show the RAT-D effects on the $\rm 870\,\mu m$ synthetic dust polarization for the large-scale map of the IK Tau envelope, assuming being observed at $\theta_{\rm incl} = 10^{\circ}$ and $\theta_{\rm incl} = 90^{\circ}$, respectively. We examine the dependence of RAT-D on the magnetic properties of grains in both Ideal and Realistic models for PM and SPM materials. We consider the disruption for grains with $S_{\rm max} = 10^6\,\erg\,\cm^{-3}$ (second row), $S_{\rm max} = 10^8\,\erg\,\cm^{-3}$ (third row) and $S_{\rm max} = 10^{10}\,\erg\,\cm^{-3}$ (last row), while the non-disrupted case is added for comparison (first row).

One can obtain that large grains are fragmented by RAT-D from stellar radiation, resulting in the enhancement of smaller grains and a narrower GSD of silicate dust. The overall polarization degree is then reduced with lower $p$ in all cases of grain alignment models. From the edge-on observations (Figure \ref{fig:Dustpol_RATD_90}), the effects of RAT-D remove most large grains in the polar region, while in the equatorial region, the RAT-D is inefficient due to $\bf k \perp B$ (see Section \ref{sec:Result_amaxJB}). Consequently, the polarization degree tends to be lower in the polar region than in the equatorial region.

The decrease in polarization fraction also depends on the internal structure of grains. Porous grains undergo stronger disruption by stellar radiation, decreasing the abundance of large grains but increasing the abundance of smaller grains (i.e., narrower GSD and smaller $\eta$). Hence, the polarization degree becomes lower for grains with lower $S_{\rm max}$. In contrast, compact grains with higher $S_{\rm max}$ are hard to disrupt by RAT-D, so large grains can survive (i.e., broader GSD and larger $\eta$) and produce higher $p$. This effect occurs in the entire envelope for all cases of Ideal and Realistic models of grain alignment in both Figure \ref{fig:Dustpol_RATD_10} and \ref{fig:Dustpol_RATD_90}.

\subsection{The p-I relationship}
\label{sec:pvsI}
\begin{figure*}
    \centering
    \includegraphics[width = 0.48\textwidth]{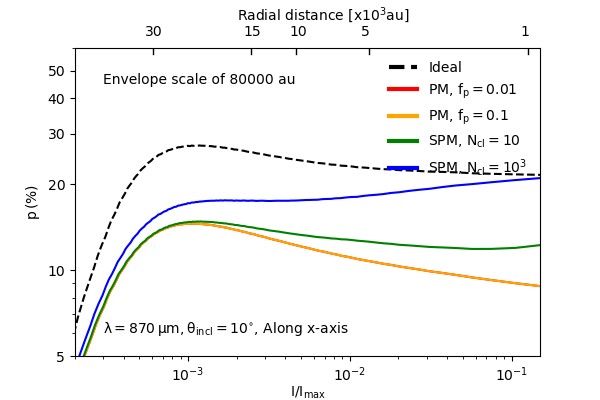} 
    \includegraphics[width = 0.48\textwidth]{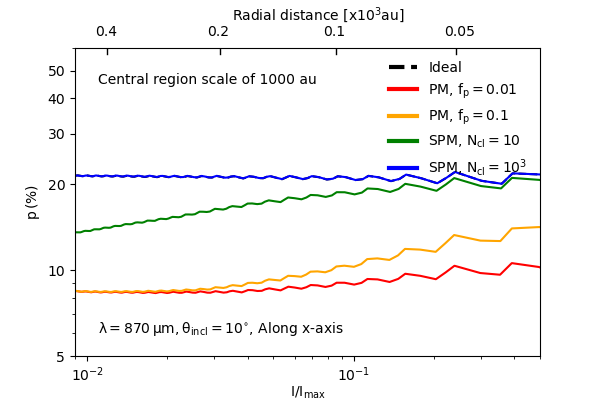
    }
    \includegraphics[width = 0.48\textwidth]{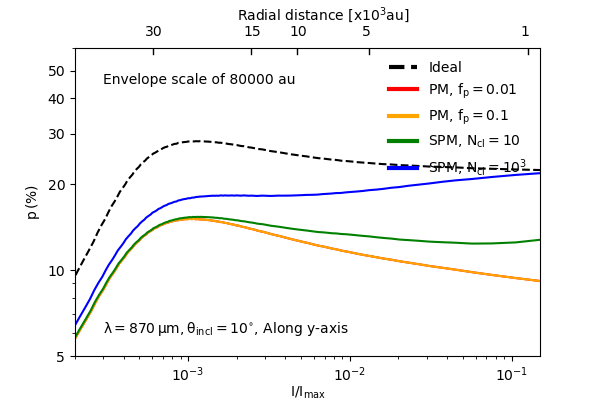} 
    \includegraphics[width = 0.48\textwidth]{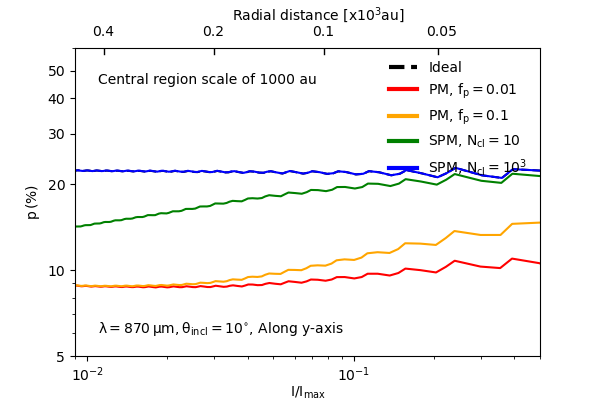}
    \caption{The variation of polarization fraction $p\,(\%)$ at $\rm 870\,\mu m$ with the normalized intensity $I/I_{\rm max}$ for both Ideal (dashed black line) and Realistic cases of PM and SPM grains (solid color lines) in the face-on view (i.e., $\theta_{\rm incl} = 10^{\circ}$), considering the observations in the envelope scale of 80000 au (left panels) and the central region scale of 1000 au (right panels). Both observations along $x$- (upper panels) and $y$-axis (bottom panels) show the increase in $p$ within the region of $r \sim 10000 - 30000\,\rm au$ as a result of the B-field components lying in the plane-of-sky. Additionally, $p$ increases for PM and SPM grains with higher levels of iron resulting from higher magnetic susceptibility and the extended MRAT effects.}
    \label{fig:pvsI_nodisr_10_fig}
\end{figure*}

Figure \ref{fig:pvsI_nodisr_10_fig} shows the variation of the $870\,\rm\mu m$ polarization fraction $p(\%)$ concerning the normalized intensity $I/I_{\rm max}$, where $I_{\rm max}$ is the maximum intensity produced by thermal dust emission in the innermost region of the IK Tau CSE, assuming observed at $\theta_{\rm incl} = 10^{\circ}$ (i.e., face-on view). The normalized intensity $I/I_{\rm max}$ is calculated at each radial distance $r$ along the $x$-axis (upper panels) and $y$-axis (lower panels) of the plane-of-sky. We analyze both observations in the large-scale envelope of 80000 au (left panels) and the small-scale central region of 1000 au (right panels). Solid color lines represent the polarization degree $p$ for Realistic models of PM and SPM grains, while that for the Ideal case is plotted by a dashed black line for comparison.

For both calculations along the $x$- and $y$-axis, the polarization degree is significantly low with $p < 5\,\%$ in the outermost envelope of $r > 30000\,\rm au$. Toward the central star, $p$ increases to the uniform values of $\sim 20\,\%$ as previously illustrated in the polarization maps in the upper panels of Figure \ref{fig:Dustpol_wave_fig}, which is mainly enhanced by the projected magnetic fields lying in the plane-of-sky observed in the $xy$-plane (see the upper left panel of Figure \ref{fig:Mag_geometry}).

\begin{figure*}
    \centering
    \includegraphics[width = 0.46\textwidth]{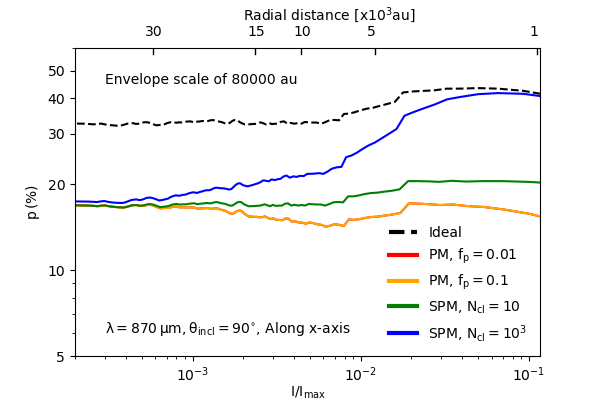} 
    \includegraphics[width = 0.46\textwidth]{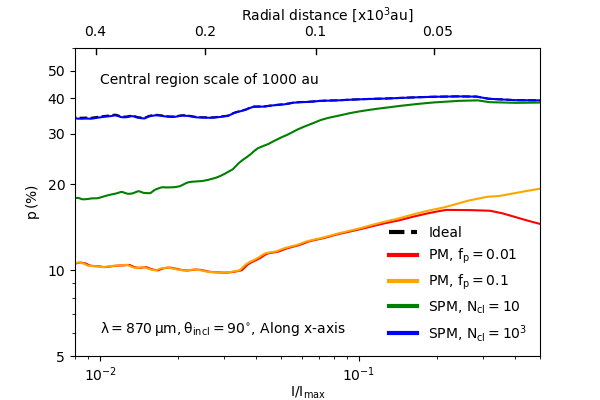
    }
    \includegraphics[width = 0.46\textwidth]{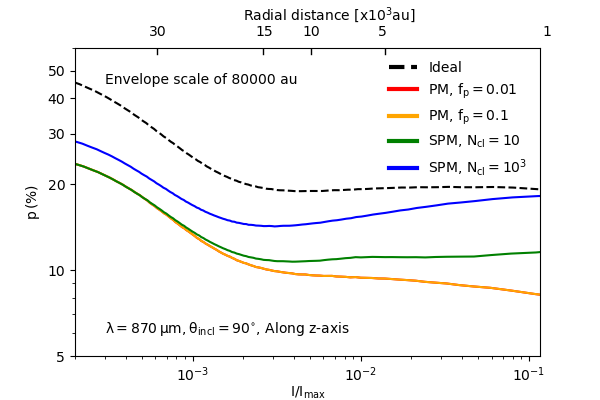} 
    \includegraphics[width = 0.46\textwidth]{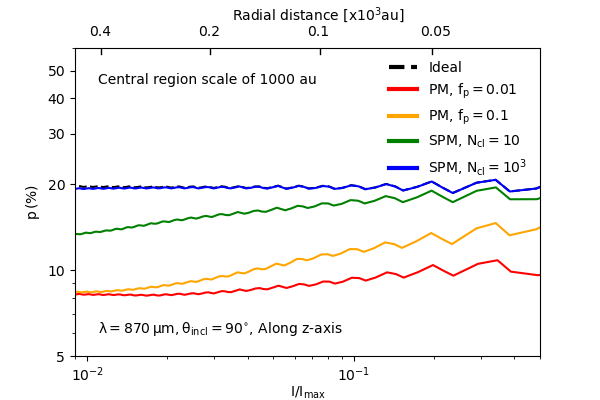}
    \caption{Similar results of Figure \ref{fig:pvsI_nodisr_10_fig} but considering the observations in the edge-on view (i.e., $\theta_{\rm incl} = 90^{\circ}$). Along the $x$-axis (upper panels), the polarization fraction increases toward the central star caused by the increased grain alignment by RATs. On the other hand, along the $z$-axis (lower panels), $p\,(\%)$ decreases significantly, resulting from the change in the curvature of the stellar dipole magnetic fields.}
    \label{fig:pvsI_nodisr_90_fig}
\end{figure*}

Figure \ref{fig:pvsI_nodisr_90_fig} shows similar, but in the edge-on view (i.e., $\theta_{\rm incl} = 90^{\circ}$), taking the calculation along the $x$-axis (i.e., along the equator region, upper panels) and the $z$-axis (i.e., along the polar region, lower panels). At the equator, the radiation fields are perpendicular to the magnetic fields; therefore, the effect of magnetic field geometry on dust polarization is negligible. Hence, for the Ideal model, the main result is produced by the alignment of large grains with $a_{\rm align} \sim 0.05\,\rm\mu m$ in the equator region of IK Tau envelope (the lower panel of Figure \ref{fig:abar_maxJB_fig}) with $p \sim 40\,\%$. Along the $z$-axis, the polarization is affected by the geometry of the dipole magnetic fields (see the upper right panel of Figure \ref{fig:Mag_geometry}). The B-field vectors become perpendicular to the plane-of-sky when moving to the inner region, causing the decrease from $p$ $\sim 45\,\%$ at $r > 30000\,\rm au$ to $p \sim 20\,\%$ at  $r < 30000\,\rm au$.

One crucial factor is the effects of the magnetic properties of dust grains on the relationship $p - I/I_{\rm max}$. Both Figure \ref{fig:pvsI_nodisr_10_fig} and \ref{fig:pvsI_nodisr_90_fig} show that the polarization degree produced by PM grains is considerably lower with $p < 10\,\%$ due to the lower magnetic susceptibility. Within $r \sim 500 - 10000\,\rm au$, the polarization fraction decreases slightly as a result of a large proportion of grains at low-J attractors having low internal relaxation (right panel of Figure \ref{fig:abar_maxJa_fig}). In the inner region of $r < 500 \,\rm au$, the contribution of magnetic relaxation in enhancing RAT alignment is significant with a high portion of grains at high-J attractors, leading to the increases in $p$ with increasing iron fraction $f_{\rm p}$. For SPM grains with iron inclusions, the profiles $p - I/I_{\rm max}$ illustrate the extension of polarization degree to the outer envelope of $r > 10000\,\rm au$ corresponding to the enhanced contribution of the MRAT mechanism with increasing $N_{\rm cl}$.

\begin{figure*}
    \centering
    \includegraphics[width = 0.46\textwidth]{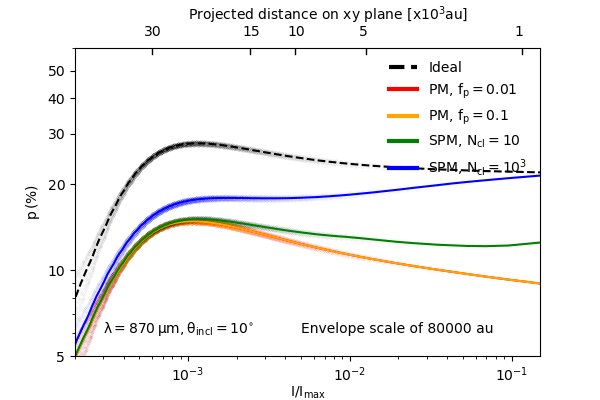} 
    \includegraphics[width = 0.46\textwidth]{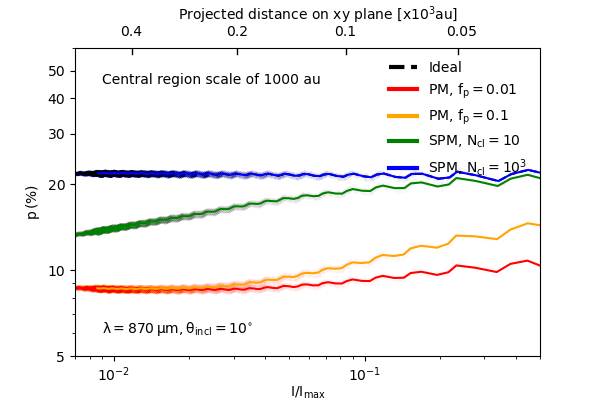}
    \includegraphics[width = 0.46\textwidth]{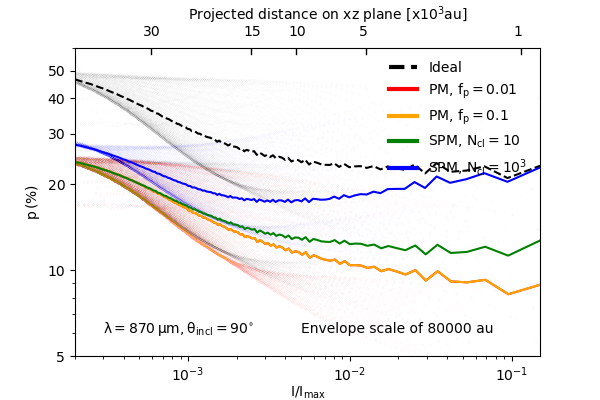} 
    \includegraphics[width = 0.46\textwidth]{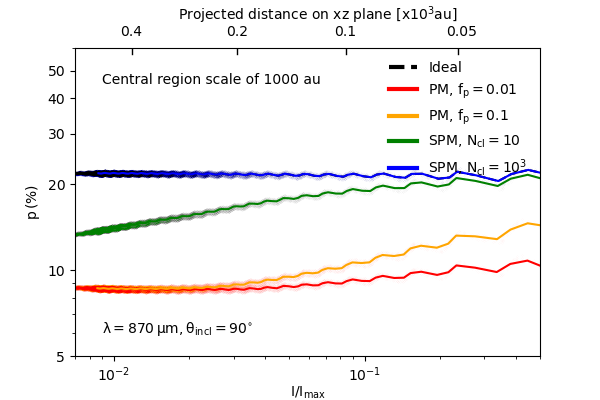}
    \caption{The averaged intensity-dependent polarization fraction $p - I/I_{\rm max}$ at each projected distance $d_{\rm proj}$ in the observed plane-of-sky, assuming observed at $\theta_{\rm incl} = 10^{\circ}$ (upper panels) and $\theta_{\rm incl} = 90^{\circ}$ (lower panels). The effects of stellar magnetic field geometry on dust polarization mostly in the outer envelope of $d_{\rm proj} > 10000 \,\rm au$, while in the inner region of $d_{\rm proj} < 500\,\rm au$, the increase in polarization degree is caused mainly by the MRAT mechanism.}
    \label{fig:pvsI_nodisr_fig}
\end{figure*}

Figure \ref{fig:pvsI_nodisr_fig} shows the mean polarization degree $p$ obtained at $870\,\rm\mu m$ with varying $I/I_{\rm max}$ calculated within the ring located at the projected distance $d_{\rm proj}$, with the radius of $d_{\rm resol} = 320\,\rm au$  for the large-scale map of the envelope (left panels) and $d_{\rm resol} = 4\,\rm au$  for the small-scale map of the central region of 1000 au (right panels). We consider face-on (upper panels) and edge-on observations (lower panels). Generally, the magnetic field morphology impacts the polarization degree mainly in the outer envelope of $d_{\rm proj} > 10000\,\rm au$. In the inner envelope of $d_{\rm proj} < 500\,\rm au$, the variation of $p$ is affected by the alignment properties of grains, with the enhancement of polarization degree induced by MRAT in this region.

\begin{figure*}
    \centering
    \includegraphics[width = 0.46\textwidth]{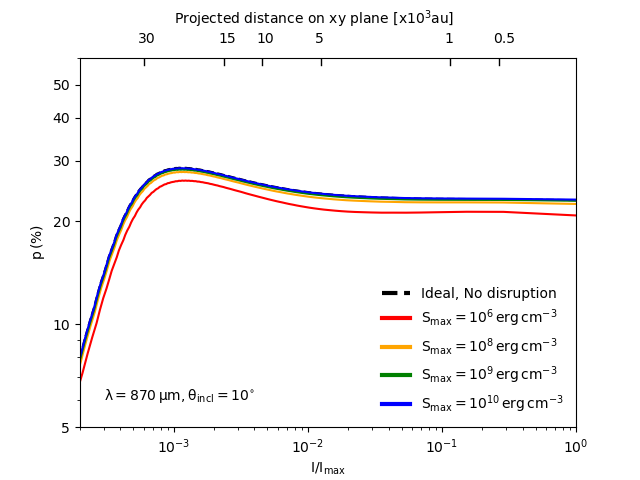} 
    \includegraphics[width = 0.46\textwidth]{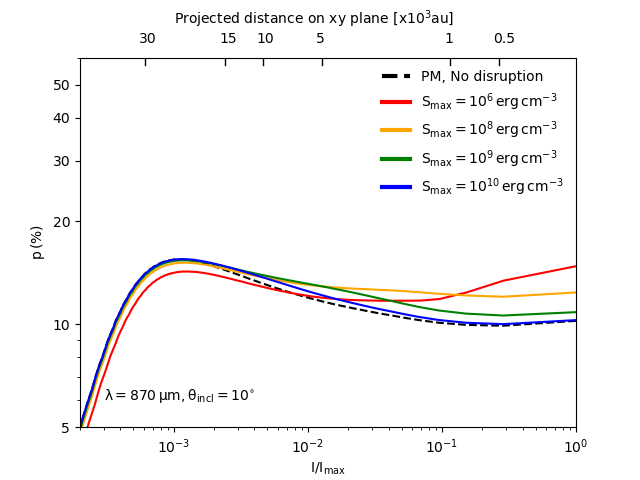}
    \includegraphics[width = 0.46\textwidth]{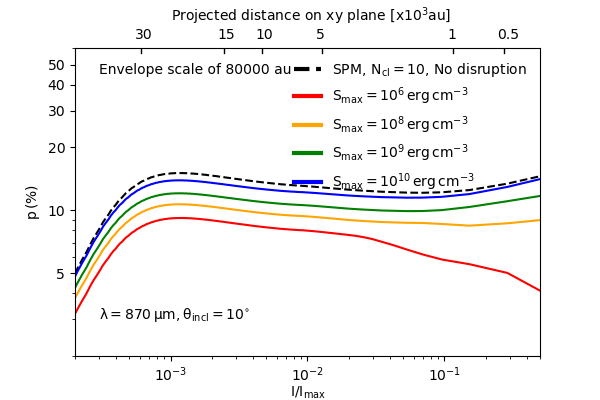}
    \includegraphics[width = 0.46\textwidth]{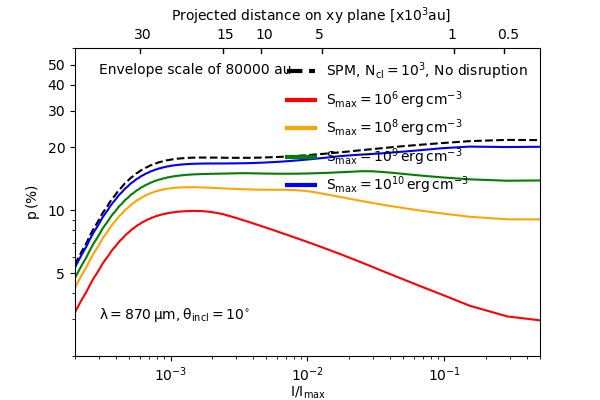}
    \caption{The effects of RAT-D on the intensity-dependent polarization degree observed in the face-on view (i.e., $\rm \theta_{incl} = 10^{\circ}$) for grains with various values of $S_{\rm max}$ (solid color lines). The model for non-disrupted grains is added for comparison (dashed black line). The polarization is reduced for porous grains (i.e., lower $S_{\rm max}$) fragmented by RAT-D. For PM grains, the impact of RAT-D is less effective for grains at low-J, resulting in the decreased polarization degree by a factor of 2.5. For SPM grains, the population of grains at high-J is enhanced by MRAT, leading to the efficient disruption by RATs and the reduction of polarization degree by a factor of 6.}
    \label{fig:pvsI_disr_incl10_fig}
\end{figure*}

The RAT-D mechanism is taken into consideration in the analysis of the averaged $p$ with increasing $I/I_{\rm max}$ observed at $\theta_{\rm incl} = 10^{\circ}$, as illustrated in Figure \ref{fig:pvsI_disr_incl10_fig}. We consider the RAT-D effect on grains with varying $S_{\rm max} = 10^6 - 10^{10}\,\erg\,\cm^{3}$ (solid color lines), while the case without disruption is plotted in the dashed black line for comparison. One can see in the Ideal model (upper left panel) that the polarization degree $p$ decreases due to the reduction of large grains by RAT-D. The polarization degree becomes lower with $p  < 5\%$ for grains with porous structures (i.e., low $S_{\rm max}$), compared with grains having compact structures (i.e., high $S_{\rm max}$) and higher $p > 10\%$.

In the Realistic models, the rotational properties of grains at high-J and low-J induced by MRAT can directly impact the RAT-D efficiency. For PM grains, the population of grains at high-J is only enhanced by MRAT at $d_{\rm proj} < 500\,\rm au$, while grains at low-J attractors have inefficient alignment with B-fields in the outer envelope. The impact of RAT-D is then less significant with the decreased polarization degree produced by mostly large grains at low-J attractors by only a factor of 2.5 from $S_{\rm max} = 10^{10}\,\erg\,\cm^{-3}$ to $S_{\rm max} = 10^6\,\erg\,\cm^{-3}$. For SPM grains, the extension of the MRAT active region enhances the abundance of grains at high-J attractors. The RAT-D effect becomes important and reduces the polarization degree by a factor of 6. This feature shifts from the inner to the outer envelope as increasing levels of iron inclusions from $10$ to $10^3$.

\subsection{The polarization spectrum \texorpdfstring{$p(\lambda)$}{plambda}}
\label{sec:pvslambda}
\begin{figure*}
    \centering
    \includegraphics[width = 0.46\textwidth]{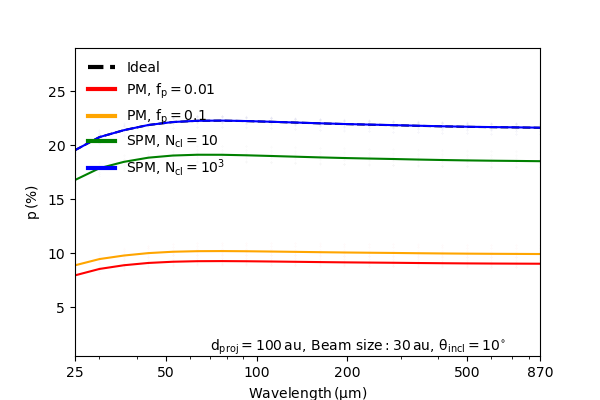} 
    \includegraphics[width = 0.46\textwidth]{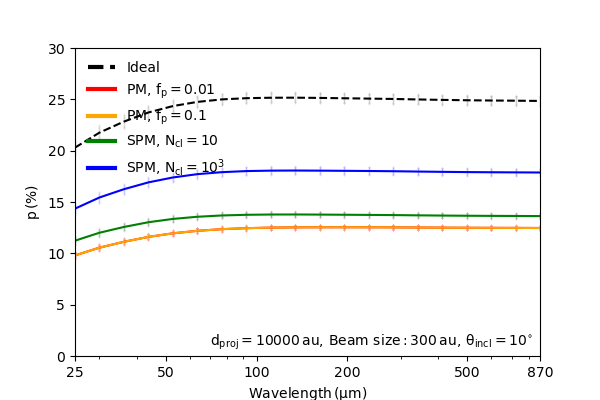}
    \caption{The face-on polarization spectrum at the wavelength from $\rm 25\,\mu m$ to $\rm 870\,\mu m$ induced by circumstellar dust at $d_{\rm proj} = 100\,\rm au$ (left panel) and $d_{\rm proj} = 10000\,\rm au$ (right panel) with different magnetic properties. The spectrum is produced mainly by large aligned grains at the equator, and the degree of polarization increases at the longer wavelengths $\lambda > 50\,\rm\mu m$.}
    \label{fig:Polspec_nodisr_incl_10}
    \includegraphics[width = 0.46\textwidth]{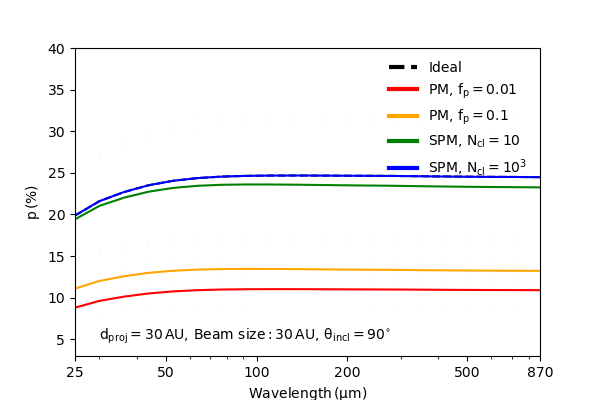}
    \includegraphics[width = 0.46\textwidth]{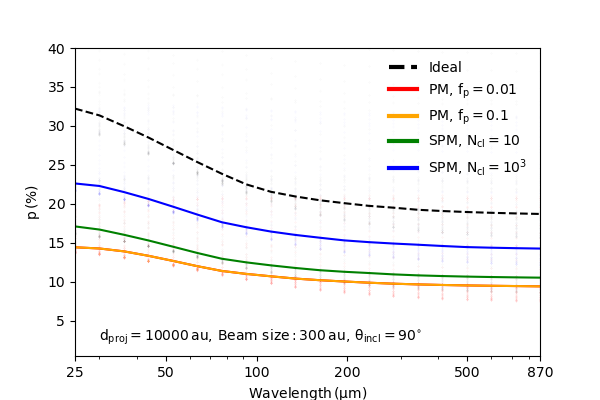}
    \caption{Same as Figure \ref{fig:Polspec_nodisr_incl_10} but considering the edge-on observations at $d_{\rm proj} = 30\,\rm au$ (left panel) and $d_{\rm proj} = 10000\,\rm au$ (right panel). At the outer region of the AGB envelope $d_{\rm proj} > 10000\,\rm au$, the pattern of the polarization spectrum changes to the increase in $p$ at the shorter wavelengths $\lambda < 50\,\rm\mu m$ due to the alignment of small grains in the polar region.}
    \label{fig:Polspec_nodisr_incl_90}
    
\end{figure*}

We investigate the polarization spectrum produced by aligned circumstellar grains in AGB envelopes in the range of optically thin wavelengths from $25\,\rm\mu m$ to $870\,\rm\mu m$, as shown in Figures \ref{fig:Polspec_nodisr_incl_10} and \ref{fig:Polspec_nodisr_incl_90}. The results are calculated in the IK Tau CSE's inner and outer regions, assuming the beam sizes of $30\,\rm au$ and $300\,\rm au$, respectively. Compared with the model Ideal plotted by the dashed black line, we consider the spectrum produced by both PM and SPM grains as illustrated in solid color lines.

Figure \ref{fig:Polspec_nodisr_incl_10} shows the results of the polarization spectrum in the face-on observation (i.e., $\theta_{\rm incl} = 10^{\circ}$) at $d_{\rm proj} = 100\,\rm au$ (left panel) and $d_{\rm proj} = 10000\,\rm au$ (right panel). In general, the spectrum arises from the thermal emission of aligned grains with $a_{\rm align} \sim 0.05\,\rm\mu m$ in the equator region with $\bf k \perp B$ (the lower panel of Figure \ref{fig:abar_maxJB_fig}). The spectrum calculated in the inner envelope of $d_{\rm proj} = 100\,\rm au$ reaches a peak at the short wavelength of $\lambda \sim 50\,\rm\mu m$ and decreases gradually, which arises from the thermal emission of warm dust near the central star. At $d_{\rm proj} = 10000\,\rm au$, the spectrum feature shows the continuous increase of $p$ to uniform values at $\lambda > 50\,\rm\mu m$, mainly from the thermal emission by cold dust located in the outer region of the IK Tau CSE.

Figure \ref{fig:Polspec_nodisr_incl_90} shows the same but in the edge-on view (i.e., $\theta_{\rm incl} = 90^{\circ}$). In the inner envelope of $d_{\rm proj} = 30\,\rm au$ (left panel), the alignment of large grains of $a_{\rm align} \sim 0.05\,\rm\mu m$ at the equator region strongly influences the polarization spectrum with the increase in $p(\%)$ increases toward the long wavelength regimes of $\lambda > 50\,\rm\mu m$. In the outer region of $d_{\rm proj} = 10000\,\rm au$ (left panel), the spectrum is dominantly induced by the alignment of small grains of $a_{\rm align} \sim 7 - 10\,\rm nm$ in the polar region due to the effect of stellar magnetic field geometry with $\bf k \parallel B$ that directly enhances the RAT alignment efficiency (Figure \ref{fig:abar_maxJB_fig}). As a result, the polarization fraction $p(\%)$ tends to increase toward the short wavelength regimes of $\lambda < 50\,\rm\mu m$.

The effects of iron inclusions on the polarization spectrum are considered in  Figures \ref{fig:Polspec_nodisr_incl_10} and \ref{fig:Polspec_nodisr_incl_90}. In the Realistic model of PM grains, the polarization spectrum is significantly lower than that in the Ideal case with $p < 15\%$ due to low magnetic susceptibility. The polarization degree tends to increase to $p \sim 20-35\%$ for SPM grains with increasing $N_{\rm cl}$ to $10^3$, which is a result of the enhanced magnetic susceptibility caused by the presence of embedded iron inside dust grains.

\begin{figure*}
    \centering
    \includegraphics[width = 0.45\textwidth]{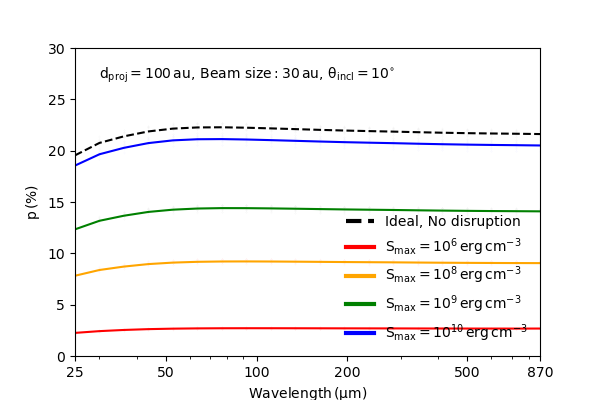} 
    \includegraphics[width = 0.45\textwidth]{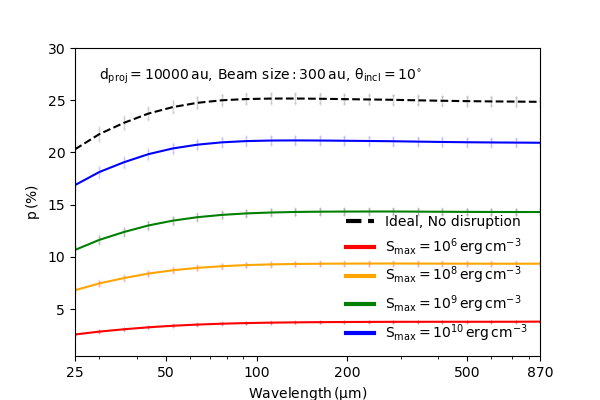}
    \includegraphics[width = 0.45\textwidth]{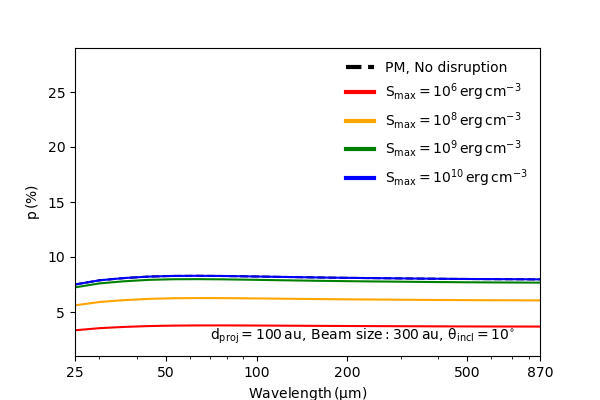}
    \includegraphics[width = 0.45\textwidth]{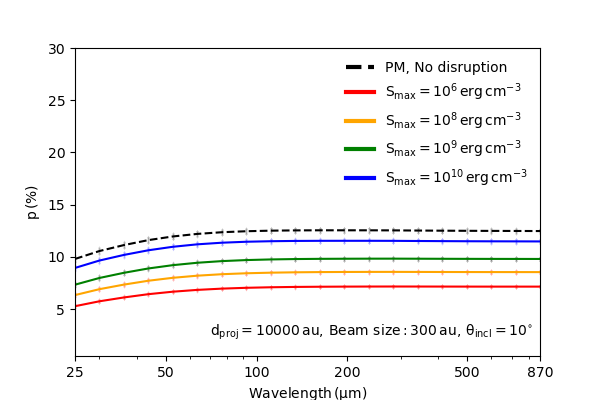}
    \includegraphics[width = 0.45\textwidth]{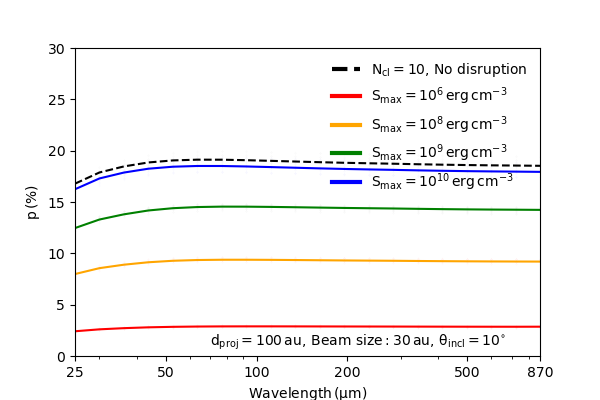}
    \includegraphics[width = 0.45\textwidth]{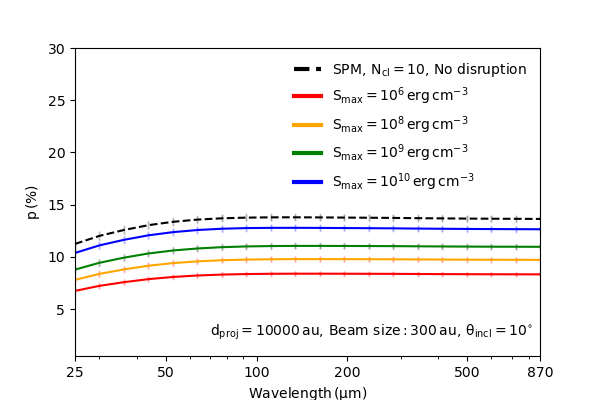}
    \includegraphics[width = 0.45\textwidth]{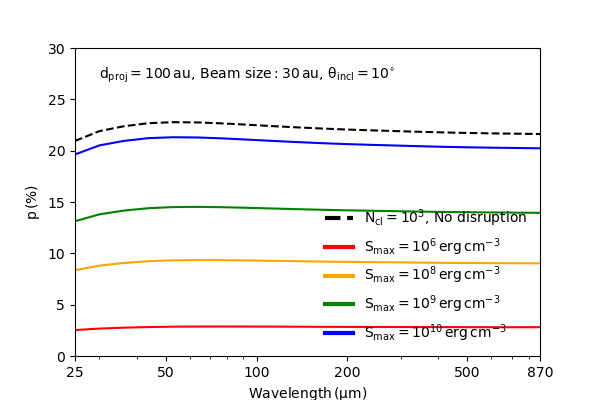}
    \includegraphics[width = 0.45\textwidth]{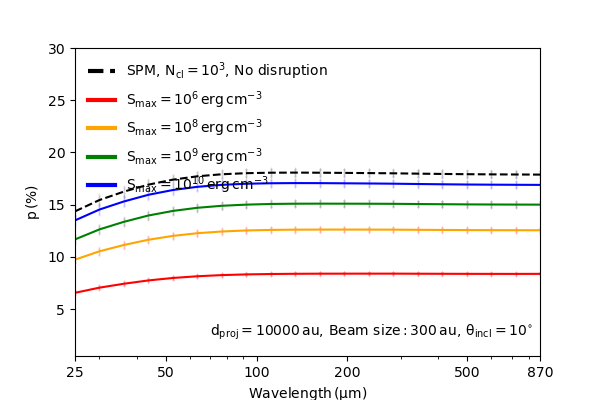}
    \caption{The impacts of RAT-D on the far-IR/sub-mm polarization spectrum at $d_{\rm proj} = 100\,\rm au$ (left panels) and $d_{\rm proj} = 10000\,\rm au$ (right panels) for various values of $S_{\rm max}$ (solid color lines) in comparison with the non-disrupted case (dashed black line), considering the face-on observation with $\theta_{\rm incl} = 10^{\circ}$. The level of polarization degree decreases with decreasing $S_{\rm max}$. The impact of RAT-D for PM grains mostly rotating at low-J is less effective, leading to a slight decrease in polarization degree by $5\%$. For SPM grains, they are populated by grains at high-J due to the MRAT effect, resulting in a considerable decrease in $p(\%)$ by $10-15\%$.}
    \label{fig:Polspec_disr_fig}
\end{figure*}

Figure \ref{fig:Polspec_disr_fig} concentrates on the effects of the RAT-D mechanism in modifying the polarization spectrum, calculating for grains with different values of $S_{\rm max}$. The polarization curve for the non-disrupted grains is added for comparison. In the Ideal case (first row), one obtains the significant reduction of polarization degree in the calculation at both $d_{\rm proj} = 100\,\rm au$ (left panel) and $d_{\rm proj} = 10000\,\rm au$ (right panel), resulting from a large number of small grains produced by RAT-D. The RAT-D mechanism is more efficient for porous grains (i.e., low $S_{\rm max}$); consequently, the polarization degree calculated in the entire spectrum is lower than that produced by large compact grains (i.e., high $S_{\rm max}$).

For PM grains (second row), the polarization spectrum is dominantly produced by grains at low-J attractors aligned with B-fields with a low alignment degree. The RAT-D impact is less effective with the slight decrease in polarization degree with decreasing $S_{\rm max}$ by only $5\%$. The RAT-D efficiency is enhanced for SPM grains with high levels of iron inclusion $N_{\rm cl} > 10$ (third and last rows) due to the increased population of grains at high-J attractors by MRAT (Figure \ref{fig:a_MRAT_fig}). The polarization degree then decreases significantly by $10-15\%$.

% Figure \ref{fig:Polspec_disr_fig} illustrates the difference in the resulting polarization spectrum at the inner and the outer region of the IK Tau CSE. At $d_{\rm proj} = 100\,\rm au$, the results show the flipping in the polarization spectrum with higher $p$ for lower values of $S_{\rm max}$ due to the combined effects of the RAT-D and MRAT mechanism on enhancing grain alignment efficiency of small porous grains constrained by RAT-D (Figure \ref{fig:a_MRAT_fig} and \ref{fig:Disr_fig}). Meanwhile, at $d_{\rm proj} = 10000\,\rm au$, the RAT-D mechanism mainly contributes to reducing the polarization degree, resulting in lower $p$ with decreasing $S_{\rm max}$.

% Turning to the models of SPM grains, the results show the modification from the flipping of polarization spectrum in the inner region of $d_{\rm proj} = 100\,\rm au$ for SPM grains with $N_{\rm cl} = 10$ (third row) to the flipping observed at the outer region of $d_{\rm proj} = 10000\,\rm au$ for SPM grains with $N_{\rm cl} = 10^3$ (last row) due to the extended contribution of MRAT regions to the outer envelope for SPM grains with high levels of iron (Figure \ref{fig:a_MRAT_fig}). And as increasing $N_{\rm cl}$ to $10^3$, the polarization spectrum calculated in the inner region of $d_{\rm proj} = 100$ tends to be reduced mainly by the RAT-D effect with lower $p$ for porous grains with low $S_{\rm max}$. 

\section{Discussion}
\label{sec:discussion}
Here, we summarize the main results and discuss the implications of our numerical results for studying dust and magnetic fields in the envelopes of AGB stars.
\subsection{Grain alignment properties}
\label{sec:discusson_mag_alignment}
The study of internal and external grain alignment by RATs is vital for examining how we reliably extract the morphology of stellar magnetic fields by using thermal dust polarization in astronomical environments. Both the internal and external alignment efficiencies of grains depend greatly on the environmental conditions (e.g., radiation, gas density, and magnetic fields) and the grain properties (e.g., mineralogy, size distribution, and magnetic properties). In the diffuse ISM environment and dense MCs, small PM grains of $a < 1\,\rm\mu m$ could sufficiently achieve perfect internal alignment with the Barnett relaxation faster than gas randomization and efficient external alignment by RATs (\citealt{Lazarian2007}; \citealt{Andersson2015}). However, this is not the case in dense environments such as protostellar cores and disks in which VLGs of $a > 10\,\rm\mu m$ with PM materials formed and grown inside could not be effectively aligned with B-fields due to the strong gas damping with high density (i.e., $n_{\rm gas} > 10^6\,\cm^{-3}$). The detailed numerical studies of  \cite{Hoang2022} showed that the presence of iron in the form of clusters could enhance the external alignment by magnetic relaxation, and therefore, only SPM grains could be sufficiently aligned with the local B-fields. The authors also showed that the embedded iron could help grains at high-J attractors have efficient internal alignment rather than those at low-J attractors with slow thermal rotation, which was confirmed by the modeling in low-mass protostellar cores using the updated POLARIS by \cite{Giang2023}. The characterization of internal and external grain alignment is important in AGB environments due to high gas density and the presence of grain drift relative to the gas within the outflows.

Our detailed modeling of RAT alignment in the IK Tau envelope shows that, in terms of internal alignment (Section \ref{sec:Result_amaxaJ}), silicate grains at high-J attractors can easily have fast internal Barnett relaxation by RATs from the strong radiation from the central AGB star, even for PM and SPM materials. Hence, they can sufficiently have the perfect internal alignment. Nevertheless, for grains at low-J attractors (dashed color lines in Figure \ref{fig:abar_maxJa_fig}), the increasing gas randomization with decreasing envelope distance causes the reduction of the Barnett relaxation efficiency. As a result, the fast internal relaxation of PM grains rotating subthermally by gas collision can only happen in the outer envelope of $r > 10000\,\rm au$, while the existence of embedded iron inclusions in SPM grains helps them achieve effective internal relaxation in the inner envelope of $r < 500\,\rm au$.

 In terms of external alignment (Section \ref{sec:Result_amaxJB} and \ref{sec:Result_aDG}), we find that the effect of grain drift can slow down the rotation of silicate grains, especially for large grains $a > 0.05\,\rm\mu m$ moving at supersonic velocities $s_{\rm d} \gg 1$ (Figure \ref{fig:vd_grain} and \ref{fig:RAT_vd}). Despite this effect, silicate grains can have effective external alignment by RATs induced by luminous stellar radiation with small $a_{\rm align} \sim 0.007 - 0.05\,\rm\mu m$ (Figure \ref{fig:abar_maxJB_fig}). Additionally, we find that the presence of strong magnetic fields of $B \sim 10\,\rm mG$ to 1 G is responsible for the efficient magnetic relaxation for PM silicate grains, which enhances the population of grains at high-J with fast internal relaxation in the inner region of $r < 500\,\rm au$ and strengthens the external alignment by RATs (Figure \ref{fig:a_MRAT_fig}). This effect is more significant for PM and SPM grains with a higher abundance of embedded iron (i.e., high $f_p$ and $N_{\rm cl}$), allowing them to have perfect alignment with the ambient B-fields in the outer envelope of $r > 10000\,\rm au$ with $f_{\rm high-J} = 1$.

\begin{figure}
    \includegraphics[width = 1\linewidth]{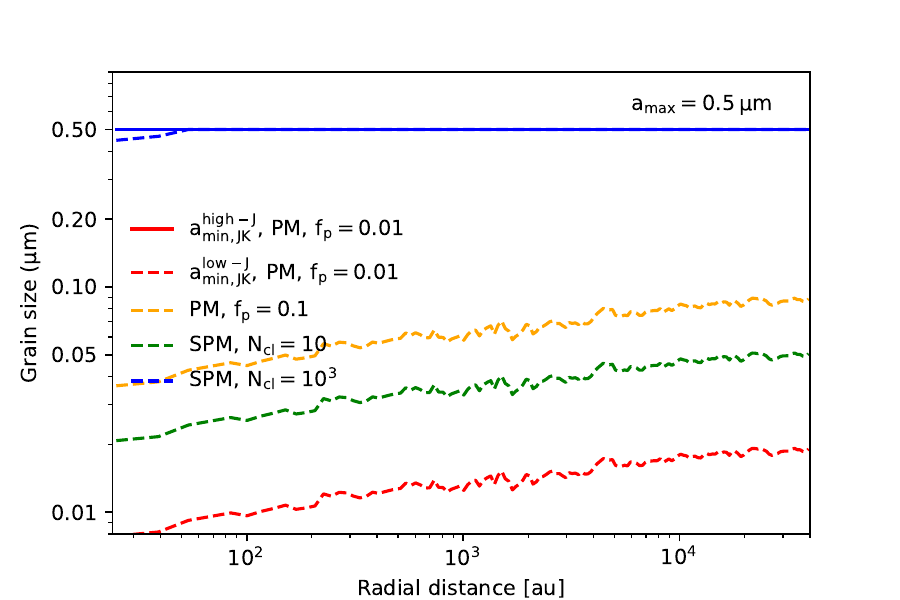}
    \caption{The minimum grain size for radiative alignment as a function of the envelope distance for grains at high-J (solid lines) and low-J attractors (dashed lines), denoted by $a^{\rm high-J}_{\rm min, JK}$ and $a^{\rm low-J}_{\rm min, JK}$, assuming the magnetic properties of PM and SPM grains. Grains at high-J attractors tend to achieve efficient magnetic alignment with $a^{\rm high-J}_{\rm min, JK} = a_{\rm max} = 0.5\,\rm\mu m$. Grains at low-J attractors are likely to be aligned with stellar radiation fields in the inner envelope of r < 500 au with $a^{\rm low-J}_{\rm min, JK} \sim 0.01 - 0.05\,\rm\mu m$. The radiative alignment efficiency decreases in the outer envelope $r > 500\,\rm au$ due to the reduced radiation strength.}
    \label{fig:kRAT}
\end{figure}

We note that circumstellar grains in AGB envelopes may be externally aligned with respect to the radial radiation field direction via k-RAT mechanism (see, e.g., \citealt{Lazarian2007}; \citealt{Hoang2014}; \citealt{Hoang2022}; \citealt{Tram2022}). The determination of the preferred alignment direction with respect to stellar magnetic fields $\bf B$ or radiation fields $\bf k$ strongly depends on the magnetic properties and rotational rates of grains (i.e., grains at high-J or low-J attractors). We calculate the minimum size for the radiative alignment, denoted by $a_{\rm min,JK}$ (also the maximum size for the magnetic alignment $a_{\rm max,JB}$) for both grains at high-J and low-J attractors in the IK Tau envelope and shown in Figure \ref{fig:kRAT}, assuming different magnetic properties of PM and SPM grains (see Equation 33 - 36 in \citealt{Tram2022}). For grains at high-J attractors, they are efficiently aligned with stellar magnetic fields with $a^{\rm high-J}_{\rm min, JK} = a_{\rm max} = 0.5\,\rm \mu m$ (solid line). For grains at low-J attractors of PM material (dashed red line), on the other hand, they are likely to be more aligned with stellar radiation fields toward the inner envelope of $r < 500 \,\rm au$ with $U \sim 10^4 - 10^9$ (Figure \ref{fig:Rad_fig}), with low $a^{\rm high-J}_{\rm min, JK} \sim 0.01 - 0.05\,\rm\mu m$. Nevertheless, the inner envelope is populated by PM grains at high-J attractors being enhanced by the magnetic relaxation due to strong stellar magnetic fields (Figure \ref{fig:a_MRAT_fig}); as a result, the magnetic alignment is dominant in this region. Meanwhile, PM grains at low-J attractors in the outer envelope r > 500 au have a lower radiative alignment efficiency with increasing $a^{\rm low-J}_{\rm min, JK}$ owing to the reduced radiation strength ($a^{\rm low-J}_{\rm min, JK} \varpropto U^{-2/3}$, see \citealt{Hoang2022}) and tend to achieve the alignment with B-fields. Consequently, the magnetic alignment by RATs is the most significant process for PM silicate grains in AGB envelopes and can be more significant for SPM grains having high levels of iron inclusions (see also Figure \ref{fig:a_MRAT_fig}). This is strongly related to the potential for reproducing the large-scale magnetic field morphology in AGB envelopes and will be discussed in Section \ref{sec:discussion_map_Bfield}.

\subsection{Roles of embedded iron inclusions in enhancing polarization degree}
\label{sec:discussion_iron}
The existence of iron inclusions locked inside grains is a key parameter for the alignment of large grains observed in very dense environments, such as protostellar cores/disks in the early phase of stellar formation. Numerical modeling of dust polarization in the low-mass protostellar cores using updated POLARIS by \cite{Giang2023} showed that large grains with iron inclusions can produce the polarization fraction $p\,(\%)$ up to $\sim 40\,\%$, which is consistent with the SMA and ALMA observations of protostellar cores/disks (\citealt{Hull2014}; \citealt{Cox2018}). This effect was also found in the thermal dust polarization observations with $p \sim 20-30\,\%$ induced by micron-sized grains $a < 1\,\rm\mu m$ in massive filaments (\citealt{Ngoc2023}).

In evolved star environments, using the updated POLARIS, we find that the enhanced alignment by MRAT is a major cause for the increase in polarization degree produced by PM sub-micron grains $a < 0.2\,\rm\mu m$ (see Figure \ref{fig:a_MRAT_fig}) toward the inner region of $r < 500\,\rm au$ to $\sim 10\,\%$, as illustrated in the profiles of $p - I/I_{\rm max}$ (see Section \ref{sec:pvsI}). For SPM grains with high levels of iron inclusion, grains at high-J attractors are populated by the MRAT. This considerably enhances the overall polarization degree in the entire envelope, possibly up to $\sim 20 - 40\,\%$, and extends to the outer envelope of $r > 10000\,\rm au$ following the extended MRAT effect. This effect is also presented in the prediction of the polarization spectrum at far-IR/sub-mm, as illustrated in Figures \ref{fig:Polspec_nodisr_incl_10} and \ref{fig:Polspec_nodisr_incl_90}. This helps constrain the level of iron locked inside grains via dust polarimetry in AGB envelopes, which will be discussed in Section \ref{sec:discussion_constrain_iron}.

\subsection{Roles of RAT-D in modifying dust properties and polarization degree}
The rotational disruption from stellar radiation on the properties of circumstellar dust has a major impact on the observational properties of evolved stars. \cite{Tram2020} showed that the RAT-D effects are independent of grain mineralogy and could be applied in the environments of both C-rich and O-rich AGB envelopes. The authors constrained the upper limit of grain size distribution by RAT-D and found the disruption of large grains of $a < 0.25\,\rm\mu m$ into smaller species as nanoparticles (i.e., $a < 100\,$\AA), which later on induces spinning dust emission at microwave frequencies below 100 GHz. \cite{Truong2022} studied the impacts of RAT-D in the Betelgeuse envelope and found their effects on the extinction and reddening of circumstellar dust at optical/near-IR wavelengths. The RAT-D effects depend on the grain internal structures and the local conditions of radiation fields and gas density, and could be used for interpreting photometric and spectroscopic observations of evolved stars in the near-UV to near-IR regimes.

In Section \ref{sec:pvsI} and \ref{sec:pvslambda}, we show the impacts of the RAT-D mechanism on the thermal dust polarization in AGB envelopes and the polarization spectrum at far-IR/sub-mm wavelengths. In general, a dominant amount of smaller grains of $a < 0.5\,\rm\mu m$ by RAT-D (Figure \ref{fig:Disr_fig}) results in the reduction of the polarization degree. The impact of rotational disruption is more efficient for porous grains with low $S_{\rm max}$ than for compact grains with high $S_{\rm max}$. The disruption by RATs is more significant if grains are located in the inner envelope of $r < 500\,\rm au$ due to the enhanced MRAT efficiency by the strong stellar magnetic fields or SPM grains with iron inclusions. By measuring the degree of thermal dust polarization, we can retrieve further information on dust properties in terms of both the internal structures of grains and the magnetic properties of circumstellar grains.

\subsection{Effects of magnetic field morphology on synthetic dust polarization}
\label{sec:discussion_mag_effect}
So far, we have stressed the importance of grain alignment and disruption in calculating the polarization of thermal dust emission in evolved star envelopes. Yet, even the conditions of the global magnetic fields themselves (i.e., magnetic field strength and geometry) significantly impact the measurement of dust polarization in the observed plane of the sky. The variation of the angle $\psi$ between the local magnetic fields and the anisotropic radiation fields could impact the grain alignment efficiency by RATs (\citealt{Lazarian2007}; \citealt{Hoang2009}; \citealt{Andersson2011}; \citealt{Lazarian2021}), while the field strength could affect the magnetic relaxation of PM and SPM grains, which then affect the efficiency of MRAT alignment (\citealt{Hoang2016a}; \citealt{Hoang2022}). More importantly, the orientation of the inclined magnetic fields with respect to the plane-of-sky strongly impacts the dust polarization degree by the projection effect (\citealt{Reissl2017}; \citealt{Chen2019}; \citealt{Hu2023}), which has been studied in explaining the dust polarization observations in MCs and young stellar objects (YSOs) (\citealt{Yang2016}; \citealt{Harrison2019}; \citealt{Tahani2022}).

In the evolved star's envelopes with the dipole magnetic field, we show the magnetic field geometry affects the efficiency of the external alignment by RATs in Section \ref{sec:Result_amaxJB}. The difference in the angle $\psi$ between the local fields and the radial stellar radiation at the equator region (i.e., $\psi \sim 0^{\circ}$) and the polar region (i.e., $\psi \sim 90^{\circ}$) causes the difference in the distribution of $a_{\rm align}$ (Figure \ref{fig:abar_maxJB_fig}), with smaller grains of $a_{\rm align} \sim 7 - 10\,\rm nm$ being aligned by RATs in the polar region of $r > 1000\,\rm au$, while the alignment size is larger at the equator of $a_{\rm align} \sim 0.05\,\rm\mu m$. Moreover, the strong field strength toward the inner region of $r < 500\,\rm au$ is a main factor in enhancing external alignment by MRAT for both PM and SPM grains (Figure \ref{fig:a_MRAT_fig}), resulting in the increase in polarization degree as we mentioned earlier in Section \ref{sec:discussion_iron}.

The projection effect of the local magnetic field with the plane-of-sky is considerable in the calculation of the intensity-dependent polarization degree. In Section \ref{sec:Dust_pol_map} and \ref{sec:pvsI}, by taking the calculations at different inclination angles, we find the change in the curvature of the three-dimensional dipole field could lead to the variation of the inclined magnetic field with the plane-of-sky and, therefore, enhance or reduce the polarization degree at each radial envelope distance. The impact of inclined magnetic fields could subsequently explain the increase in $p$ with decreasing envelope distance when observed in the face-on view of $\theta_{\rm incl} = 10^{\circ}$ (Figure \ref{fig:pvsI_nodisr_10_fig}) and the decrease in $p$ when observed in the edge-on view of $\theta_{\rm incl} = 90^{\circ}$ (Figure \ref{fig:pvsI_nodisr_90_fig}).

The effects of magnetic field geometry are also included in the calculation of the thermal dust polarization spectrum at far-IR/sub-mm. In Section \ref{sec:pvslambda}, we show that the difference in the size distribution of aligned grains at the equator and the polar region due to magnetic field morphology (the upper panels of Figure \ref{fig:abar_maxJB_fig}) results in the difference in the spectrum of polarized thermal dust emission induced by grains in these regions. And as shown in Figure \ref{fig:Polspec_nodisr_incl_10}, the spectrum calculated in the face-on view is mostly produced by large aligned sizes in the equator region with the increasing polarization degree in longer wavelengths of $\lambda > 50\,\rm\mu m$. Meanwhile, in the calculation in the edge-on view (Figure \ref{fig:Polspec_nodisr_incl_90}), the spectrum is affected by the thermal dust emission from small aligned grains in the polar region, causing the rise in the polarization degree toward shorter wavelengths of $\lambda < 50\,\rm\mu m$.

However, as discussed by \cite{Vlemmings2019}, the possibility of stellar magnetic fields having a toroidal configuration cannot be ruled out, which was detected in the envelopes of AGB stars, post-AGB objects, and PNe via maser polarization (\citealt{Vlemmings2006}; \citealt{Ferreira2013}) and dust continuum polarization (\citealt{Greaves2002}; \citealt{Sabin2015}; \citealt{Sabin2020}). In comparison with the dipole field, the strength of the toroidal field is lower by a factor of 10 (see in Figure \ref{fig:Mag_geometry_toroidal} in the Appendix \ref{sec:appendix_toroidal}), particularly in the inner envelope of $r < 100\,\rm au$. Instead of being enhanced, the MRAT efficiency is expected to be significantly reduced due to lower magnetic strength, and being affected by the strong thermal fluctuation of electrons inside grains due to high dust temperature $T_{\rm d} > 100\,\rm K$ in the inner envelope close to the central AGB star (see in \citealt{Hoang2016a}). The inner envelope of $r < 1000\,\rm au$ is then dominated by grains at low-J attractors, with decreased alignment efficiency with increasing gas randomization (Figure \ref{fig:abar_maxJa_fig}). This could result in the `polarization hole' effect in which the polarization degree considerably decreases toward the inner envelope of AGB stars (see Appendix \ref{sec:appendix_toroidal}). Hence, the impacts of magnetic field geometry on both grain alignment properties and the polarization degree should be considered in the study of dust polarization from aligned grains in AGB envelopes.

\subsection{Implications for constraining the global stellar magnetic fields}
\label{sec:discussion_map_Bfield}
In Section \ref{sec:discusson_mag_alignment}, we found that the internal relaxation by the Barnett effect is efficient, applying to both PM silicate grains at high-J attractors enhanced by magnetic relaxation in the inner envelope of $r < 500 \,\rm au$ (Figure \ref{fig:a_MRAT_fig}) and grains at low-J in the outer envelope of $r > 10000\,\rm au$ (Figure \ref{fig:abar_maxJa_fig}). This produces an abundance of grains having fast internal relaxation (then right internal alignment with $\bf J \parallel a_1$). Additionally, silicate grains can have perfect external alignment by both RATs and MRAT with $\bf J \parallel B$ (see Section \ref{sec:discussion_iron}). This generates polarized thermal dust emission with the shortest axes parallel to the magnetic fields ($\bf P \perp B$) and reproduces the large-scale magnetic field morphology through the polarization patterns (Figures \ref{fig:Dustpol_mag_scale10000_fig} and \ref{fig:Dustpol_mag_scale1000_fig}). We can directly map the global magnetic fields derived from the polarimetric observations of far-IR/sub-mm instruments such as SOFIA/HAWC+, JCMT, CARMA, and ALMA. Combing with other measurements in the local B-fields from maser (\citealt{Boboltz2005}; \citealt{Ferreira2013}) and non-maser line emissions (\citealt{Vlemmings2012}; \citealt{Huang2020}), we can obtain a more general picture of the magnetic field geometry, which is a major step to understand further its role in the mass-loss mechanism in evolved star's envelopes.

We examined the dependence of the polarization degree on the magnetic geometry, particularly the projection effect on the plane-of-sky when calculated at different viewing inclination angles (Section \ref{sec:discussion_mag_effect}). This effect mainly happens in the observation at the outer envelope of $r > 10000\,\rm au$ where the variation of magnetic field curvature is significant, resulting in the modification of the polarization degree along the envelope distance as shown in the profile of $p - I/I_{\rm max}$ (see Figure \ref{fig:pvsI_nodisr_fig}). By interpreting the intensity-dependent polarization degree from observational data, we could roughly estimate the observed inclination angle with respect to the line-of-sight, leading to the ultimate goal of reconstructing three-dimensional global magnetic fields of AGB stars.

In addition, the observed inclination angle can be determined through the thermal polarization spectrum at far-IR/sub-mm. In the case of the dipole field, the wavelength-dependent polarization degree induced by circumstellar dust is distinguishable between the observations at the face-on and edge-on views within the wavelengths of $\lambda \sim 25 - 100\,\rm\mu m$, especially at the outer envelope of $d_{\rm proj} \sim 10000\,\rm au$ as a result of the impact of magnetic field geometry on the alignment properties (Section \ref{sec:discussion_mag_effect}). Thus, the resulting spectrum of polarized thermal emission from aligned grains could contribute to identifying the three-dimensional structure of stellar magnetic fields, which requires the future development of far-IR spectroscopic instruments in the next decades.

Note that the technique of tracing stellar magnetic fields via thermal dust polarization is highly applicable in the envelopes of O-rich AGB stars. In the C-rich AGB stars, \cite{Hoang2023} studied in detail the alignment of carbonaceous grains and found that the diamagnetic carbonaceous grains could have efficient internal alignment by inelastic relaxation and tend to be aligned with radiation field direction rather than magnetic field direction. Observations of thermal dust polarization from IRC+10216 by \cite{Andersson2022} reveal the radial polarization pattern, which is suggested as evidence of carbonaceous alignment along the radiation direction with wrong internal alignment due to grain drift relative to the gas. Besides, for charged carbonaceous grains with relatively small magnetic moments, they could be aligned with the electric field induced by the drift of grains relative to the ambient B-fields $\bf E_{\rm ind} = [v_{\rm grain}/c \times B]$ (see \citealt{Lazarian2020}). The alignment with respect to the induced electric field could occur for small grains at low-J attractors $a < 0.01\,\rm\mu m$ in the outermost of the C-rich AGB envelopes, as numerically studied by \cite{Hoang2023}. In future research, we will incorporate the detailed alignment physics for carbonaceous grains developed by \cite{Hoang2023} into the updated POLARIS and perform modeling of dust polarization by carbonaceous grains for different astrophysical environments.

\subsection{Implications for constraining iron abundance in circumstellar dust}
\label{sec:discussion_constrain_iron}
Observations (e.g., \citealt{Jenkins2009}) reveal that over $95\,\%$ of iron is locked inside interstellar dust, which is presented in the outflows of evolved stars (e.g., \citealt{Jones1990}; \citealt{Tielens2005}) and the ejecta of supernovae (SNe) (\citealt{Dwek2016}). The studies of gaseous depletion in the AGB envelopes showed that iron could be incorporated into the newly-formed dust at the dust formation zone about two to three stellar radii (\citealt{McDonald2011}; \citealt{Dwek2016}; \citealt{Marini2019}). However, in what form iron atoms exist in dust grains, i.e., diffusive distribution or iron clusters, remains elusive. The theoretical modeling by \cite{Karovicoiva2013} and \cite{Hofner2022} showed that the contribution of iron in olivine grains or in the metallic form in the wind-driving mechanism could potentially be detected in the signature of mid-IR interferometric spectroscopy. Yet, determining an accurate iron level and its form inside grains is challenging. \cite{Hoang2016a} studied the alignment of grains with iron inclusions and showed the possibility of constraining the level of iron inclusions in cosmic dust via dust polarization.

In Section \ref{sec:discussion_iron}, we mentioned the crucial role of embedded iron in enhancing the polarization degree produced by PM and SPM grains in AGB envelopes as a result of the enhanced efficiency of internal and external alignment with B-fields (Section \ref{sec:discusson_mag_alignment}). The effects of embedded iron for different magnetic properties of grains are more prominent in the inner envelope of $r < 500\,\rm au$ in which the alignment is mainly impacted by the MRAT mechanism, as illustrated in the results of $p - I/I_{\rm max}$ regardless of the configuration of magnetic field geometry (see Figure \ref{fig:pvsI_nodisr_fig} and \ref{fig:pvsI_nodisr_toroidal}). By measuring the dependency of polarization degree induced by circumstellar dust on the intensity at sub-mm bands, we could first identify the configuration of iron atoms distributed inside grains (i.e., PM or SPM grains) and, later on, determine the level of embedded iron through the level of polarization degree. This provides proper insight into circumstellar dust properties and helps advance our understanding of their formation and evolution within the evolved star outflows.

\subsection{Effects of grain drift on grain alignment}
\label{sec:discussion_drift}
Up to now, we have studied the magnetic alignment by RATs and MRAT for circumstellar grains when the effect of grain drift relative to the gas on the rotational damping is taken into consideration. Our numerical calculations showed that the presence of relative grain drift can reduce the rotation rate of grains; nevertheless, the RAT and MRAT alignment is still efficient in AGB envelopes because circumstellar grains are exposed to strong stellar radiation fields (see Section \ref{sec:discusson_mag_alignment}).

It is worth noting that the relative grain drift itself can contribute to the spin-up and alignment processes of circumstellar grains in AGB envelopes by METs \citep{LazHoang.2007b,HoangChoLazarian2018} - analogous to the effects of RATs from stellar radiation. For nano-sized grains $a < 10 \,\rm nm$ firstly formed in the inner envelopes, the rotational excitation of spinning grains by {\it stochastic} METs can spin up the grain rotation rate to suprathermal speed (see, e.g., \citealt{HoangTram2019}; \citealt{Tram2020}). For large grains $a > 0.05\,\rm\mu m$, with irregular shapes, the {\it regular} METs by the relative grain drift can bring them to suprathermal rotation and achieve the alignment with stellar magnetic fields (see \citealt{HoangChoLazarian2018}; \citealt{Hoang2022}). Here, we take analytical calculations of the suprathermal rotation parameter $St$ induced by the {\it regular} METs from the strong grain drift, which is given by (\citealt{Hoang2022})
\begin{equation}
\begin{split}
    St_{\rm MET} &= \frac{\Omega_{\rm MET}}{\Omega_{\rm ther}} = \left(\frac{s_d^2 v_{\rm ther}}{a}\right)\left(\frac{5s\sqrt{\pi}Q_{\rm spinup}}{8\Gamma_{\parallel}}\right)\left(\frac{I_{\parallel}}{kT_{\rm gas}}\right)^{1/2} \\
    &\simeq 0.86 \hat{\rho}^{1/2}s_{\rm d,-1}^2 a_{-5}^{3/2}\left(\frac{s^{3/2}Q_{\rm spinup,-3}}{\Gamma_{\parallel}}\right),
\end{split}
\end{equation}
where $s_{\rm d,-1} = s_{\rm d}/0.1, Q_{\rm spinup,-3} = Q_{\rm spinup}/10^{-3}$ is the spin-up efficiency by METs (\citealt{HoangChoLazarian2018}) and $\Gamma_{\parallel}$ is the geometric factor due to gas collision (for $s = 0.5, \Gamma_{\parallel} \simeq 0.62$, see \citealt{Roberge1993}).

\begin{figure}
    \includegraphics[width = 1\linewidth]{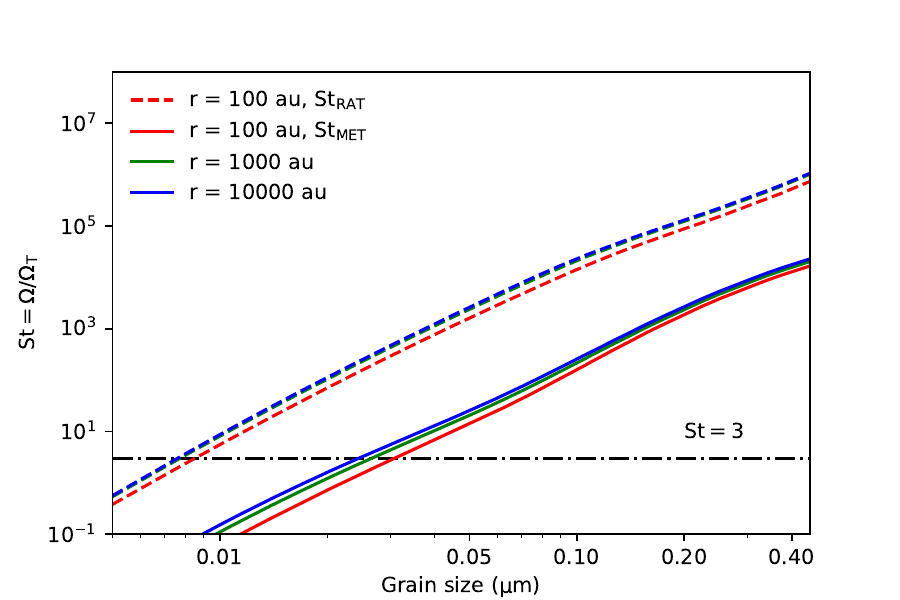}
    \caption{The analytical calculations of the suprathermal rotation parameter $St$ for grain sizes at r = 100, 1000, and 10000 au produced by RATs from stellar radiation (dashed lines) and METs from the grain drift relative to the gas (solid lines). The METs can bring grains $a > 0.02 - 0.03\,\rm \mu m $ to suprathermal rotation and achieve efficient magnetic alignment. However, their rotation rate is much lower than the rotation rate induced by RATs from luminous stellar radiation.}
    \label{fig:RAT_MET}
\end{figure}

Figure \ref{fig:RAT_MET} shows the comparison of the suprathermal rotation $St$ over grain sizes produced by RATs and METs at $r = 100, 1000$, and 10000 au in the IK Tau envelope. The METs from the relative grain drift by radiation pressure can efficiently spin up grain sizes $a > a_{\rm align} \sim 0.02 - 0.03\,\rm\mu m$ to suprathermal speed and help them be aligned with B-fields. However, the rotation rate induced by the METs is lower by a factor of $\sim 100$, compared to the RATs due to intense radiation from the central AGB star with $U_{\rm rad} \sim 10^4 - 10^9$ (Figure \ref{fig:Rad_fig}). Therefore, the RAT and MRAT alignment are the more important processes of grain alignment in AGB circumstellar environments.  The MET alignment could be more significant for an environment with weaker radiation fields and stronger gas flow, for instance, in protostellar disks where the radiation fields are strongly reddened by VLGs, and these grains experience strong grain drift due to differential Keplerian rotation with respect to the gas (see \citealt{Hoang2022}). In the special case of AGB envelopes, the direction of grain drift and radiation fields are parallel to each other with $\bf k \parallel v$. Then, the combined contribution of RATs and METs is expected to enhance the alignment efficiency and could result in a much higher polarization degree. In follow-up studies, we will implement the detailed physics of grain alignment by both {\it stochastic} and {\it regular} METs in the recent version of the POLARIS code to investigate the MET alignment for different conditions of astrophysical environments.

\subsection{Effects of stellar mass-loss rate}
Throughout this paper, we performed the numerical modeling of thermal dust polarization from aligned grains for a specific case of the IK Tau envelope. We discussed the implications of our results for mapping stellar magnetic fields (Section \ref{sec:discussion_map_Bfield}) and constraining the abundance of iron locked inside grains (Section \ref{sec:discussion_constrain_iron}). Our numerical modeling method can also be applied to the CSEs of other O-rich AGB and RSG stars having different physical properties, such as stellar mass-loss rates.

The mass-loss rate is expected to directly affect the grain alignment efficiency of circumstellar grains and the results of thermal dust polarization. Consider the case in which circumstellar matter is being pushed away by radiation pressure with the mass-loss rate proportional to the luminosity as $\dot{M} \varpropto L_{\ast}R_{\ast}/M_{\ast}$(\citealt{Reimers1977}, see also in \citealt{deJager1988}; \citealt{vanLoon2005}; \citealt{Danilovich2015}; \citealt{Prager2022}). The increased mass-loss rate increases the gas density (Equation \ref{eq:gas_density}) and the rate of gas randomization, while the increased stellar luminosity increases the local radiative energy density as $u_{\lambda}\propto L_{\star}/(4\pi r^{2}c)$, which produces stronger RATs onto dust grains (Equation \ref{eq:radiative_torque}). The stellar luminosity and mass-loss rate also affect the grain drift relative to the gas and could produce much stronger drift for AGB stars with higher $L_{\ast}$ and lower $\dot{M}$, which enhances the rotational damping of spinning grains (Equation \ref{eq:vd} and \ref{eq:F_sd}). Even that, the efficiency of RAT alignment is expected to not significantly change due to the competition between enhanced gas randomization and enhanced spin-up by stronger RATs. However, the increased gas randomization due to higher mass loss could decrease the internal alignment efficiency by Barnett relaxation and the external alignment efficiency by MRAT. The enhanced mass-loss rate could then lead to a decrease in the degree of thermal dust polarization (see Figure \ref{fig:pvsI_Luminosity} in Appendix \ref{sec:appendix_massloss}). The effect of stellar mass-loss rate is essential for grain alignment and thermal dust polarization in AGB environments and will be investigated further in future works.

\section{Summary}
\label{sec:summary}
Our study concentrates on predicting the polarization of thermal dust emission from aligned grains in the CSE of IK Tau using the physics of grain alignment and rotational disruption by magnetically enhanced RATs, as well as the effect of grain drift relative to the gas within the outflows. Using the updated version of POLARIS developed by \cite{Giang2023}, we then perform the numerical modeling of thermal dust polarization induced by aligned grains and discuss the main implications for investigating magnetic fields and dust properties in evolved star outflows via far-IR/sub-mm polarimetry. The main findings are summarized as follows:

\begin{enumerate}
    \item Despite grain drift relative to the gas, paramagnetic (PM) silicate grains at high-J attractors can have suprathermal rotation induced by RATs from the strong stellar radiation and achieve efficient internal and external alignment. They are effectively aligned with their shortest axis perpendicular to the ambient B-fields with the minimum alignment size of $a_{\rm align} \sim 0.007 - 0.05 \,\rm\mu m$. The efficiency of grain alignment is significantly increased by the magnetic relaxation induced by the strong magnetic fields of $\rm B \sim 10\,\rm mG - 1\,G$, allowing grains to have efficient alignment in the inner envelope of $r < 500\,\rm au$ with $n_{\rm gas} > 10^6 \,\cm^{-3}$.  Meanwhile, grains at low-J attractors can have efficient alignment only in the outer envelope of $r > 10000\,\rm au$ with $n_{\rm gas} < 10^3 \,\cm^{-3}$. The presence of iron inclusions incorporated into superparamagnetic (SPM) grains can enhance grain alignment efficiency via the MRAT mechanism and help them to be aligned with stellar magnetic fields in the entire envelope.

    \item Both PM and SPM silicate grains with efficient internal and external alignment can produce the thermal dust polarization with the pattern of $\bf P \perp B$ at far-IR/sub-mm wavelengths. Consequently, the global structure of stellar magnetic fields in AGB envelopes can be obtained by rotating the polarization vectors to 90 degrees.

    \item The polarization degree produced by PM grains can increase up to $\sim 10\%$ toward the central region of $r < 500\,\rm au$ because of the enhanced grain alignment by MRAT. The polarization degree is considerably higher for SPM grains up to $\sim 20 - 40\,\%$, and this feature can be extended to the outer envelope of $r > 10000\,\rm au$. 

    \item The rotational disruption by RATs (RAT-D) can constrain the maximum size of silicate grains in AGB envelopes and affect the polarization degree. The increasing abundance of small grains of $a < 0.5\,\rm\mu m$ and steeper slope of the grain size distribution $\eta < -3.5$ due to RAT-D decreases the polarization degree. The effect of RAT-D is more prominent for porous grains with lower tensile strength $S_{\rm max}$. This mechanism is more efficient for grains aligned at high-J by MRAT, particularly for SPM grains with high levels of iron inclusions.

    \item The stellar magnetic field strength and geometry can directly influence the properties of grain alignment and dust polarization due to the projection effect of magnetic fields with respect to the observed plane of the sky. The magnetic field morphology can cause the enhancement or reduction of the degree of polarization, depending strongly on the variation of the magnetic fields along the line-of-sight when being observed at different inclination angles. 
    
    \item Thermal dust polarization from aligned grains is essential for determining the global magnetic field morphology in the evolved star's envelopes. The main structure of stellar magnetic fields can be mapped through the polarization pattern in the plane-of-sky, while the observed inclination angle can be identified through the $p - I$ relationship and the spectrum of polarized thermal dust emission. This provides a key tool for reconstructing the three-dimensional geometry of the magnetic fields of AGB stars.

    \item The properties of thermal dust polarization from aligned grains can potentially help us constrain the level of iron inclusions locked inside circumstellar dust. This offers a unique way to constrain the magnetic properties of circumstellar dust and understand the role of iron in the evolution of circumstellar dust in AGB envelopes. 

    \item The grain drift relative to the gas due to radiation pressure could spin up and align grains with the magnetic fields via the MEchanical Torques (METs). However, the effect of METs is less efficient than RATs in this intense radiation environment around the central AGB star.

    \item Our present numerical modeling of thermal dust polarization from aligned grains can also be applied to other O-rich evolved stars. However, the envelope properties, such as stellar mass-loss rates and luminosity, could affect our modeling results of grain alignment and thermal dust polarization. Our follow-up study will consider these effects to investigate dust properties and magnetic field morphology via thermal dust polarization in various conditions of evolved star envelopes.

\end{enumerate}

\section*{Acknowledgements}
We thank anonymous referees for helpful comments that improved our paper. This work was partly supported by a grant from the Simons Foundation to IFIRSE, ICISE (916424, N.H.). Dieu D. Nguyen is grateful to the LABEX Lyon Institute of Origins (ANR- 10-LABX-0066) Lyon for its financial support within the program “Investissements d’Avenir” of the French government operated by the National Research Agency (ANR). N.B.N. was funded by the Master, Ph.D. Scholarship Programme of Vingroup Innovation Foundation (VINIF), code VINIF.2022.TS083. We thank members of Vietnam Astrophysics Research NETwork (VARNET) for their contribution to commenting and improving the quality of this paper.

\section*{Data Availability}
The data underlying this article will be shared on reasonable request to the corresponding author.

%%%%%%%%%%%%%%%%% BODY OF PAPER %%%%%%%%%%%%%%%%%%

%%%%%%%%%%%%%%%%%%%% REFERENCES %%%%%%%%%%%%%%%%%%

% The best way to enter references is to use BibTeX:

\bibliographystyle{mnras}
\bibliography{AGB_Dust} % if your bibtex file is called example.bib

%%%%%%%%%%%%%%%%%%%%%%%%%%%%%%%%%%%%%%%%%%%%%%%%%%

%%%%%%%%%%%%%%%%% APPENDICES %%%%%%%%%%%%%%%%%%%%%

\appendix

\section{Synthetic polarization observations for the toroidal magnetic field}
\label{sec:appendix_toroidal}
\begin{figure*}
    \centering
    \includegraphics[width = 1\textwidth]{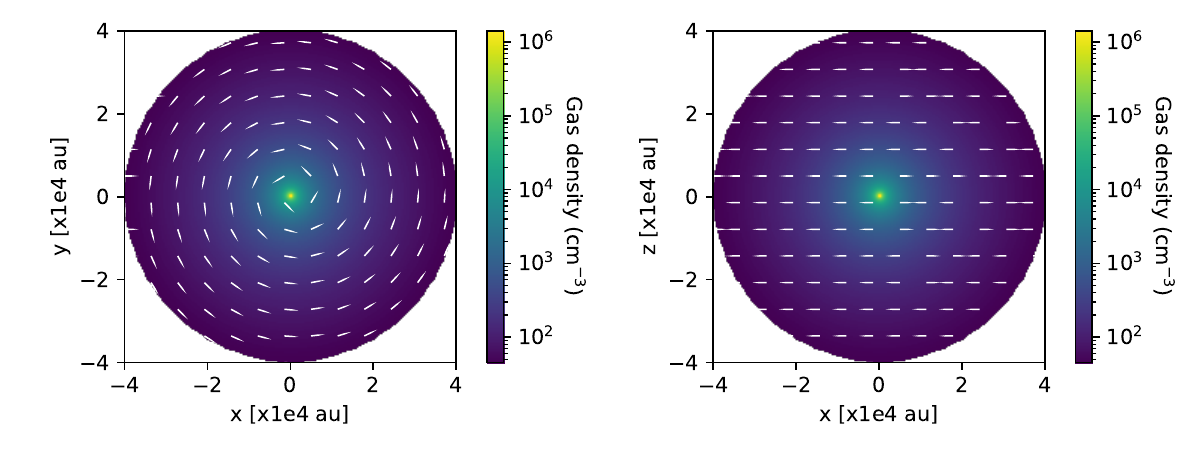}
    \includegraphics[width = 0.48\textwidth]{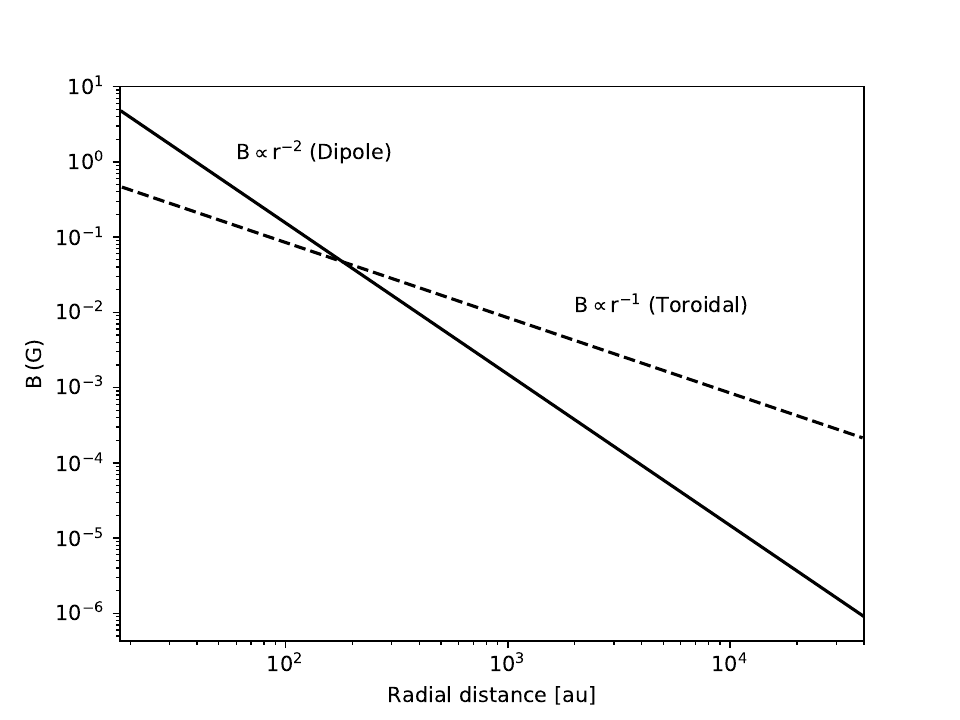}
    \caption{The morphology of the toroidal magnetic field in the $xy$- (upper left panel) and $xz$- planes (upper right panel). The lower panel illustrates the strength of the toroidal field with respect to the envelope distance with $B \sim r^{-1}$ (dashed black line), compared with the field strength of the dipole field (solid black line).}
    \label{fig:Mag_geometry_toroidal}
\end{figure*}

In this section, we quantify the effects of the toroidal magnetic field on the measurement of dust polarization in the O-rich AGB envelopes. We adopt the toroidal structure of stellar magnetic fields as illustrated in the upper panels of Figure \ref{fig:Mag_geometry_toroidal}. The field strength varies with $B \varpropto r^{-1}$ as plotted in a dashed black line followed by the radial profile of the toroidal field strength from maser observations (\citealt{Vlemmings2019}). In comparison with the dipole field (solid black line), the strength is reduced by a factor of 10 in the inner region of $r < 100\,\rm au$, while it increases up to 1 mG at $r \sim 10000\,\rm au$.

\begin{figure*}
    \centering
    \includegraphics[width = 1\linewidth]{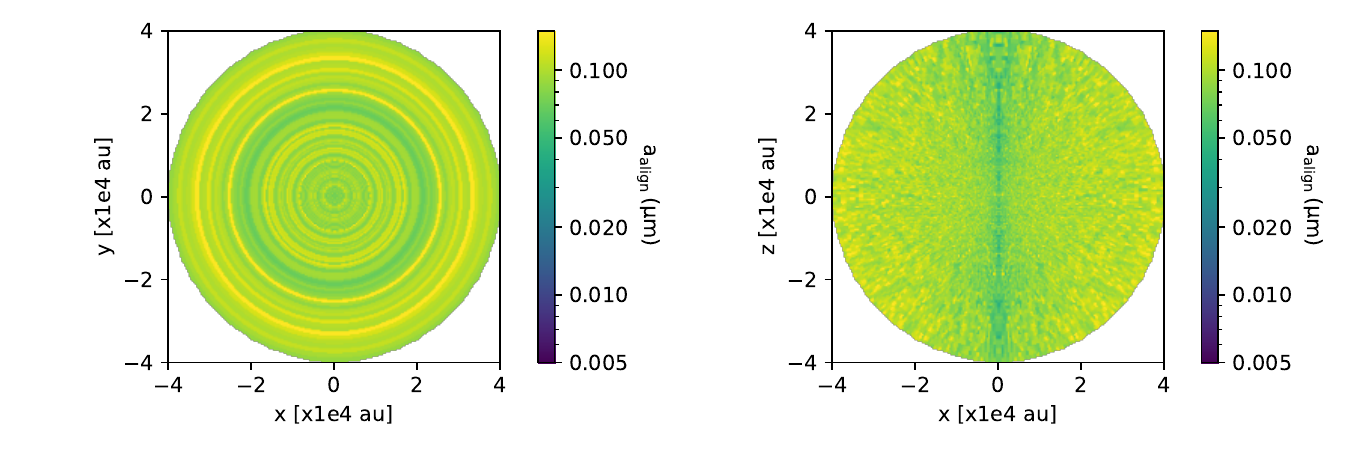}
    \includegraphics[width = 0.48\linewidth]{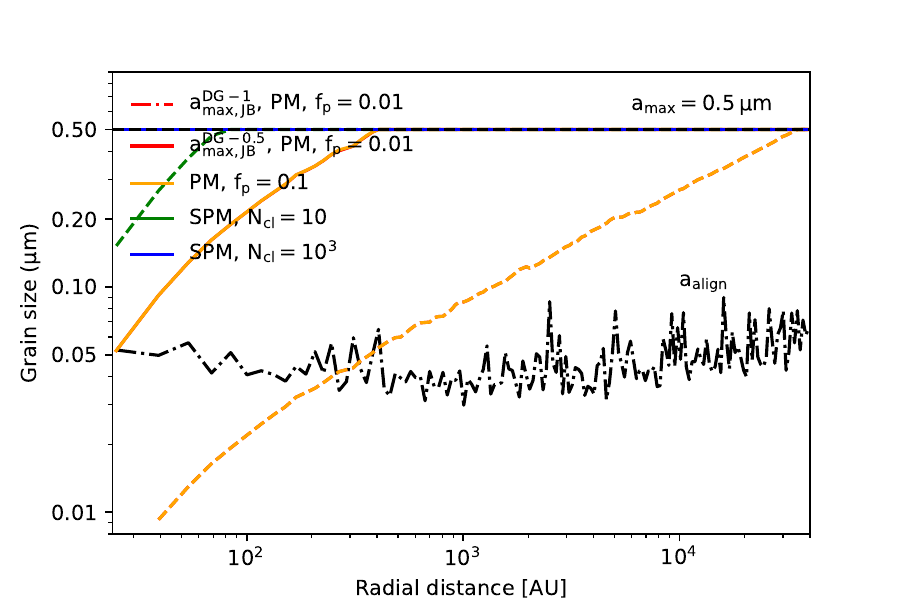}
    \caption{Upper panels: The distribution of aligned size $a_{\rm align}$ in $xy$- and $xz$- planes affected by the morphology of the toroidal field. The RAT alignment efficiency is reduced with large $a_{\rm align} \sim 0.05\,\rm\mu m$ due to the dominance of B-field components perpendicular to the radiation fields. Lower panel: The variation of the aligned size induced by RATs ($a_{\rm align}$, dashed black line) and the critical sizes having efficient alignment by MRAT ($a_{\rm max, JB}^{\rm DG-0.5}$ and $a_{\rm max, JB}^{\rm DG-1}$, color lines) with increasing envelope distance. The lower strength of the toroidal field strongly reduces the impact of magnetic relaxation. The MRAT alignment is then reduced in the inner envelope of $r < 1000\,\rm au$.}
    \label{fig:abar_maxJB_toroidal}
\end{figure*}

The upper panels of Figure \ref{fig:abar_maxJB_toroidal} show the maps of the minimum size of the external alignment by RATs $a_{\rm align}$ in both the $xy$- (left panel) and the $xz$-plane (right panel) under the impact of toroidal field. The lower panel of Figure \ref{fig:abar_maxJB_toroidal} shows the profiles of the minimum aligned size by RATs $a_{\rm align}$ (dashed black line) and critical grain sizes having external alignment by MRAT $a_{\rm max, JB}^{\rm DG-0.5}$ and $a_{\rm max, JB}^{\rm DG-1}$ (color lines). In terms of RAT alignment, the local magnetic fields are mostly perpendicular to the radiation fields (i.e., $\psi \sim 90^{\circ}$), which reduces the RAT alignment efficiency and results in $a_{\rm align} \sim 0.05\,\rm\mu m$. The MRAT mechanism, on the other hand, is insignificant due to the lower strength of the toroidal field in the inner envelope (Figure \ref{fig:Mag_geometry_toroidal}). Furthermore, the grain's magnetic susceptibility is suppressed by the strong thermal fluctuation of electrons inside grains with increasing dust temperature $T_{\rm d} > 100\,\K$ (Figure \ref{fig:Rad_fig}, see also in \citealt{Hoang2016a}; \citealt{Giang2023}), leading to the lower efficiency of magnetic relaxation and the loss of MRAT alignment toward the inner envelope $r < 1000\,\rm au$ with decreasing $a_{\rm max, JB}^{\rm DG-0.5}$ and $a_{\rm max, JB}^{\rm DG-1}$. The alignment of circumstellar grains by MRAT can be enhanced if they have high levels of iron inclusion $N_{\rm cl} > 10$ with $a_{\rm max, JB}^{\rm DG-1} = a_{\rm max, JB}^{\rm DG-0.5} = a_{\rm max} = 0.5\,\rm \mu m$ in the entire envelope.

\begin{figure*}
    \centering
    \includegraphics[width = 1\textwidth]{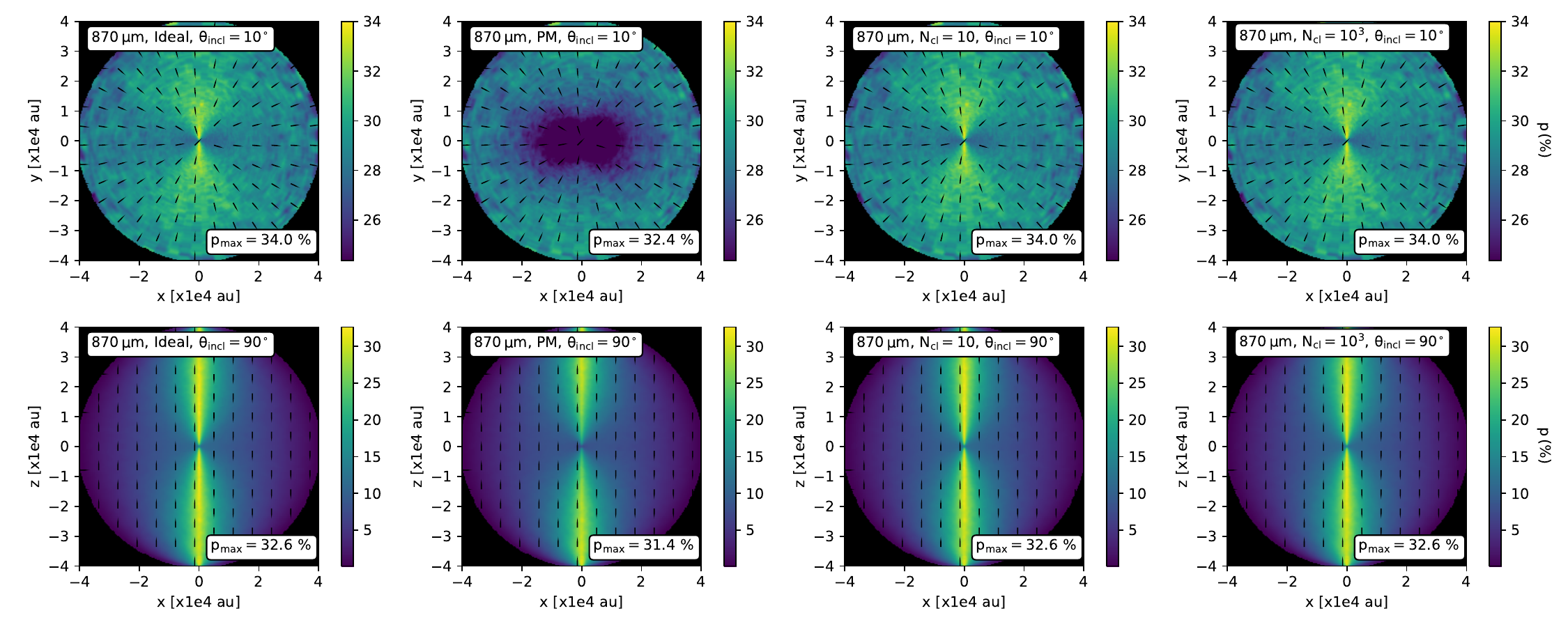}  
    \caption{Polarization maps at $870\,\rm\mu m$ produced by circumstellar dust with different magnetic properties on the large-scale of the envelope in the face-on (upper panels) and edge-on views (lower panels), indicating the toroidal field geometry through the polarization pattern with $\bf P \perp B$. The loss of MRAT alignment toward the inner envelope (see Figure \ref{fig:abar_maxJB_toroidal}) produces the `polarization hole' observed in the case of PM grains. The polarization degree is enhanced if grains contain high levels of iron inclusions.}
    \label{fig:Dustpol_mag_scale10000_toroidal}
\end{figure*}

Figure \ref{fig:Dustpol_mag_scale10000_toroidal} illustrates the large-scale maps of thermal dust polarization when the toroidal field is taken into consideration, assuming $\theta_{\rm incl} = 10^{\circ}$ (upper panels) and $\theta_{\rm incl} = 90^{\circ}$ (lower panels). In the face-on view, the polarization degree tends to have a uniform value of $\sim 40\,\%$ owing to the projection effect of the toroidal field lying to the plane-of-sky. In the edge-on view, the change in the curvature of the toroidal field lines results in a lower polarization degree to $\sim 5\%$ toward the outer region of $r > 10000\,\rm au$. Besides, the reduction of MRAT alignment for PM grains causes the loss of polarization degree in the inner region of $r < 500\,\rm au$ (i.e., `polarization hole' effect). The issue can be solved when grains are superparamagnetic with embedded iron inclusions.

\begin{figure*}
    \centering
    \includegraphics[width = 0.48\textwidth]{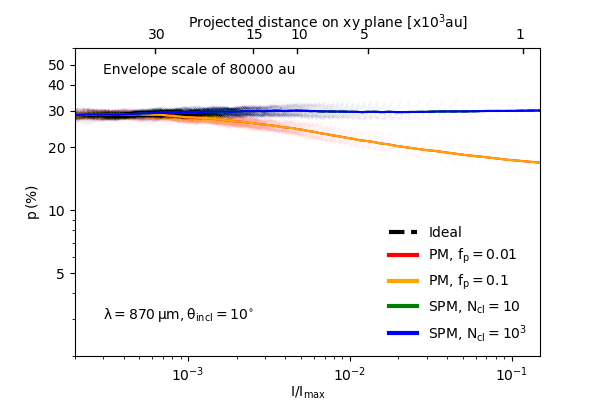} 
    \includegraphics[width = 0.48\textwidth]{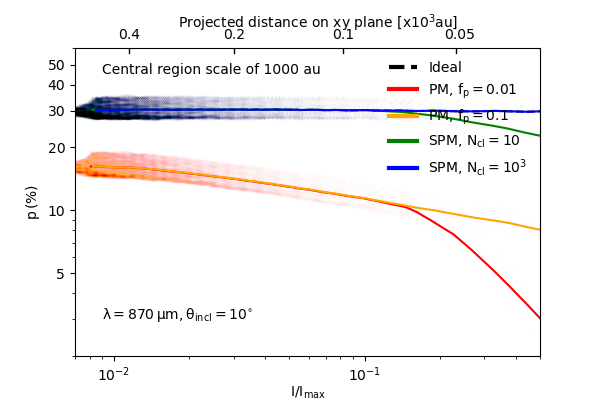}
    \includegraphics[width = 0.48\textwidth]{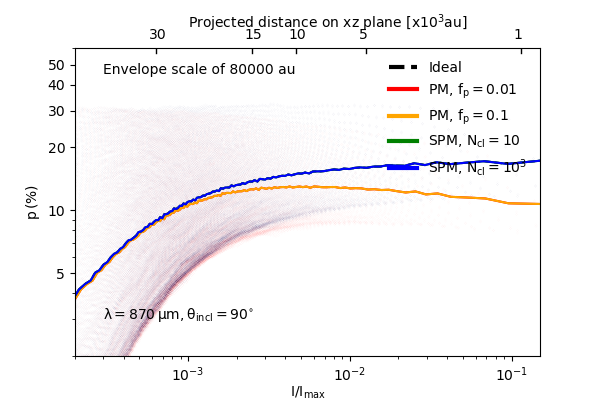} 
    \includegraphics[width = 0.48\textwidth]{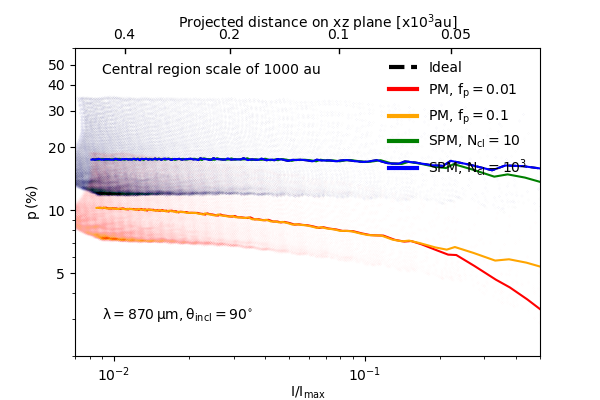}
    \caption{The mean polarization fraction $p\,(\%)$ with respect to the normalized intensity $I/I_{\rm max}$ calculating for the case of the toroidal field. The polarization degree is considerably affected by the projection of the toroidal field in the outer envelope of $d_{\rm proj} > 10000\,\rm au$. The polarization degree produced by PM grains decreases significantly in the inner region of $d_{\rm proj} < 1000\,\rm au$ because of the reduction of MRAT alignment. A higher abundance of iron inclusions in SPM grains can enhance the degree of polarization.}
    \label{fig:pvsI_nodisr_toroidal}
\end{figure*}

Figure \ref{fig:pvsI_nodisr_toroidal} shows the profile $p - I$ for the case of the toroidal field. Similar to Figure \ref{fig:pvsI_nodisr_fig}, the projection of the toroidal field in the plane-of-sky mostly affects the degree of dust polarization in the outer envelope of $d_{\rm proj} > 10000\,\rm au$. Meanwhile, the magnetic properties of grains impact the resulting polarization fraction in the inner region of $d_{\rm proj} < 500\,\rm au$, with a significant decrease to $p < 5\%$ owing to the loss of MRAT alignment efficiency for PM grains. The polarization degree is enhanced up to $\sim 20 - 30\%$ for SPM grains with increasing $N_{\rm cl}$.

\section{Synthetic polarization observations for different mass-loss rates}
\label{sec:appendix_massloss}
This section investigates the dependence of alignment properties and thermal dust polarization on the gaseous mass-loss rate $\dot{M}$ in evolved star's envelopes. We perform the numerical modeling of grain alignment and thermal dust polarization for different mass-loss rates, assuming the values of $\dot{M}$ followed by the original de Jager prescription as (see \citealt{deJager1988})
\begin{equation}
    \log{\dot{M}} = 1.769\log(L_{\ast}/L_{\odot}) - 1.676\log(T_{\rm eff}) - 8.158,
\end{equation}
where $L_{\ast}$ and $T_{\rm eff}$ are the stellar luminosity and the effective temperature, respectively. The mass-loss rate is positively correlated with the stellar luminosity with $\dot{M} \varpropto L_{\ast}^{1.769}$. We consider an increase in luminosity of evolved stars from $10^3$ to $10^5\,L_{\odot}$, corresponding to an increase in stellar mass-loss rate from $10^{-8}$ to $10^{-5}\, \rm M_{\odot} \, \rm yr^{-1}$.

\begin{figure*}
    \centering
    \includegraphics[width = 0.48\textwidth]{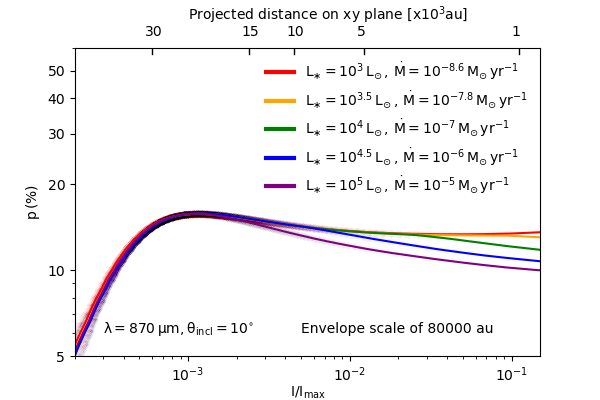} 
    \includegraphics[width = 0.48\textwidth]{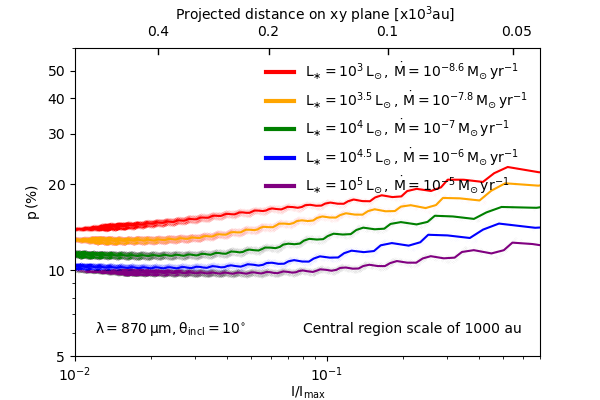}
    \caption{The averaged intensity-dependent polarization fraction $p - I/I_{\rm max}$ calculating for evolved stars with different stellar luminosities and mass-loss rates, considering they are positively correlated followed by the de Jager prescription (\citealt{deJager1988}). The thermal dust polarization is produced by PM grains with $f_{\rm p} = 0.1$. Under the effect of the gaseous mass-loss rate, the polarization degree is reduced due to the enhanced gas randomization with increasing mass-loss rate $\dot{M}$.}
    \label{fig:pvsI_Luminosity}
\end{figure*}

Figure \ref{fig:pvsI_Luminosity} illustrates the results of thermal dust polarization for different stellar luminosities and mass-loss rates, assuming the alignment of PM grains with $f_{\rm p} = 0.1$ and $\theta_{\rm incl} = 10^{\circ}$. The RAT efficiency is unchanged due to the competitive impact between the enhanced rotational damping by gas randomization with increasing $\dot{M}$ and the enhanced spin-up by RATs from stellar sources with higher luminosity $L_{\ast}$. However, the enhanced gas randomization reduces the internal alignment efficiency by Barnett relaxation and the external alignment efficiency by MRAT. As a result, they contribute to a lower alignment degree and produce a lower polarization degree by $\sim 10\%$.

% \begin{figure*}
%     \centering
%     \includegraphics[width = 0.48\textwidth]{AGB_POLARIS/Figure/Stellar_effect/align_Luminosity.pdf} 
%     \includegraphics[width = 0.48\textwidth]{AGB_POLARIS/Figure/Stellar_effect/aDGmax_Luminosity.pdf}
%     \caption{.}
%     \label{fig:abar_JB_Luminosity}
% \end{figure*}

\section{Effect of grain size distribution on the thermal dust polarization in AGB envelopes}
\label{sec:appendix_GSD}

\begin{figure*}
    \centering
    \includegraphics[width = 0.48\textwidth]{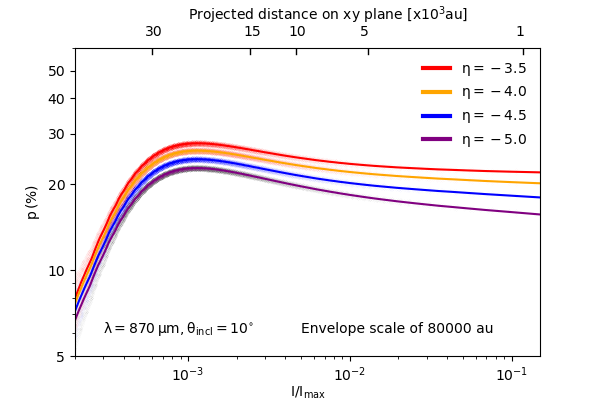} 
    \includegraphics[width = 0.48\textwidth]{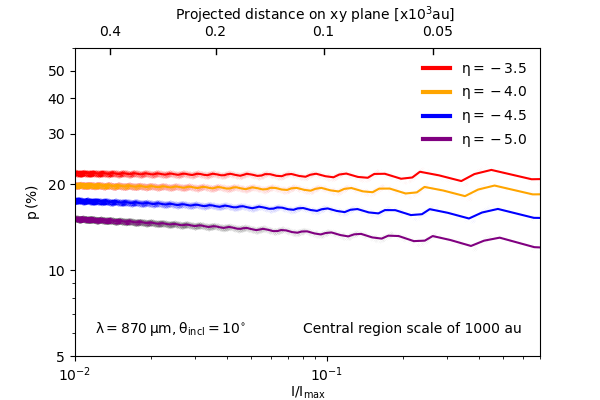}
    \caption{The resulting polarization degree $p(\%) - I/I_{\rm max}$ considering the variation of the grain size distribution from $\eta = -3.5$ to $\eta = -5$. The increasing abundance of smaller grains (i.e., smaller $\eta$) with lower RAT alignment efficiency leads to a lower thermal dust polarization degree in the entire envelope.}
    \label{fig:pvsI_GSD}
\end{figure*}

There are possibilities that the AGB circumstellar envelopes could originally contain a higher population of small grains with a steeper slope of GSD $\eta < -3.5$, rather than the MRN-like distribution as we adopted in the main numerical calculations (\citealt{Dominik1989}; \citealt{VandeSande2020}). Here, we present the dependence of thermal dust polarization degree at $870\,\rm\mu m$ on the GSDs, which is illustrated in Figure \ref{fig:pvsI_GSD} for the variation of the slope of grain size distribution from $\eta = -3.5$ to $\eta = -5$. The Ideal RAT alignment is taken into consideration. For GSDs with a higher abundance of small grains (i.e., smaller $\eta$), they have a lower RAT alignment efficiency. The total polarization degree is consequently reduced by a factor of 2 as the GSD slope $\eta$ decreases to $-5$.

Note that for circumstellar grains being close to the central AGB star, the original size distribution can be modified by the RAT-D effect with lower $\eta < -3.5$ and $a_{\rm max} < 0.5\,\rm\mu m$ (see in Figure \ref{fig:Disr_fig}). The enhancement of smaller grains by RAT-D contributes to the same effect of the reduction of the overall polarization degree in AGB environments (see Figure \ref{fig:Dustpol_RATD_10}, \ref{fig:Dustpol_RATD_90} and \ref{fig:pvsI_disr_incl10_fig}).

%%%%%%%%%%%%%%%%%%%%%%%%%%%%%%%%%%%%%%%%%%%%%%%%%%

% Don't change these lines
\bsp	% typesetting comment
\label{lastpage}
\end{document}